\documentclass{jfm}
\usepackage{amsmath}
\usepackage{natbib}
\usepackage{graphicx}
\usepackage{epstopdf, epsfig}
\usepackage{subfigure}
\usepackage{figsize}

\usepackage{soul}
\usepackage{xcolor}
\definecolor{change}{rgb}{0,0,1}

\newcommand{\corr}[1]{{\color{black}#1}} 

\usepackage{physics}

\shorttitle{Direct Calculation of Eddy Viscosity of Turbulent Channel Flow}
\shortauthor{D. Park and A. Mani}

\title{Direct Calculation of the Eddy Viscosity Operator in Turbulent Channel Flow at Re$_\tau$=180}

\author{Danah Park\aff{1}
  \corresp{\email{danah12@stanford.edu}}
 \and Ali Mani\aff{1, 2}}

\affiliation{
	\aff{1}Department of Mechanical Engineering, Stanford University, Stanford, CA 94305, USA
	\aff{2}Center for Turbulence Research, Stanford University, Stanford, CA 94305, USA
}

\begin{document}

\maketitle

\renewcommand{\thefootnote}{\textsuperscript{\textcolor{change}{\alph{footnote}}}}

\begin{abstract}

This study aims to quantify how turbulence in a channel flow mixes momentum in the mean sense. We applied the macroscopic forcing method (Mani and Park, Physical Review Fluids, 2021, p.054607) to direct numerical simulation (DNS) of a turbulent channel flow at Re$_\tau$=180 using two different forcing strategies that are designed to separately assess the anisotropy and nonlocality of momentum mixing. In the first strategy, the leading term of the Kramers-Moyal expansion of the eddy viscosity is quantified, revealing all 81 tensorial coefficients that essentially characterize the local-limit eddy viscosity. The results indicate: (1) the eddy viscosity has significant anisotropy, (2) Reynolds stresses are generated by both the mean strain rate and mean rotation rate tensors associated with the momentum field, and (3) the local-limit eddy viscosity generates asymmetric Reynolds stress tensors. In the second strategy, the eddy viscosity is quantified as an integration kernel revealing the nonlocal influence of the mean momentum gradient at each wall-normal coordinate on all nine components of the Reynolds stresses over the channel width. Our results indicate that while the shear component of the Reynolds stress is reasonably \corr{reproduced} by the local mean gradients, other components of the Reynolds stress are highly nonlocal. These results provide an understanding of anisotropy and nonlocality requirements for closure modeling of momentum transport in \corr{attached} wall-bounded turbulent flows. 

\end{abstract}


\section{Introduction}

Many of the turbulence models in use today are based on the Boussinesq approximation \citep{Boussinesq1877} in which the Reynolds stresses are assumed to be a linear function of the local mean velocity gradients. This approximation furthermore assumes isotropy of the tensor representing the coefficients of this linear relation, which is commonly referred to as eddy viscosity. The two simplifications offered by the Boussinesq approximation reduce the job of turbulence modeling to a determination of a scalar eddy viscosity field from which local Reynolds stresses can be determined algebraically without the need to solve any additional equations. For cases in which a single component of the Reynolds stress plays the dominant role, such as in parallel flows, a scalar eddy viscosity can be tuned to yield acceptable Reynolds stress fields \citep{Pope2001}. However, most turbulence models utilize this approximation even for multi-dimensional flows \citep{Spalart1994, Chien1982, Menter1994, Durbin1993, Hanjalic1972, Wilcox2008}. While some models allow anisotropic eddy viscosities \citep{Spalart2000, Mani2013, Rumsey2020}, they still retain the locality of the Reynolds stress dependence on the mean velocity gradient. 

Experimental measurements, as well as DNS data suggest that the isotropy and locality assumptions of the Boussinesq approximation are not strictly valid. Several studies have shown significant misalignment between the principal axis of the Reynolds stress and strain rate tensors indicating non-negligible anisotropy of the eddy viscosity operator \citep{Rogallo1981, Rogers1987, Champagne1970, Harris1977, Moin1982, Coleman1996}. Furthermore, the assumption of Reynolds stress locality is often not true because turbulent mixing may exist from the history of the straining in a given region of a turbulent flow. For instance, the experiment conducted by \citet{Warhaft1980} showed that the Reynolds stress can arise from the history effects of straining, even with a locally zero mean strain rate. In this case, the Reynolds stress should incorporate temporal or spatial nonlocality of the strain rate tensor.

Given these pieces of evidence, various modeling techniques have attempted to relax both locality and isotropy assumptions via development of second-order closure models \citep{Wilcox1998, Speziale1991, Cecora2015, Launder1975, Gerolymos2012} often using the Reynolds stress transport equation as a framework to identify the needed closures. Each of these models, provides a specific way in which Reynolds stresses could depend nonlocally or anisotropically on the velocity gradient field. However, standard data of turbulent flows, either from DNS or experiments, does not provide sufficient information to allow proper discrimination between these models. While these data reveal anisotropy of the Reynolds stresses, they do not uniquely determine the anisotropy or nonlocality of the closure operators that express their dependence on the mean velocity gradient. Closing this gap would require quantification of the eddy viscosity as an operator acting on the mean velocity gradient. With this goal in mind, this study presents a direct quantification of the eddy viscosity operator in a canonical turbulent flow via utilization of the macroscopic forcing method (MFM), developed by \citet{Mani2021}.

Prior to the description of our work, we start by reviewing generalized forms of the eddy diffusivity and eddy viscosity operators for scalar and momentum transport in turbulent flows. Firstly, one way of generalizing the Boussinesq approximation is to allow for the anisotropy of the eddy viscosity. \citet{Batchelor1949} suggested using a second-order tensor replacing the diffusion coefficient in the Fickian model to describe the mean transport of a scalar quantity. Later, a similar concept was suggested by \citet{Rogers1989}, where the mean turbulent flux of a passive scalar was approximated with an algebraic model expressed in a second-order tensor eddy diffusivity. This anisotropic eddy diffusivity model can be written as the following: $-\overline{u'_i c'}=D_{ij} \pdv*{C}{x_j}$ where $\overline{(\cdot)}$ represents ensemble-average, $u'_i$ represents the fluctuation of the velocity, $C$ and $c'$ represent the mean and the fluctuation of the scalar quantity being transported, $x_i$ represents the \corr{spatial} Cartesian coordinate, and $D^0_{ij}$ represents the second-order eddy diffusivity tensor that is local. 

Similarly, for the turbulent momentum flux, one method of generalizing the Boussinesq approximation is to use a tensorial representation of the eddy viscosity. \citet{Hinze1959} has suggested the use of the fourth-order tensor as the eddy viscosity. Later, \citet{Stanivsic1965} conducted a systematic investigation of the tensorial character of the eddy viscosity coefficient and revealed that the eddy viscosity tensor has to be at minimum fourth-order. In parallel to the anisotropic eddy diffusivity model, the anisotropic eddy viscosity model for momentum transport can be written as: $-\overline{u'_i u'_j} = D^0_{ijkl} \pdv*{U_l}{x_k}$ where $U_l$ represents mean velocity field. Here, the Reynolds stresses $\overline{u'_i u'_j}$ is locally closed in terms of the fourth-order tensorial eddy viscosity $D^0_{ijkl}$ and the mean velocity gradient.

An even more general form of the eddy viscosity can be used to incorporate not only anisotropy but also nonlocality. \citet{Hamba2005, Hamba2013} suggested writing the closure of the Reynolds stress in terms of the mean velocity gradient at remote times and locations. This form of eddy viscosity involves a fourth-order tensorial kernel, which we refer to as the eddy viscosity kernel. For statistically stationary flows, this relation can be expressed as 
\begin{equation}
    -\overline{u_i'u_j'} (\vb{x})
        = \int D_{ijkl}(\vb{x}, \vb{y}) \left. \pdv{U_l}{x_k} \right\vert_{\vb{y}} \dd[3]{\vb{y}},
    \label{eq:generalform}
\end{equation}
where $D_{ijkl}(\vb{x}, \vb{y})$ is the eddy viscosity kernel indicating how mean gradients at location $\vb{y}$ result in Reynolds stresses at location $\vb{x}$. \corr{When written in dimensional form, the eddy viscosity kernel does not have the same dimension as the kinematic viscosity. Instead, the kernel represents increment of viscosity per unit volume of the nonlocality dimension. In the case of Equation~\ref{eq:generalform} temporal nonlocality is not considered due to the system's statistical stationarity, but full three dimensional spatial nonlocality is considered. As a result, the dimensional kernel would have unit of diffusivity per unit volume or $m^{-1}s^{-1}$}.

\citet{Hamba2005} reported the first quantification of the eddy viscosity kernel for a turbulent channel flow using a Green's function formulation approach based on an earlier work by \citet{Kraichnan1987}. However, their study focuses on a subset of the tensorial coefficients, i.e., $D_{ij21}$. This choice is motivated since the mean velocity profile in the channel flow is insensitive to other components of the eddy viscosity kernel, given the mean velocity gradient, $\partial U_l / \partial x_k$ shown in Equation~\ref{eq:generalform}, is nonzero only for $(k,l)=(2,1)$. Nevertheless, quantification of other components of eddy viscosity in this canonical setting would provide significant insights about momentum mixing in the broader context of wall-bounded shear flows. Aside from this shortcoming, \citet{Hamba2005} chose to manually enforce the symmetry $D_{ijkl}=D_{jikl}$ by performing arithmetic averaging of the respective components (i.e., $ij$ and $ji$) of the output data from their simulations. This choice was made given the expectation that the Reynolds stress tensor as the output of Equation~\ref{eq:generalform} must always be symmetric, while the raw kernels did not follow this symmetry. 

Recently, \citet{Mani2021} presented an alternative interpretation of Equation~\ref{eq:generalform} in the context of the generalized momentum transport (GMT) equation. GMT can be derived by applying the Reynolds Transport Theorem to momentum transport without constraining the momentum field to be identical to the velocity field. In this context, the Reynolds stress, expressed as $\overline{u_i'v_j'}$, is interpreted as the mean product of two conceptually different fields, with $u_i$ representing the kinematic displacement of volume acting as a transporter of momentum, and $v_j$ representing momentum per unit mass, the quantity of interest that results in friction and pressure. Navier-Stokes (NS) is rendered as a special solution to GMT in which the two fields are constrained to be equal. Specifically, when GMT is supplied with the same boundary conditions and forcing conditions as those in NS, the solution to NS is the only attractor solution to GMT, as shown theoretically and numerically by \citet{Mani2021}. With this interpretation, Equation (\ref{eq:generalform}) is in fact a closure operator to the ensemble-averaged GMT and not the Reynolds Averaged Navier-Stokes (RANS) equation. Therefore, $D_{ijkl}$ and $D_{jikl}$ are not required to be equal, since $\overline{u_i'v_j'}\ne \overline{u_j'v_i'} $. The present study addresses this issue, by examining the raw eddy viscosity operator without any symmetry averaging. We confirm that while the eddy viscosity kernel of channel flow is not symmetric, it still results in symmetric Reynolds stresses when it acts on the mean velocity gradient of the same flow \corr{from which the eddy viscosity data are obtained.}

As previous mentioned, \cite{Mani2021} provide a statistical technique called the macroscopic forcing method (MFM), which allows direct measurement of a flow's eddy viscosity $D_{ijkl}$ with data gathered from direct numerical simulations (DNS) of the Navier-Stokes equation and GMT. More generally speaking, MFM allows precise computation of RANS closure operators via applying various macroscopic forcing to the GMT equations which can be utilized to extract the eddy viscosity operator. It is worth noting that macroscopic forcing is not limited to delta functions, which reveal Green's functions as outputs. For instance, \citet{Shirian2022} employed harmonic forcing to efficiently unveil the eddy diffusivity operator for homogeneous isotropic turbulence. They successfully fitted this operator with an analytical expression. An alternative approach by \citet{Mani2021} , which is more relevant to this study, is the inverse macroscopic forcing method (IMFM), in which forcing to constrain mean polynomial fields was shown to reveal nonlocal moments of the underlying eddy diffusivity operator in an economical way compared to the Green's function approach. 
We examine a systematic procedure for obtaining a local operator approximation of the full eddy viscosity operator by considering a Kramers-Moyal expansion \citep{Van1992} of the eddy viscosity operator and quantifying its leading term. This approach does not only enable estimation of the eddy viscosity in an economical fashion, but it also separates out the easy-to-comprehend local eddy viscosity by utilizing this established expansion, which we believe was a missing piece in the analysis of \citet{Hamba2005}. 

The rest of this paper is organized as follows. In Section 2, we define the flow system and the model used which involves the fourth-order tensorial eddy viscosity kernel, and review the computational methodology. In Section 3, we begin with evaluating the isotropy assumption in the Boussinesq's approximation. For simplicity, we conduct the leading-order (local-limit) approximation to the eddy viscosity kernel to solely focus on the anisotropy of the eddy viscosity. With the measured local eddy viscosity tensor, we discuss the following: the standard eddy viscosity, the quantified anisotropy, the dependency of Reynolds stress on the rate of rotation, the leading-order Reynolds stress, and the positive definiteness of the leading-order eddy viscosity operator. In Section 4, we extend our study to nonlocal effects, by computing the full eddy viscosity kernel representing the nonlocal effects in the wall-normal direction. In Section 5, we summarize our results and discuss potential extensions to this study.

\section{Problem Setup and Governing Equations}



\begin{figure}
  \vspace{0.3cm}
  \centerline{\includegraphics[width=0.5\linewidth]{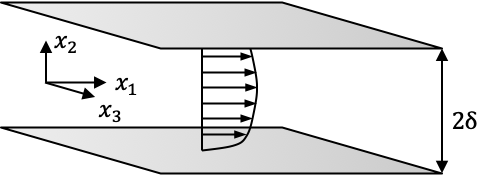}}
  \caption{Schematics of the channel flow.}
\label{fig:channel}
\end{figure}

Figure~\ref{fig:channel} shows the schematics of the channel flow and its coordinate system where the flow is bounded by top and bottom walls spaced $2\delta$ apart. We denote the Cartesian coordinates $x_i$, where $x_1$ is the streamwise direction, $x_2$ is the wall-normal direction, and $x_3$ is the spanwise direction. The dimensionless equations expressing mass and momentum conservation are as follows:
\begin{gather}
    \pdv{u_i}{t} + \pdv{u_j u_i}{x_j} = -\pdv{p}{x_i}+\frac{1}{\mathrm{Re}} \pdv{u_i}{x_j}{x_j} + r_i ,
    \label{eq:NS}   \\
    \pdv{u_j}{x_j} = 0 ,
    \label{eq:continuity}
\end{gather}
where $u_i$ is the flow velocity, $p$ is the pressure normalized by the density, $t$ is time, and $r_i=(1,0,0)$ represents normalized mean pressure gradient. The dimensionless spatial coordinates are normalized by $\delta$, and $\mathrm{Re}$ represents the Reynolds number defined based on $\delta$ and the friction velocity $u_\tau = \sqrt{\tau_w / \rho}$ where $\tau_w$ is the mean wall shear stress balancing the force due to the mean pressure gradient and $\rho$ is the fluid density.



The RANS equations can be obtained by taking the ensemble-average of Equations~\ref{eq:NS} and \ref{eq:continuity}, yielding:
\begin{gather}
    \pdv{U_i}{t} + \pdv{U_j U_i}{x_j} = -\pdv{\overline{p}}{x_i} + \frac{1}{\mathrm{Re}} \pdv{U_i}{x_j}{x_j} - \pdv{\overline{u'_j u'_i}}{x_j} + \overline{r}_i,
    \label{eq:RANS}     \\
    \frac{\partial U_j}{\partial x_j}=0,
    \label{eq:continuityRANS}
\end{gather}
where $U_{i}$ is the mean velocity, $u'_{i}$ is the velocity fluctuation around the mean velocity, and $\overline{(\cdot)}$ implies ensemble-averaged quantities. To close this system, the divergence of Reynolds stresses, $\pdv*{\overline{u'_j u'_i}}{x_j}$, needs to be modeled in terms of the primary variable $U_i$. This can be generally expressed as an operator acting on the ensemble-averaged field, $- \pdv*{\overline{u'_j u'_i}}{x_j} \equiv  \overline{\mathcal{L}}\left( U_i \right)$. One form of such operators is expressed in Equation (\ref{eq:generalform}).

A DNS solution to the channel flow does not provide enough information to fully quantify the nonlocal eddy viscosity kernel, $D_{ijkl}(\vb{x}, \vb{y})$. A full characterization of $D$ requires quantification of Reynolds stresses in response to all possible independent flow gradients scenarios. Following \citet{Mani2021}, we next describe the procedure of obtaining $D$. 
In this paper, however, we limit the scope of our analysis to the one-dimensional RANS context, in which the wall-normal coordinate is the only independent variable since the flow is statistically homogeneous in all other space-time coordinates. 
In other words, we assume the form: $D_{ijkl}  = D_{ijkl} (x_2, y_2)$, and the other dimensions are integrated out in Equation~\ref{eq:generalform}. However, the employed macroscopic forcing methodology is in principle generalizable to multi-dimensional cases, and with higher computational expense can capture the full behavior of $D_{ijkl} (\vb{x}, \vb{y})$.


\subsection{Macroscopic Forcing Method}

In this section, we discuss details on how to use MFM to measure the eddy viscosity, starting from the generalized momentum transport (GMT) equations. 



\subsubsection{Generalized Momentum Transport equation}

In our earlier work, which was mainly on the transport of passive scalars, we briefly introduced how one can apply MFM to analyze momentum transport \citep{Mani2021}. To quantitatively determine the eddy viscosity operator, one first needs the detailed velocity field of the specific flow of interest. One method of obtaining such velocity fields is to perform a DNS simulation, which we call the donor simulation, as it donates a velocity field whose eddy viscosity is to be determined. 


To analyze momentum transport by a given flow, we will now consider GMT, which can be derived from the Reynolds Transport Theorem for a fluid system with a Fickian model for molecular viscosity.
\begin{gather}
    \pdv{v_i}{t} + \pdv{u_j v_i}{x_j} = -\pdv{q}{x_i} + \frac{1}{\mathrm{Re}} \pdv{v_i}{x_j}{x_j} + s_i,
    \label{eq:GMT}  \\
    \pdv{v_j}{x_j}=0.
    \label{eq:continuityGMT}
\end{gather}
where $v_i$ represents momentum per unit mass, and is considered to be different from $u_i$, the donor velocity field. \corr{$s_i$ is the macroscopic forcing}. Also, $q$ is the generalized pressure to ensure the incompressibility of the \corr{momentum} field $v_i$.

Equations~\ref{eq:GMT} and \ref{eq:continuityGMT} then describe a passive solenoidal vector field that is transported by the background velocity field $u_j$ governed by Equation~\ref{eq:NS}. An advantage of working with GMT, as opposed to NS, is its linearity with respect to the transported quantity, $v_i$. Under such conditions, expressing the generalized eddy viscosity in the format given by Equation~\ref{eq:generalform} becomes meaningful. As discussed by \citet{Mani2021}, GMT spans a larger solution space than NS; NS is a special subset of the GMT space where $v_i=u_i$.

An important question that naturally follows is whether the computed RANS operator of GMT is the same as that of the NS equation. In our earlier work, we already showed analytically and numerically that the macroscopic operators of the GMT and NS equations are identical \citep{Mani2021}. \corr{In brief}, we showed that the solutions of GMT and NS equations become microscopically the same after sufficient time regardless of the initial conditions when we apply the same boundary conditions to both equations. The time scale at which the solutions become identical was found to be $\tau_\textrm{mix} = 16.6 \delta / u_\tau$ for a turbulent channel flow. Therefore, it is justified that the macroscopic operator of the GMT equation obtained by MFM is the same as the RANS operator of the NS equations. In sum, GMT works as an auxiliary set of equations that probes RANS operator of NS and therefore we can obtain eddy viscosity of the RANS equations by investigating that of the GMT equations.

It is important to note that \citet{Hamba2005} wrote an equation very similar to GMT equations in spite of taking a conceptually different derivation path. His passive vector equation is indeed GMT subtracted by the mean of GMT. The main difference lies in the explicit inclusion of forcing in the equations, allowing for a general macroscopic field. In contrast, \citet{Hamba2005} implicitly applies forcing by specifically considering Dirac delta function mean fields.

\subsubsection{Analysis Strategy}

We aim to study two aspects of the eddy viscosity kernel in a turbulent channel flow: the anisotropy and the nonlocality. To fully investigate such non-Boussinesq effects, it is ideal to compute every value of the full eddy viscosity kernel $D_{ijkl}$ in Equation \ref{eq:generalform}. Since the channel flow is homogeneous in $x_1$ and $x_3$ directions and statistically stationary, we integrate the mixing effect in these directions. The simplified Reynolds stress for GMT variables can be expressed as:

\begin{eqnarray}
    -\overline{u_i'v_j'}(x_2)&&=\int D_{ijkl}({x_2}, {y_2})\left.\frac{\partial V_l}{\partial x_k} \right\vert_{{y_2}}\mathrm{ d}y_2.
    \label{eq:generalformchannel}
\end{eqnarray}

Equation \ref{eq:generalformchannel} incorporates anisotropy via tensorial representation and nonlocality via the integration form. MFM has the capability to compute all the elements in the eddy viscosity kernel $D_{ijkl}\left(x_2,y_2\right)$ by tracking the influence of each entry of $dV_l/dx_k$ on the entire Reynolds stress field. It has been demonstrated by \citet{Liu2021} that such a brute force approach is theoretically equivalent to Hamba's Green's function approach \citep{Hamba2005}.

However, one caveat is that the cost of each simulation is significant and consequently it is not desirable to conduct a full nonlocal MFM analysis. To conduct computation for $D_{ijkl}$ for given $k$ and $l$, one requires as many DNS simulations as the number of degree of freedom of the RANS space. Therefore, to reduce the cost of the analysis, we conduct two separate analyses for the anisotropy and nonlocality, both using MFM.

First, we focus on studying the anisotropic nature of the eddy viscosity. However, to focus exclusively on anisotropy, we systematically construct a local approximation of the eddy viscosity operator using the Kramers-Moyal expansion \citep{Van1992}, as investigated by \citet{Mani2021}. For instance, in a parallel flow where $dV_1/dx_2$ is the only active component of the velocity gradient, the Reynolds stress $\overline{u_2^\prime v_1^\prime}$ in Equation~\ref{eq:generalformchannel} can be written as the integral of only $D_{2121}$ component of the eddy viscosity. By considering a Taylor series expansion of $dV_1/dx_2$ around $y_2=x_2$, one can re-express the eddy viscosity operator in terms of the following expansion:
\begin{align}
    -\overline{u_2'v_1'}(x_2)
        &= \int D_{2121}({x_2}, {y_2}) \left. \pdv{V_1}{x_2} \right|_{y_2} \dd{y_2} \label{eq:leadingApprox0}   \\
        &= \int D_{2121}({x_2}, {y_2}) \left( \left. \pdv{V_1}{x_2} \right|_{x_2}
            + (y_2-x_2) \left. \pdv[2]{V_1}{x_2} \right|_{x_2} + \cdots \right) \dd{y_2} \\
        & = \sum_{n=0}^{\infty} D^n_{2121}({x_2}) \pdv{^{n+1} V_1}{x_2^{n+1}}
    \label{eq:leadingApprox}
\end{align}
where $D^n_{2121}=\int{D_{2121}(x_2,y_2)(y_2-x_2)^n/n!\dd y_2}$ represents the $n$-th spatial moment of the eddy viscosity kernel.

As discussed by \citet{Mani2021}, the leading term in this expansion encapsulates the local limit eddy viscosity while the subsequent terms characterize finite moments associated with the nonlocal effects. The general form of this leading-order approximation for all components of the Reynolds stress and mean velocity gradient is as below:
\begin{equation}
    -\overline{u_i'v_j'} (x_2) = D^0_{ijkl} (x_2) \pdv{V_l}{x_k},
    \label{eq:leadingApproxGeneral}
\end{equation}
where $D^0_{ijkl}(x_2)$ is called the leading-order eddy viscosity tensor:
\begin{equation}
    D^0_{ijkl}(x_2)=\int{D_{ijkl} \dd y_2}.
    \label{eq:leadingordertensor}
\end{equation}
Equation~\ref{eq:leadingApproxGeneral} would be exact only when $D_{ijkl}(x_2, y_2)$ is local, i.e. $D_{ijkl}(x_2, y_2) = D^0_{ijkl}(x_2) \delta(y_2 - x_2)$ where $\delta(x)$ is a Dirac delta function.

The local eddy viscosity in Equation~\ref{eq:leadingApproxGeneral} is no longer a scalar value varying in space; it is a fourth-order tensor with 81 coefficients. The tensor representation was suggested by previous researchers including \citet{Batchelor1949}, but the full quantification has not been conducted to the authors' knowledge. As presented in Appendix B, using only 9 MFM simulations, we computed all 81 coefficients of the eddy viscosity tensor. The resulting tensor elements are provided in Appendix C.

The next investigation focuses on the nonlocality of the eddy viscosity. As conducting MFM to measure the full kernel can be costly for complex turbulent flow systems, we focus on calculating a subset of tensorial kernel components, specifically the kernel components that are multiplied to $\partial V_1/\partial x_2$ in Equation \ref{eq:generalformchannel}. The computed tensorial kernel components are $D_{ij21}\left(x_2,y_2\right)$ and they are associated with the Reynolds stresses which correspond to the velocity gradient $\partial U_1/\partial x_2$, the only velocity gradient appearing in the RANS closure for a channel flow. The detailed steps on how to measure eddy viscosity kernel using MFM is discussed in Appendix E.


\subsubsection{Application of Macroscopic Forcing Method}
\label{sec:mfm}

\begin{figure}
  \vspace{0.3cm}
  \centerline{\includegraphics[width=0.9\linewidth]{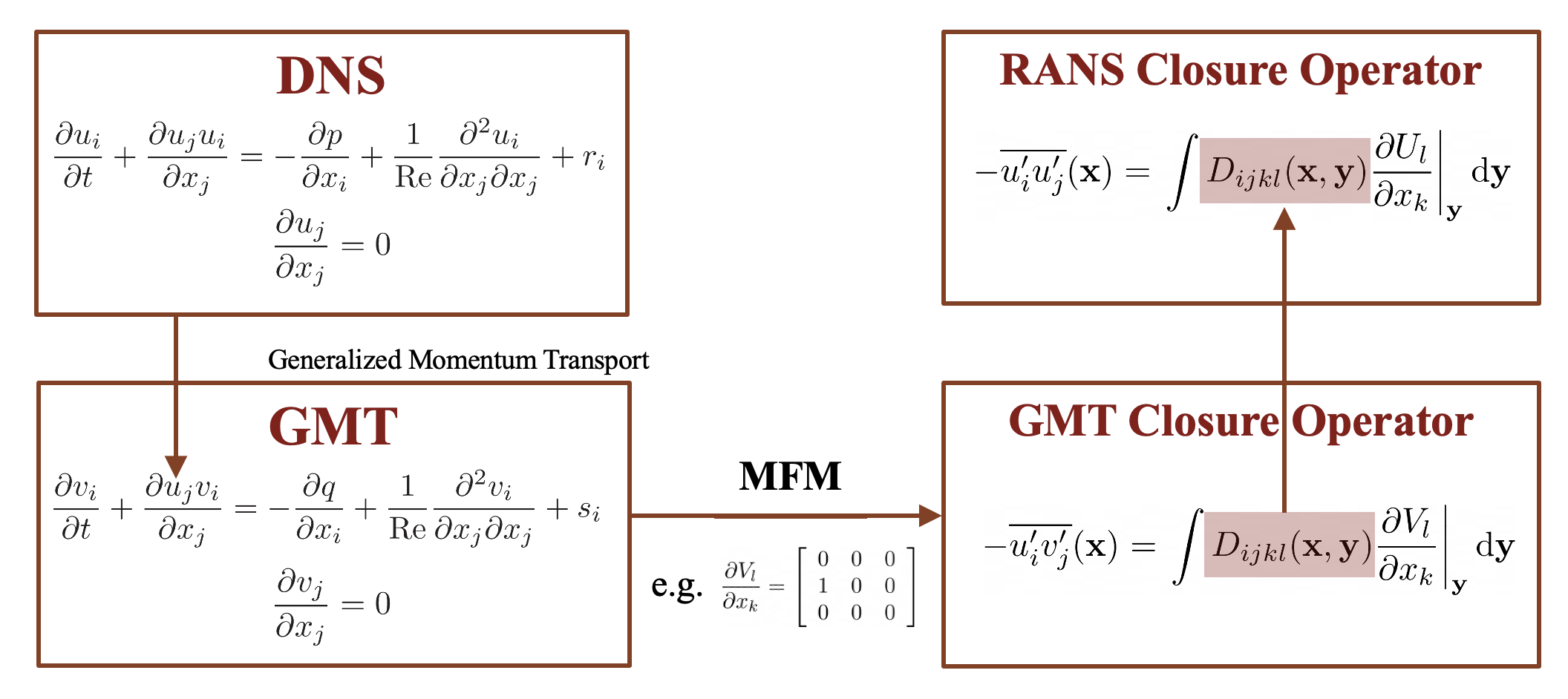}}
  \caption{Schematics of the MFM analysis.}
\label{fig:mfmanalysis}
\end{figure}

The next step involves how we actually compute the leading-order eddy viscosity tensor and the eddy viscosity kernel. Figure~\ref{fig:mfmanalysis} illustrates how we conducted our MFM analysis. To apply MFM, we start with two sets of solvers: one for the NS equations and the other for GMT. At each time step, we solve the NS equation to obtain the velocity field $u_i$ and feed it as the advecting velocity to the GMT solver. For the GMT equations, we force the Reynolds-averaged GMT variable $V_i$ to be a specific value in order to acquire certain information about the eddy viscosity. A forcing field $s_i$ that results in $V_1=x_2$ and $V_2=V_3=0$ generates GMT data from which we can extract the leading-order eddy viscosity $D^0_{ij21}$. More specifically, $D^0_{2121}$ can be obtained by post-processing $\overline{u'_2 v'_1}$ from this GMT simulation, and re-evaluating Equation~\ref{eq:leadingApprox0} - \ref{eq:leadingApprox} to observe:
\begin{equation}
    -\overline{u'_2 v'_1}(x_2) = \int D_{2121}({x_2}, {y_2}) \left. \pdv{V_1}{x_2} \right|_{y_2} \dd{y_2}
        \stackrel{V_1 = x_2}{=} \int D_{2121}({x_2}, {y_2}) \dd{y_2}
        = D^0_{2121}(x_2)
    \label{eq:MFMexample}
\end{equation}

As discussed, the macroscopic operator of the Reynolds-averaged GMT and the RANS operator of NS are identical. Therefore, $D^0_{2121}$ corresponds to the standard eddy viscosity $\nu_T$ used in the Boussinesq approximation in RANS models. Likewise, we can compute other components of the leading-order eddy viscosity tensor using different selections of the macroscopic forcing field, $s_i$, such that other components of the mean velocity gradient are activated. 

Additionally, the same setup shown in Figure~\ref{fig:mfmanalysis}, can be used to compute the full kernel of eddy viscosity. The main difference is to apply macroscopic forcings that would generate mean \corr{gradient} fields in the form of Dirac delta functions. For example, a macroscopic field, $s_i$, that sustains \corr{$\partial V_1/\partial x_2=\delta(x_2-y_2^*)$}, would result in GMT data from which we can extract \corr{$D_{ij21}(x_2,y_2^*)$}, by merely post processing the 
$\overline{u'_i v'_j}$ data. This specific choice of forcing would result in data similar to those obtained by \citet{Hamba2005}, with the difference that Hamba used only the symmetric portion of the momentum flux tensor in order to ensure symmetry of the Reynolds stresses. As we shall see, GMT does not produce symmetric eddy viscosity kernels and thus $D_{ijkl} \ne D_{jikl}$. This is intuitively understandable noting that $D_{ijkl}$ quantifies the rate of mixing of the mean $j$-momentum by the $i$-component of the velocity fluctuations while $D_{jikl}$ quantifies the rate of mixing of the mean $i$-momentum by the $j$-component of the velocity fluctuations. Since in this framework, momentum and velocity fields can be quantitatively different, the symmetry does not hold. Likewise, this asymmetry propagates to the Kramers-Moyal expansion of the eddy viscosity operator, and as we shall see, even the leading-order eddy viscosities are not symmetric. 

Lastly, we note that the macroscopic forcing procedure used in this work is an inverse forcing method as discussed by \citet{Mani2021}, since we explicitly set the desired mean momentum field $V_i$ for each GMT simulation, as opposed to setting the macroscopic forcing field. 

\subsection{Simulation Setup}
We adapt MFM solver to a three-dimensional incompressible NS solver originally developed by \citet{Bose2010} and modified by \citet{Seo2015}. The present DNS uses the fractional step method with semi-implicit time advancement \citep{Kim1985}. For the temporal difference scheme, we use second order Crank-Nicholson for the wall-normal diffusion and Adams-Bashforth for the rest of the terms. The solver uses a second-order finite spatial discretization on a staggered mesh \citep{Morinishi1998}. Also, we use a uniform grid in the streamwise and spanwise directions and grid-stretching in the wall-normal direction. The domain is periodic both in the spanwise and the streamwise directions, and the no-slip boundary condition is applied at the two walls. 

The numerical setup for the GMT solver is almost identical to that of DNS, except for two differences. The first is that GMT obtains the background velocity from the NS solver at every time step. The other difference is that GMT utilizes macroscopic forcing, in order to maintain a desired macroscopic momentum field $V_i(x_2)$. To be most rigorous, the selected macroscopic forcing, $s_i(x_2)$, must be independent of time. Likewise, the resulting mean velocity field needs to match the pre-set $V_i(x_2)$ only after time averaging. However, constraining the simulations in this fashion, would require expensive iterations over which the entire simulation must be repeated after each adjustment of $s_i(x_2)$. To avoid this cost, in our implementation, we computed ensemble averages by averaging fields only in the $x_1$ and $x_3$ directions, and we constrained $s_i(x_2)$ at each time step such that $V_i(x_2)$ is matched to the pre-set $V_i(x_2)$. \corr{In other word, one can set $s_i(x_2)$ such that the homogeneous plane average of the temporal term ${\partial v_i}/{\partial t}$ in Equation~\ref{eq:GMT} becomes zero at each timestep.}

However, in this implementation the resulting $s_i(x_2)$ is not perfectly time independent. Due to finite number of samples per time step, fluctuations in time are observed. One remedy to reduce these fluctuations is to increase the number of samples by selecting a longer domain in the $x_1$ and $x_3$ directions. We have performed such domain convergence studies in Appendix A indicating the adequacy of the selected domain size in our MFM analysis. 

There are two sets of forcings for MFM presented in this paper, each corresponding to the analysis of anisotropy and the nonlocality of eddy viscosity (Table~\ref{table:setup}). Within each set, multiple simulations are performed where the macroscopic forcings are varied to reveal different components of the eddy viscosity. The first set uses GMT simulations under different macroscopic forcings to reveal the leading-order eddy viscosity tensor $D^0_{ijkl}$. We utilize these measurements to understand the anisotropy of the eddy viscosity. The second set probes a subset of the entire eddy viscosity kernel, $D_{ij21}$, which quantifies the nonlocality of the eddy viscosity in response to the most significant velocity gradient $\partial U_1 / \partial x_2$. In addition to the analysis method and the resulting eddy viscosity, Table~\ref{table:setup} presents the number of total DNSs in each set, the domain size, the spatial resolution, and the sampling times. For the first set, only nine DNSs are needed corresponding to $k, l \in \{1, 2, 3\}$, and for the second set, MFM analyses require a set of simulations with the number of the macroscopic degrees of freedom. The results of each set are discussed in Sections 3 and 4 respectively, and the detailed simulation setup of each set is discussed in Appendices B and C, respectively. Also, the measured eddy viscosities, $D^0_{ijkl}$ and $D_{ij21}$, are provided as the supplementary data.

\begin{table}
\centering
\begin{tabular}{ c c c c c c }
    Analysis   & Eddy Viscosity    & Number of DNS's   & $L_1 \times L_2 \times L_3$   & $N_1 \times N_2 \times N_3 $   & $T u_\tau/L_2$    \\ 
    Anisotropy     & $D^0_{ijkl} (x_2)$    & 9     & $2 \pi \times 2 \times \pi$   & $144\times144\times144$    & 750    \\  
    Nonlocality   & $D_{ij21} (x_2, y_2)$ & 145   & $2 \pi \times 2 \times \pi$   & $144\times144\times144$    & 500    \\
\end{tabular}
\caption{\label{table:setup} Simulation setup for anisotropy analysis and nonlocality analysis of the eddy viscosity. $L_1 \times L_2 \times L_3 $ is the domain size, $N_1 \times N_2 \times N_3 $ is the number of grid, and $T u_\tau/L_2$ is the total simulation time in wall units. }
\end{table}

\section{Anisotropy Analysis}

In this section, we compute the leading-order eddy viscosity tensor $D^0_{ijkl}$ and focus on the analysis on anisotropy of the eddy viscosity and specifically contrast it to the standard eddy viscosity implied by the Boussinesq model. In addition, we assess dependency of Reynolds stresses on the rate of rotation, examine reconstruction of Reynolds stresses using the leading-order eddy viscosity, and lastly discuss positive definiteness of the leading-order eddy viscosity tensor.

\subsection{Standard Eddy Viscosity}
\label{standardeddyviscosity}

\begin{figure}
  \vspace{0.3cm}
  \centerline{\includegraphics[width=0.5\linewidth]{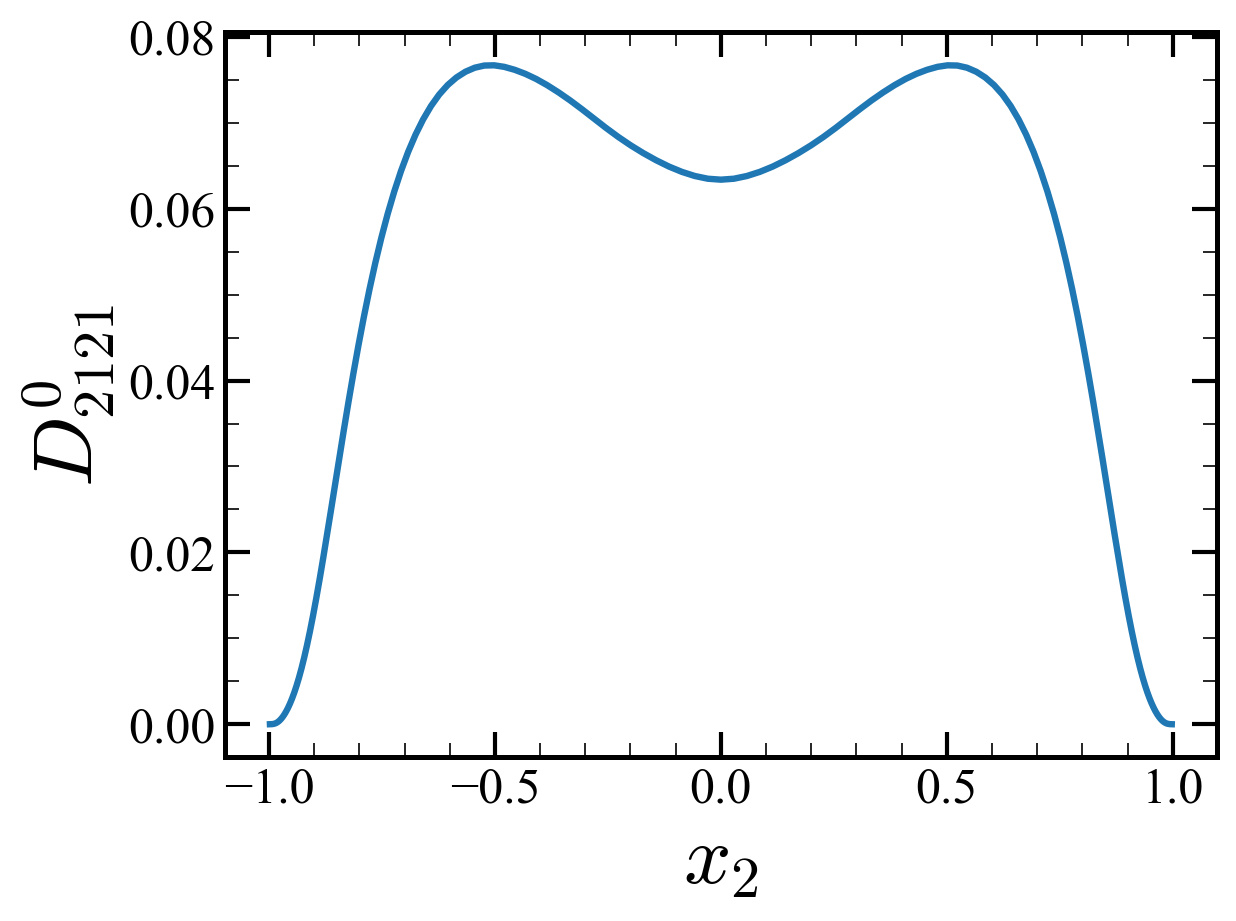}}
  \caption{Eddy viscosity element $D^0_{2121}$.}
\label{fig:D02121}
\end{figure}

In parallel flows, among all the components of the eddy viscosity tensor, by far the most important component is $D^0_{2121}$ which represents the mixing effect by $\pdv*{U_1}{x_2}$. This component also corresponds to the standard eddy viscosity $\nu_T$.
Figure~\ref{fig:D02121} shows the MFM-measured $D^0_{2121}$ across the wall-normal dimension $x_2$. An important observation here is that the MFM allows us to measure the eddy viscosity at the channel center plane $x_2 = 0$, where the velocity gradient $\pdv*{U_1}{x_2}$ is zero due to the symmetry of the mean velocity profile. This value is hard to obtain in typical approaches---tuning $\nu_T$ to $\overline{u'_2 u'_1} / \left(\pdv*{U_1}{x_2}\right)$. \corr{The eddy viscosity can be numerically determined by analyzing the sampling points converging towards the centerline. However, in the vicinity of the centerline, both the numerator and denominator of the expression are significantly influenced by statistical noise. To achieve a reliable estimate, an extensive period of time integration is necessary to reduce the noise to acceptably low levels.}

\begin{figure}
\vspace{0.3cm}
\centering
  \subfigure[\corr{Normalized $u'_1$} in $x_1$-$x_3$ plane ]{\includegraphics[height=0.2\linewidth]{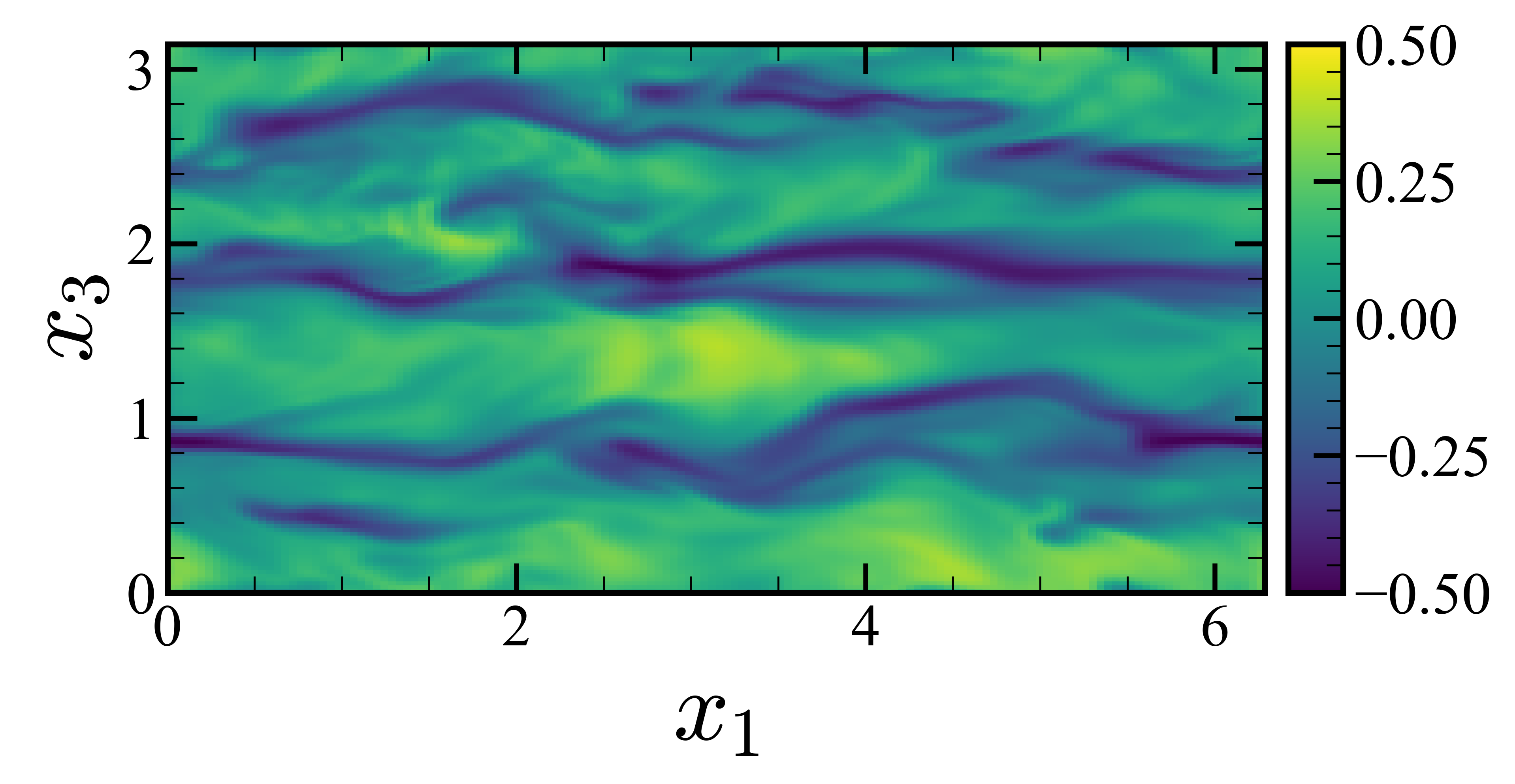}} 
  \subfigure[\corr{Normalized $v'_1$} in $x_1$-$x_3$ plane]{\includegraphics[height=0.2\linewidth]{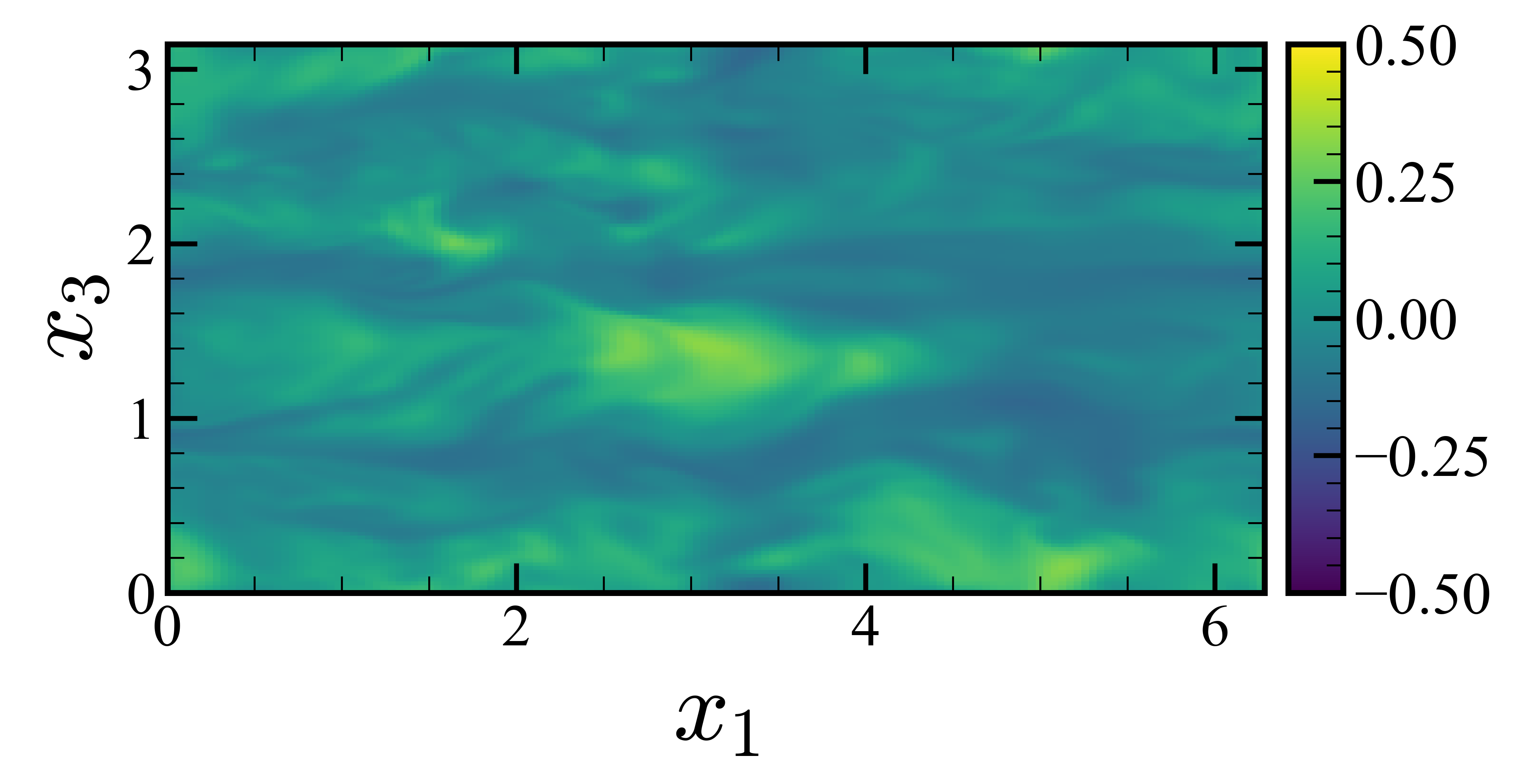}} \\
  \subfigure[\corr{Normalized $u'_1$} in $x_3$-$x_2$ plane]{\includegraphics[height=0.22\linewidth]{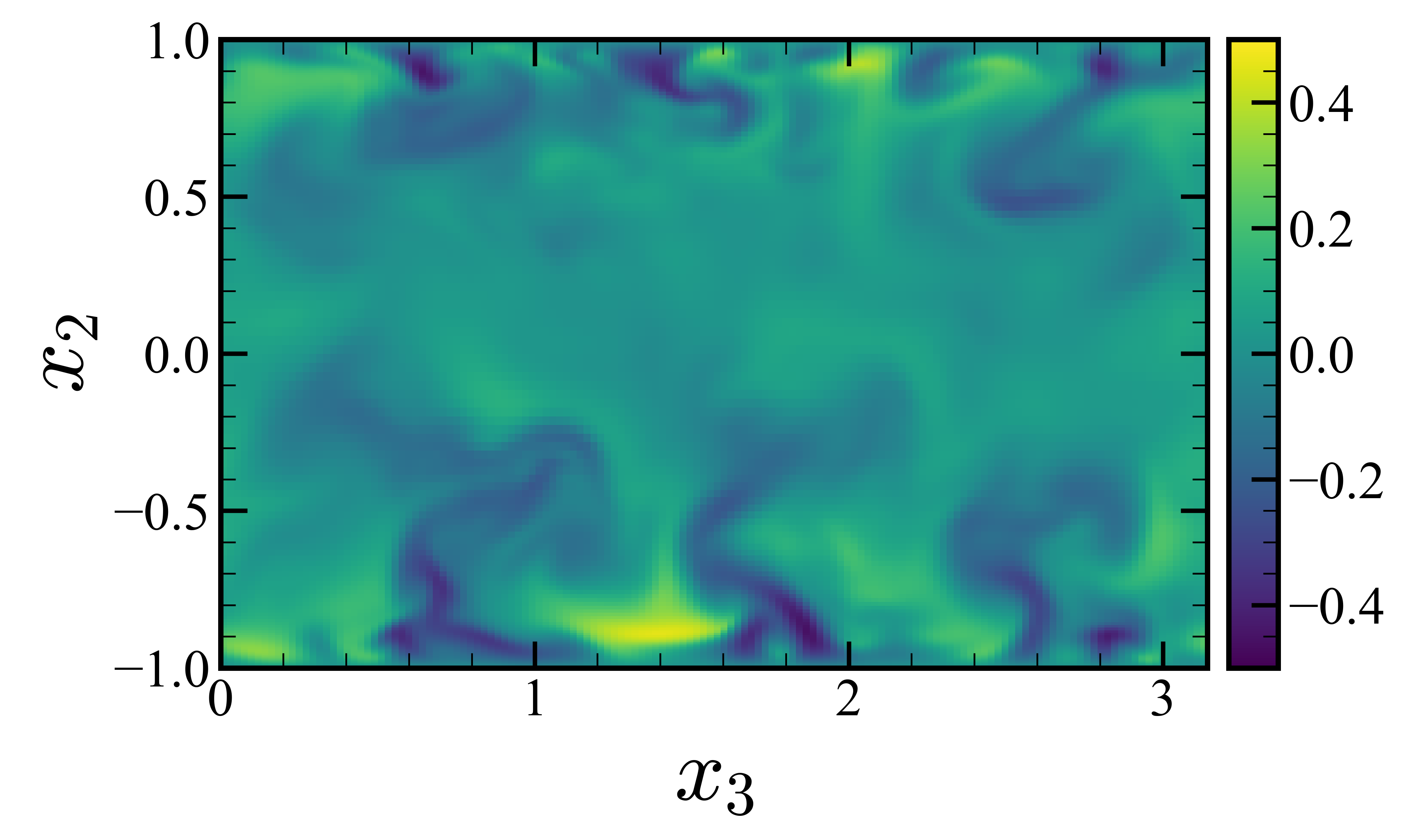}}
  \subfigure[\corr{Normalized $v'_1$} in $x_3$-$x_2$ plane]{\includegraphics[height=0.22\linewidth]{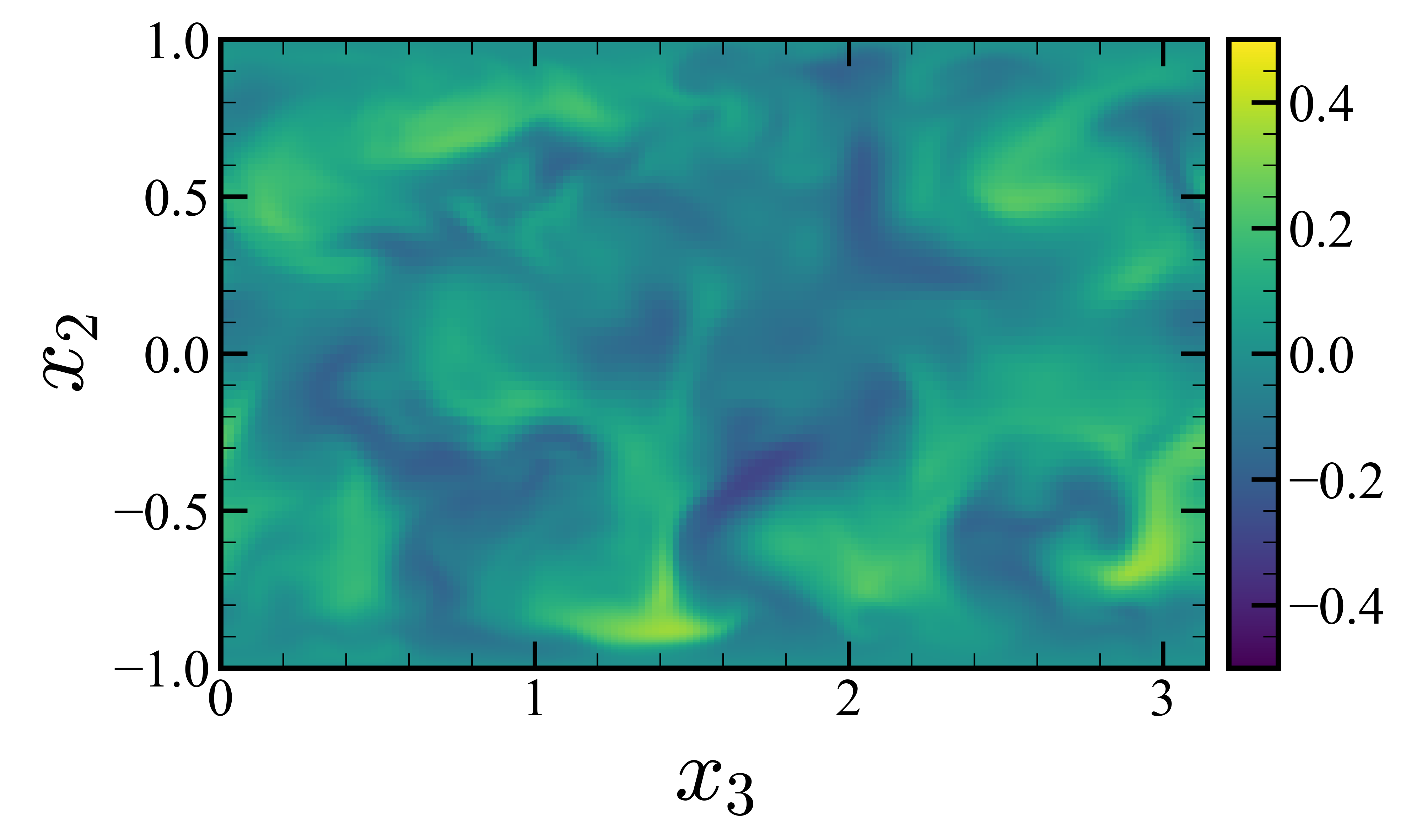}}
  \caption{\corr{Instantaneous velocity contours $u'_1$ and $v'_1$ normalized with each maximum value, $u'_1/(2\max(u'_1))$ and $v'_1/(2\max(v'_1))$}: (a) and (b) correspond to the cross-section taken at $x_2 = -0.8492$ and (c) and (d) correspond to the cross-section taken at $x_1=3.08$. The shown vector field $v_i$  corresponds to a leading-order MFM in which GMT equation is macroscopically forced to achieve $V_1=x_2$ and $V_2=V_3=0$.} 
\label{fig:instantaneous}
\end{figure}
 
Figure~\ref{fig:instantaneous} shows \corr{instantaneous field data for the normalized streamwise velocities $u_1'$ and $v_1'$} of the MFM simulation for evaluation of $D^0_{2121}$ at the same instantaneous time. Figures~(a) and (b) show the velocity profile over $(x_1, x_3)$ cross-section taken at $x_2 = -0.8492$ ($x_2^+ = 27$) and Figures~(c) and (d) show the velocity profile over $(x_3, x_2)$ cross-section taken at $x_1 = 3.08$. The key feature shown is that even though the forcings for the NS vector field $u_i$ and the GMT vector field $v_i$ are completely different macroscopically, MFM leads to similar features in the $u'_1$ and $v'_1$. The same qualitative observation holds across all three components of the $u'_i$ and $v'_i$ fields. Furthermore, while for $x_2<0$ we observe positive correlation between $u'_i$ and $v'_i$ fields, the sign of correlation flips for $x_2>0$. For this specific MFM analysis, the sole difference between $u'_i$ and $v'_i$ fields is in the enforced mean velocity profile. As shown by \citet{Mani2021} without forcing, GMT would result in $v$-fields identical to $u$-fields after a few flow through times regardless of the choice of initial conditions. The case shown in Figure~\ref{fig:instantaneous} corresponds to a forced GMT in which the mean velocity gradient is kept constant $\partial V_1/\partial x_2=1$ in order to examine mixing by the leading-order (local limit) eddy viscosity. The observation in Figure~\ref{fig:instantaneous} suggests that mixing of \corr{the streamwise momentum} in turbulent channel flow is \corr{substantially influenced} by the leading-order effects.  


\begin{figure}
\vspace{0.3cm}
\centering
  \subfigure[$U_1$]{\includegraphics[width=0.35\linewidth]{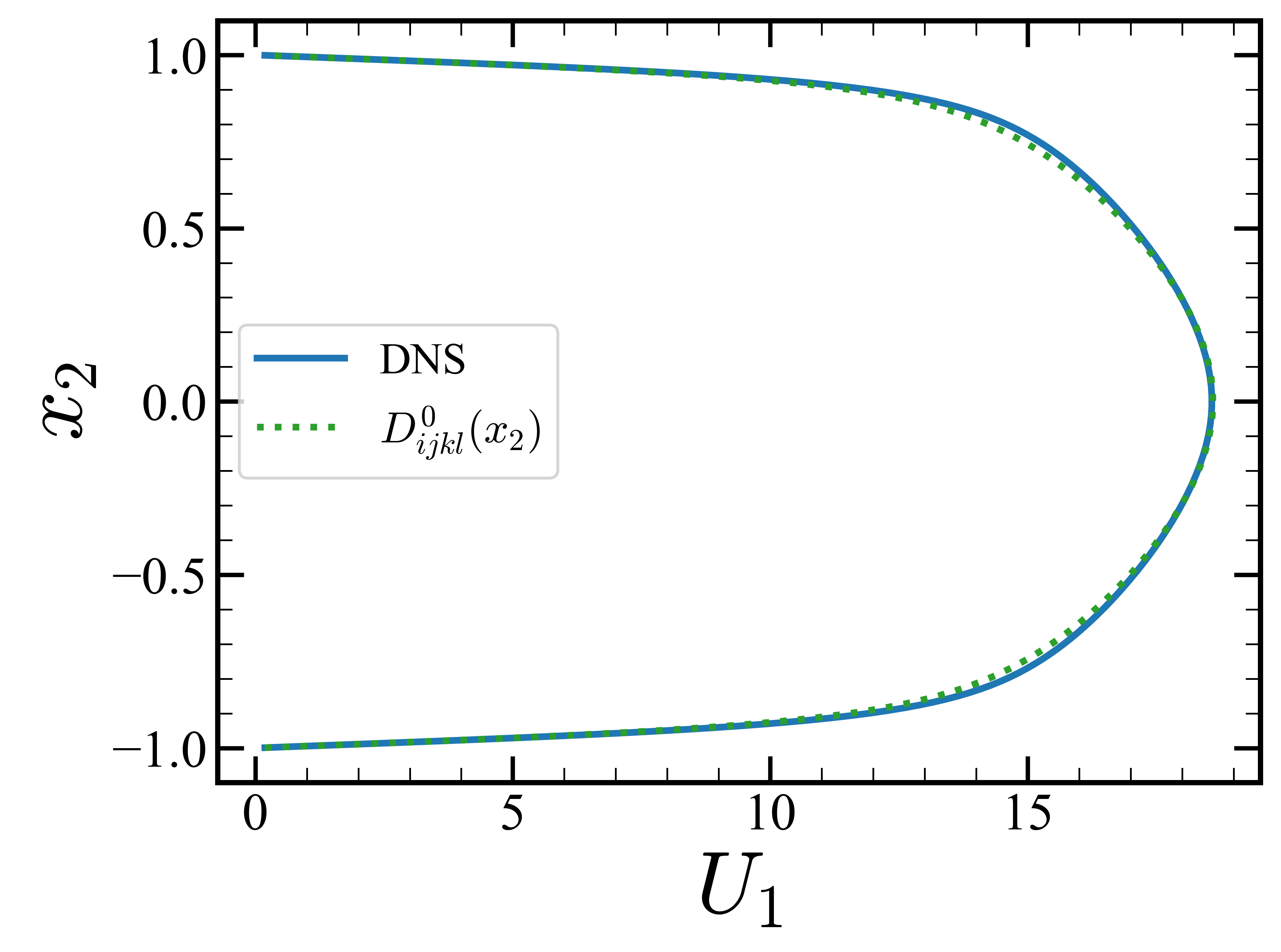}} 
  \subfigure[$D^{0+}_{2121}$]{\includegraphics[width=0.35\linewidth]{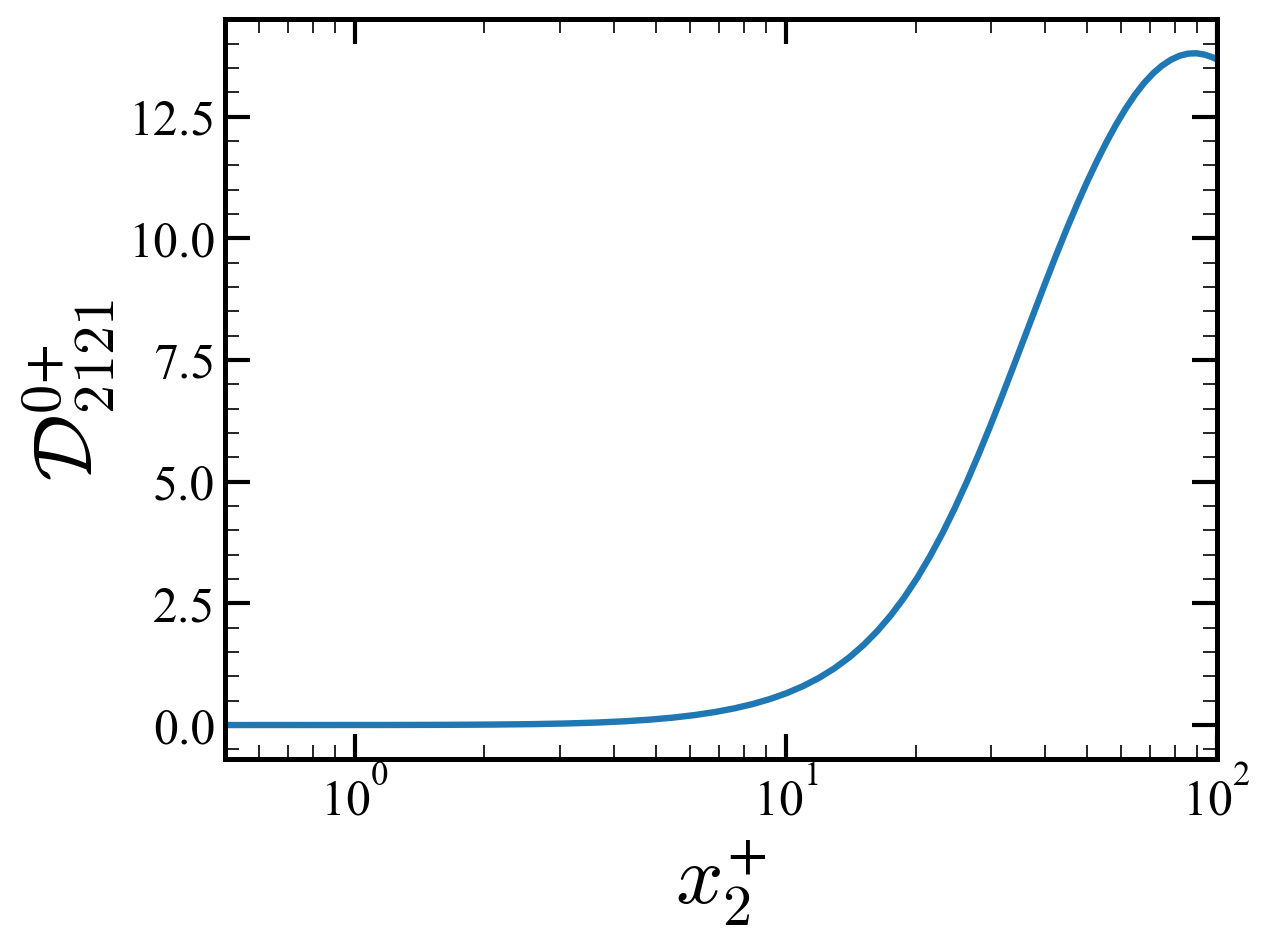}} 
  \caption{\corr{$U_1$ reconstructed from $D^0_{2121}$ and $D^{0}_{2121}$ in wall units: (a) RANS prediction using $D^0_{2121}$ where the dotted green line is the mean velocity prediction using $D^0_{2121}$ and the blue solid line is its comparison to the DNS data and (b) $D^{0}_{2121}$ in wall-units where $D^{0+}_{2121}=D^{0}_{2121}\mathrm{Re}$ and $x_2^+=x_2\mathrm{Re}/\delta$.}}
\label{fig:rans}
\end{figure}

To assess this conclusion quantitatively we next obtain the RANS solution using the measured $D^0_{2121}$ to examine how accurate the leading-order eddy viscosity performs for the prediction of the mean velocity profile. Since the prediction of the mean channel flow only requires one component in the Reynolds stress, we conduct the RANS simulation using $D^0_{2121}$ and compare the predicted solution with that of the DNS. As shown in 
Figure~\ref{fig:rans}, the MFM-based leading-order RANS solution predicts the DNS solution very accurately with an accuracy of 99\%. The accuracy is computed with max absolute error $\max\left(U_1^{\mathrm{DNS}}-U_1^{\mathrm{MFM}}\right)/\max\left(U_1^{\mathrm{DNS}}\right)$, where $U_1^{\mathrm{DNS}}$ is the streamwise velocity from DNS and $U_1^{\mathrm{MFM}}$ is the streamwise velocity predicted from RANS using MFM-measured eddy viscosity $D^0_{2121}$. 
\corr{The accuracy of this local RANS prediction indicates that the mean momentum mixing in the turbulent channel flow might be local. To assess this conclusion with more certainty we will later directly examine the nonlocality of the eddy viscosity kernel.} 



\citet{Hamba2005} also reported a small subset of the components of the leading-order (local limit) eddy viscosity, through a more expensive method of first computing the full eddy viscosity kernel for those components and then performing integration as in Equation \ref{eq:leadingordertensor}. Our result in Figure~\ref{fig:rans} regarding \corr{accuracy of the leading-order eddy viscosity} is in contrast to his result (see Figure 4 in \citet{Hamba2005}). We attribute this difference to the fact that Hamba used the average of $D^0_{2121}$ and $D^0_{1221}$ as the representative local eddy viscosity. This averaging was motivated to enforce symmetric Reynolds stresses. However, conceptually these two eddy viscosities represent different mixing rates: the former represents mixing of the streamwise momentum in the wall normal direction, while the latter represents mixing of the wall-normal momentum in the streamwise direction. As we shall see, while a full eddy viscosity kernel reproduces symmetric Reynolds stresses, the leading-order eddy viscosity causes errors not only in magnitude but also in symmetry of Reynolds stresses. 

We next use MFM to quantify other components of $D^0_{ijkl}$. While these components do not affect prediction of the mean velocity profile in purely parallel flows, they provide an understanding of momentum mixing by this parallel flow, if hypothetical mean momentum gradients were imposed in other directions. \corr{One motivation to study these additional components of $D^0_{ijkl}$ is to provide reference data of closure operators, as opposed to closure terms, for models that offer anisotropic eddy viscosity.} 
Our analysis is \corr{additionally} motivated by observation of spatially developing \corr{attached} turbulent boundary layers, where weak momentum gradient could exist in both streamwise and spanwise directions. These mean gradients induce additional Reynolds stresses, due to components of $D_{ijkl}$ other than $D_{2121}$. Additionally, it has been observed that turbulent boundary layers have similar hairpin structures in their velocity field as those seen in turbulent channel flows \citep{Eitel2015}, and thus are expected to mix momentum in manners \corr{qualitatively} similar to that of a turbulent channel flow. While quantitative differences are expected between the two flows, we expect anisotropy in eddy viscosity observed in turbulent channel flow be at least qualitatively representative of anisotropy encountered in wall-attached turbulent boundary layers in the absence of substantial wall curvature. \corr{Some of these qualitative similarities, such as components in $D_{ijkl}$  with the highest magnitude, can already be confirmed from the study of \citet{Park2022} with a specific focus to their analysis of pre-separation zone of turbulent boundary layers.} However, given the stringent statistical convergence requirements for MFM simulations, e.g. at least an order of magnitude longer simulations needed than commonly reported DNS, compared to turbulent boundary layers, turbulent channel flows have the advantage of cheaper runtime per time step and availability of additional homogeneous direction for statistical convergence. 



\subsection{Quantifying Anisotropy}

We computed all other components of the anisotropic eddy viscosity tensor $D^0_{ijkl}$, a total of 81 coefficients as a function of the wall-normal coordinate. All the data are shown in Appendix C. Out of 81 components, 41 are non-zeros and 40 are inevitably zero due to the symmetry in spanwise direction.

Out of all the elements, the largest eddy viscosity component is $D^0_{1111}$, with a maximum value of 1.318, and the smallest nonzero eddy viscosity component is $D^0_{2331}$ with the maximum value of 0.00248. After comparing these values to a maximum value of the nominal eddy viscosity $D^0_{2121}$, which is 0.0767, we determined that the largest coefficient in the eddy viscosity tensor is one order of magnitude larger than the nominal eddy viscosity and three orders of magnitude larger than the smallest coefficient, indicating a significant anisotropy. When we examine these ratios locally at each $x_2$, the differences are more drastic and may go up to a few orders of magnitude.
After $D^0_{1111}$, the largest eddy viscosity components are $D^0_{1212}$ and $D^0_{1313}$ with maximum values of 0.573 and 0.407, respectively. All three eddy viscosities have their first and third index represented by the streamwise direction. These indices respectively represent the component of the velocity field that mixes momentum and the direction of the mean-momentum gradient. This observation coincides with the fact that $u'_1$ is the largest fluctuating velocity component in channel flow. Combining the two observations, we conclude that $u'_1$ is the strongest mixer of momentum and is most effective in mixing in the $x_1$ direction, as intuitively expected. Specifically, \corr{the rate of momentum mixing} in the streamwise direction is substantially faster than the standard eddy viscosity which characterises \corr{the rate of (streamwise) momentum mixing} in the wall-normal direction.

Additionally, all three dominant eddy viscosity components have repeated second and fourth indices. These indices respectively represent the momentum component that is being mixed and the momentum component whose mean gradient is responsible for mixing. Based on this observation, we conclude that within $D^0_{1j1l}$, mean gradient of component $l$ most effectively contributes to the generation of $\overline{u'_1 v'_j}$ when $j=l$. In other words, gradient of each momentum component most effectively generates fluxes of the same momentum component at least in the leading-order limit. This latter observation is extendable to $D^0_{ijil}$ components, \corr{and is not a surprising outcome given that the production term in the transport equation for $\overline{u'_i v'_j}$ involves the mean gradient of $V_j$.}

\begin{figure}
\vspace{0.3cm}
\centering
  \subfigure[MFM]{\includegraphics[width=0.35\linewidth]{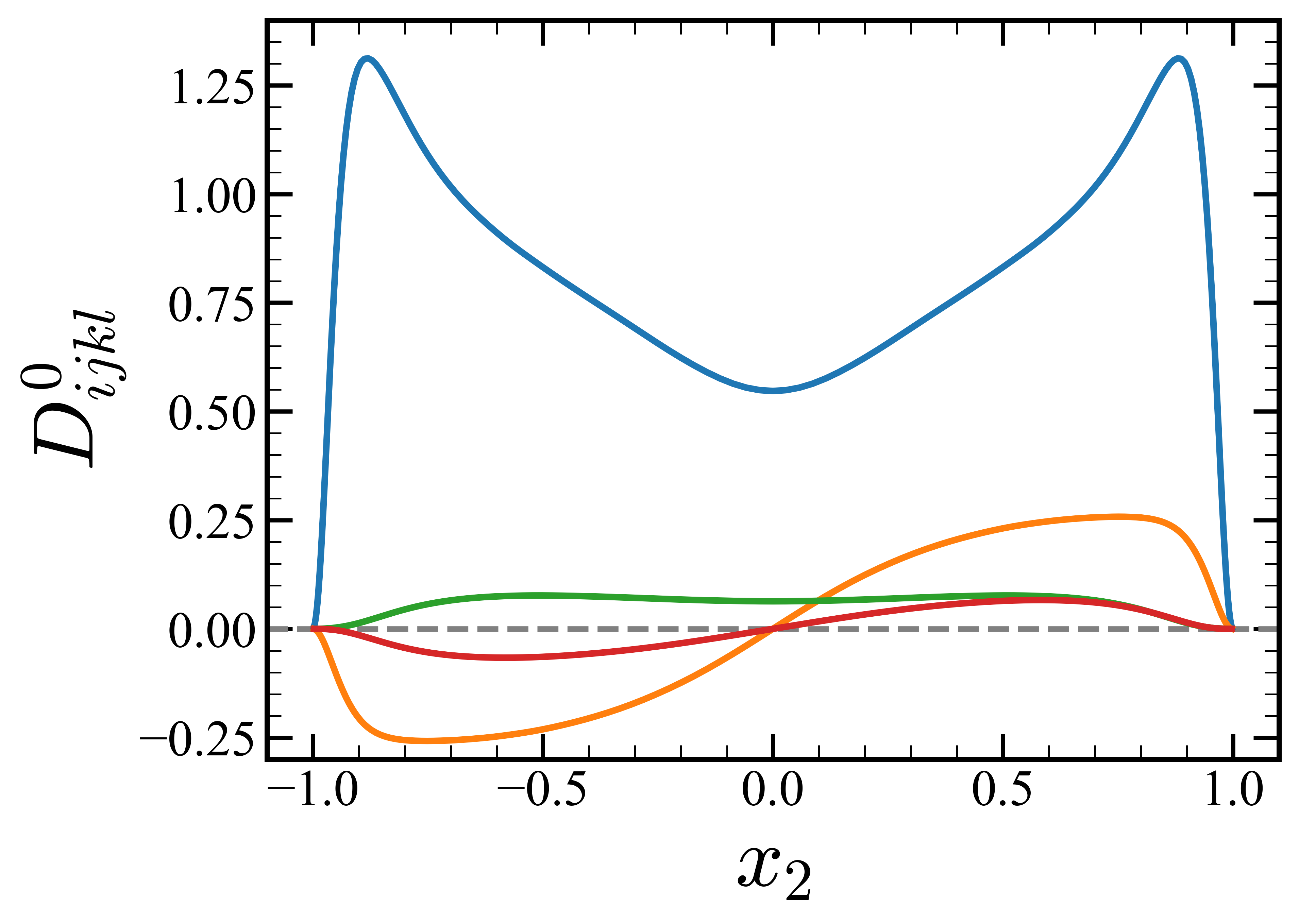}} 
  \subfigure[Boussinesq]{\includegraphics[width=0.35\linewidth]{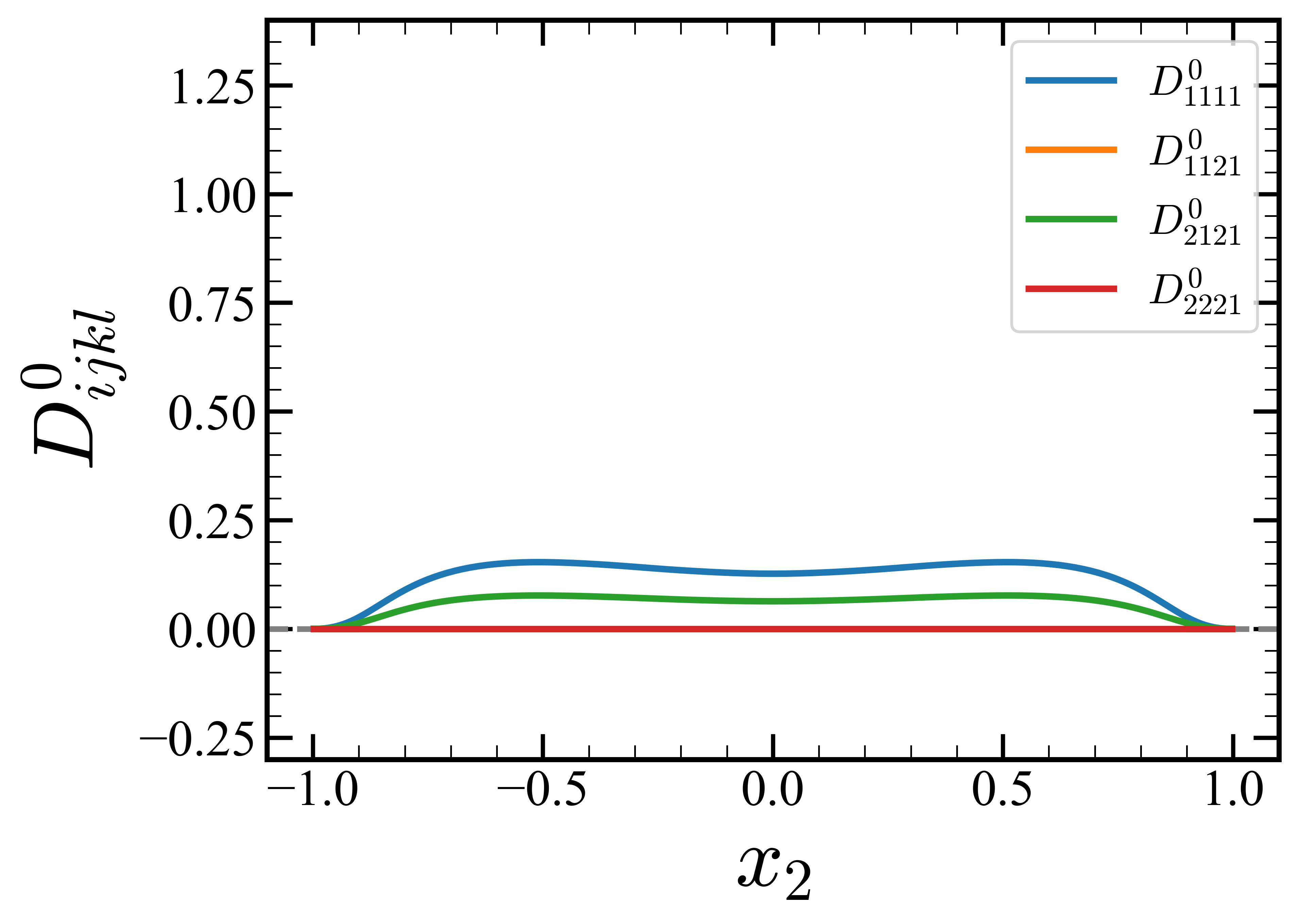}} 
  \caption{Comparison of the measured eddy viscosity elements $D^0_{1111}$ (blue line), $D^0_{1121}$ (orange line), $D^0_{2121}$ (green line), and $D^0_{2221}$ (red line) to the Boussinesq approximation.}
\label{fig:asymptotics}
\end{figure}

As we discussed since flow structures and thus momentum mixing is similar between the channel flow and the attached boundary layers, we can use the measured eddy viscosity anisotropy in the former setting to identify important eddy viscosity components for the latter setting. To this end, we present in Appendix D a scaling analysis of various gradients contributing to the Reynolds stress tensor. Combining this analysis with the measured order of magnitude of each eddy viscosity component that acts as a pre-factor multiplying components of the velocity gradient tensor, we identify the key eddy viscosity components that contribute dominantly to the Reynolds stress tensor budget. Based on our analysis we identify $D_{1111}$, $D_{1121}$, $D_{2121}$, and $D_{2221}$ as the key four, out of 16, dominant eddy viscosity components for 2D spatially developing turbulent boundary layers.

Motivated by this example, we next examine the identified anisotropy against the Bousinesq approximation. When we cast the Boussinesq approximation to our tensorial representation, the components in the eddy viscosity tensor are in ratio of 0, 1, or 2 to the standard eddy viscosity $\nu_T$. For instance, the four elements are prescribed with following ratios; $D_{1111}=2\nu_T$, $D_{1121}=0$, $D_{2121}=\nu_T$, and $D_{2221}=0$. Figure~\ref{fig:asymptotics} shows the comparison of these eddy viscosity components to the Boussinesq approximation. In Figure~\ref{fig:asymptotics}(a), we show the measured four elements using our MFM calculation. In Figure~\ref{fig:asymptotics}(b) we set the standard eddy viscosity to the MFM-measured leading-order value, $\nu_T=D^0_{2121}$ and prescribe the other components with the ratio to $\nu_T$. As shown in the figure, a huge anisotropy is observed not only among all elements but specificallty among these four critical elements, and the ratio of these plots can locally go up to hundreds. We conclude that, while $D_{2121}$ is the most important eddy viscosity component for parallel and semi-parallel flows, the presence of small non-parallel effects could lead to significant influence of anisotropy in momentum transport in wall bounded flows.

Lastly, we point that there have been attempts to include the anisotropy in RANS such as Spalart-Allmaras model with quadratic constitutive relation (SA-QCR) \citep{Spalart2000, Mani2013, Rumsey2020}. However, examining our results suggest that these models do not captured the level of the anisotropy that MFM measured. For instance, SA-QCR still prescribes $D_{1111}=2\nu_T$ and the anisotropy is not yet introduced in needed directions.

\subsection{Dependence on the rate of rotation}

\begin{figure}
  \vspace{0.3cm}
  \centerline{\includegraphics[width=0.5\linewidth]{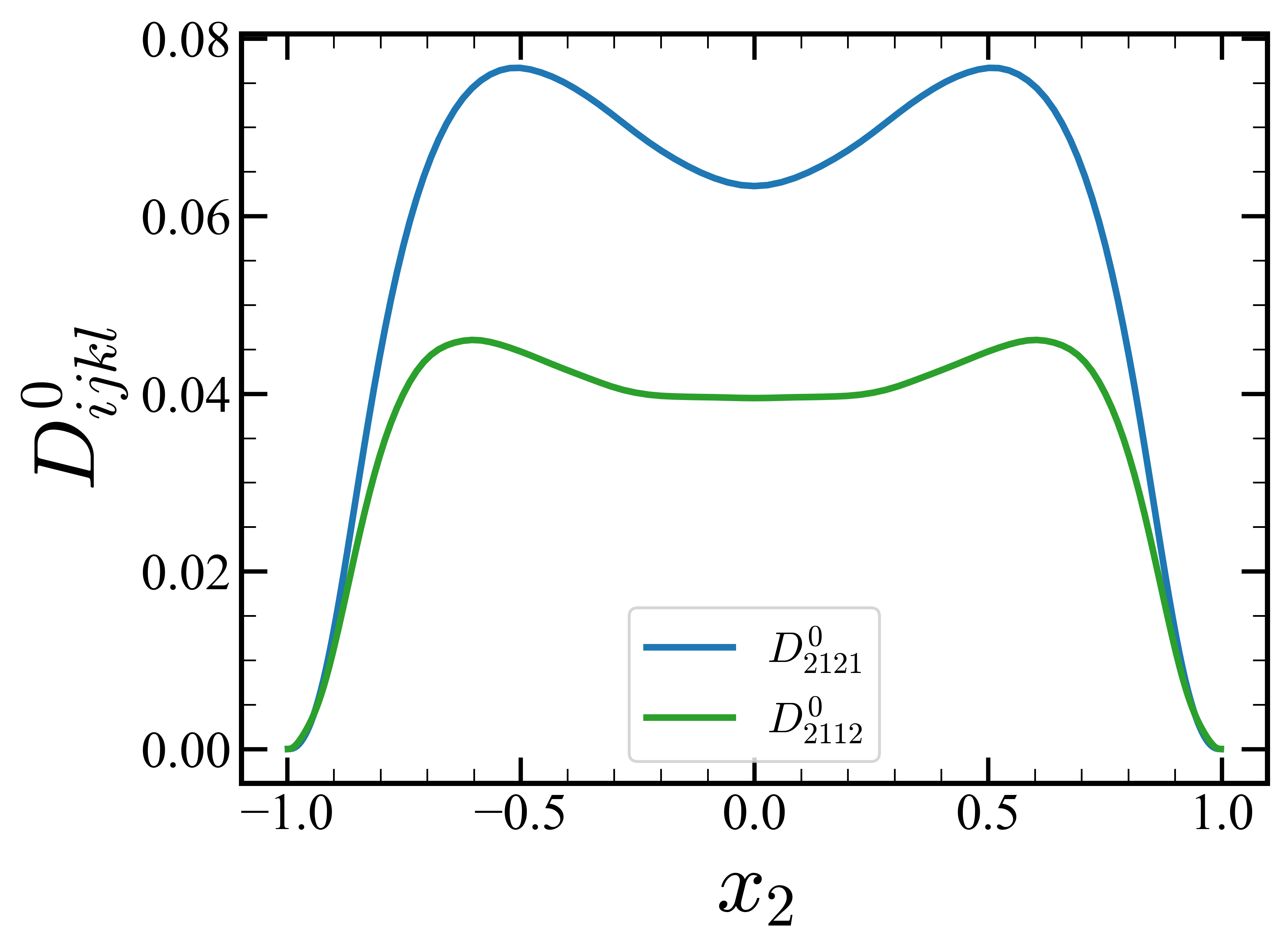}}
  \caption{Comparison of $D^0_{2121}$ (blue line) and $D^0_{2112}$ (green line).}
\label{fig:D02121_D02112}
\end{figure}

\corr{The dependence of Reynolds stress on rate of rotation is studied previously in the
literature. Key main examples are 1) investigation of environmental flows that include the
Coriolis term \citep{Speziale1992} and 2) Corrections to the Boussinesq eddy viscosity based on Cayley-Hamilton
theorem \citep{Pope1975}, which are also incorporated in Quadratic Constitutive Relation
(QCR) models \citep{Spalart1994}. However, these models incorporate the mean rotation effects as higher-order nonlinear corrections to the Boussinesq eddy viscosity.} 

\corr{With our eddy viscosity tensor notation, insensitivity of Reynolds to the mean rotation implies that $D^0_{ijkl}$ must be equal to $D^0_{ijlk}$} because under this condition, each Reynolds stress component, $\overline{u'_i u'_j}$, would be equally sensitive to both $\partial U_l/\partial x_k$ and $\partial U_k/\partial x_l$, and thus is a function of the summation $\partial U_l/\partial x_k + \partial U_k/\partial x_l$, which is $2S_{kl}$. 
However, our measurement of the leading-order eddy viscosity tensor invalidates the relation $D^0_{ijkl} D^0_{ijlk}$. Figure~\ref{fig:D02121_D02112} shows the comparison between $D^0_{2121}$ and $D^0_{2112}$. These two components have the same sign and their qualitative shape is similar, but the magnitudes are drastically different. 
This highlights an important conclusion: sensitivity of Reynolds stresses on mean rotation is not \corr{a secondary or higher-order effect and is present even at the leading-order term of the eddy viscosity expansion.} Likewise, we reach the same conclusion with the case of $D^0_{ij13}$ and $D^0_{ij23}$.

\subsection{Leading-order Reynolds stress}

\begin{figure}
\vspace{0.3cm}
\centering
  \subfigure[$D^0_{1121}$]{\includegraphics[width=0.3\linewidth]{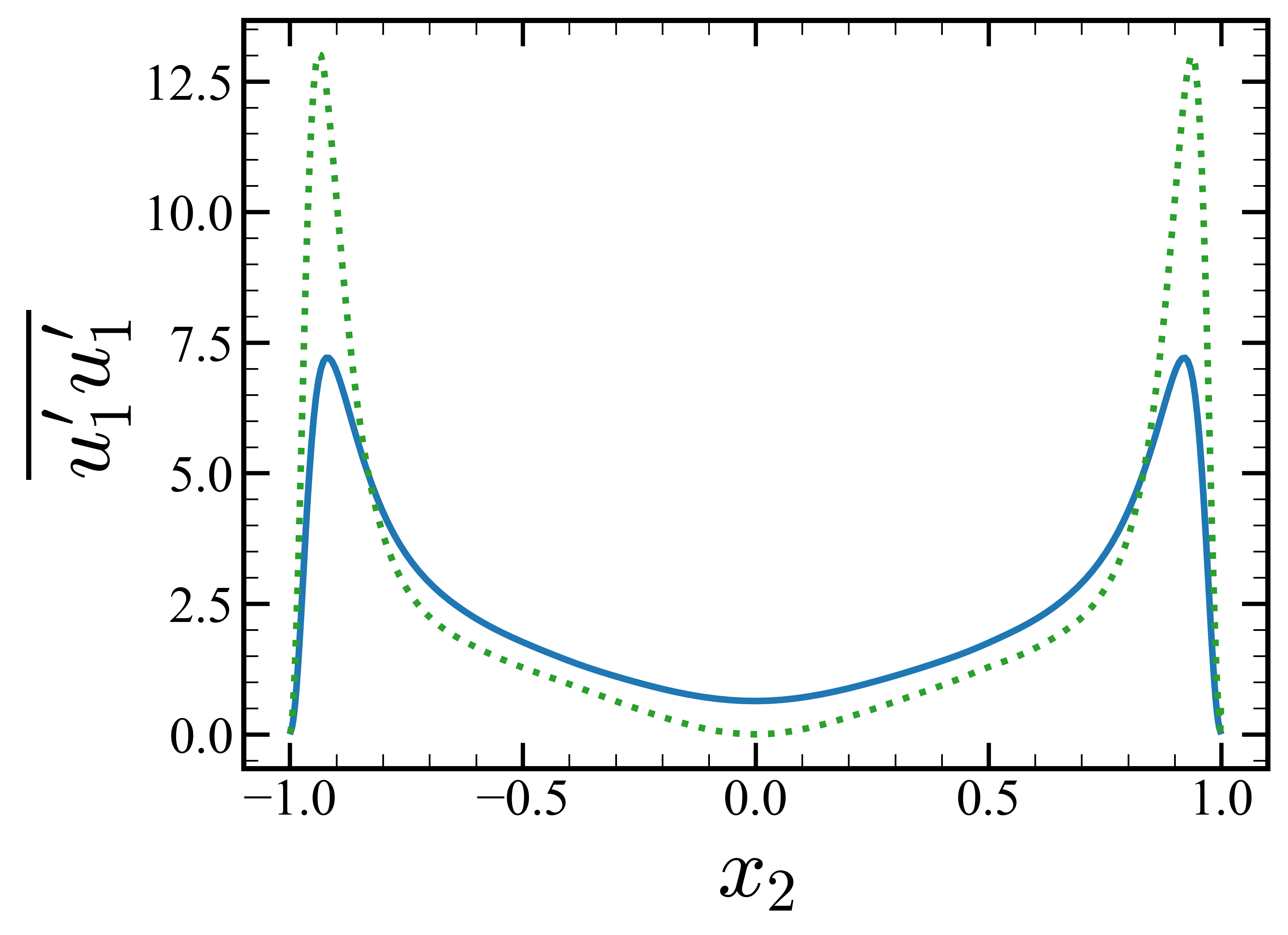}} 
  \subfigure[$D^0_{1221}$]{\includegraphics[width=0.3\linewidth]{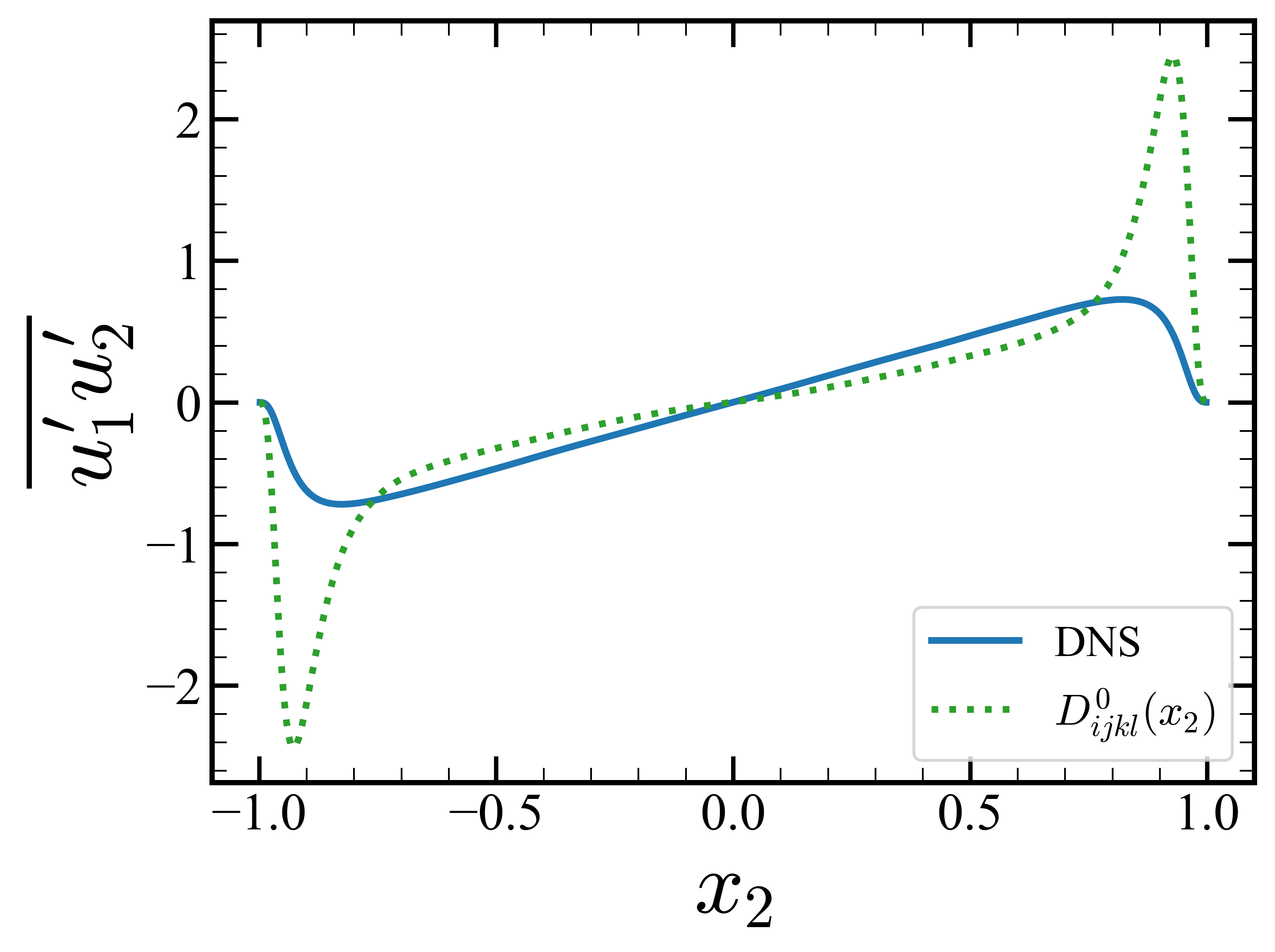}} \\
  \subfigure[$D^0_{2121}$]{\includegraphics[width=0.3\linewidth]{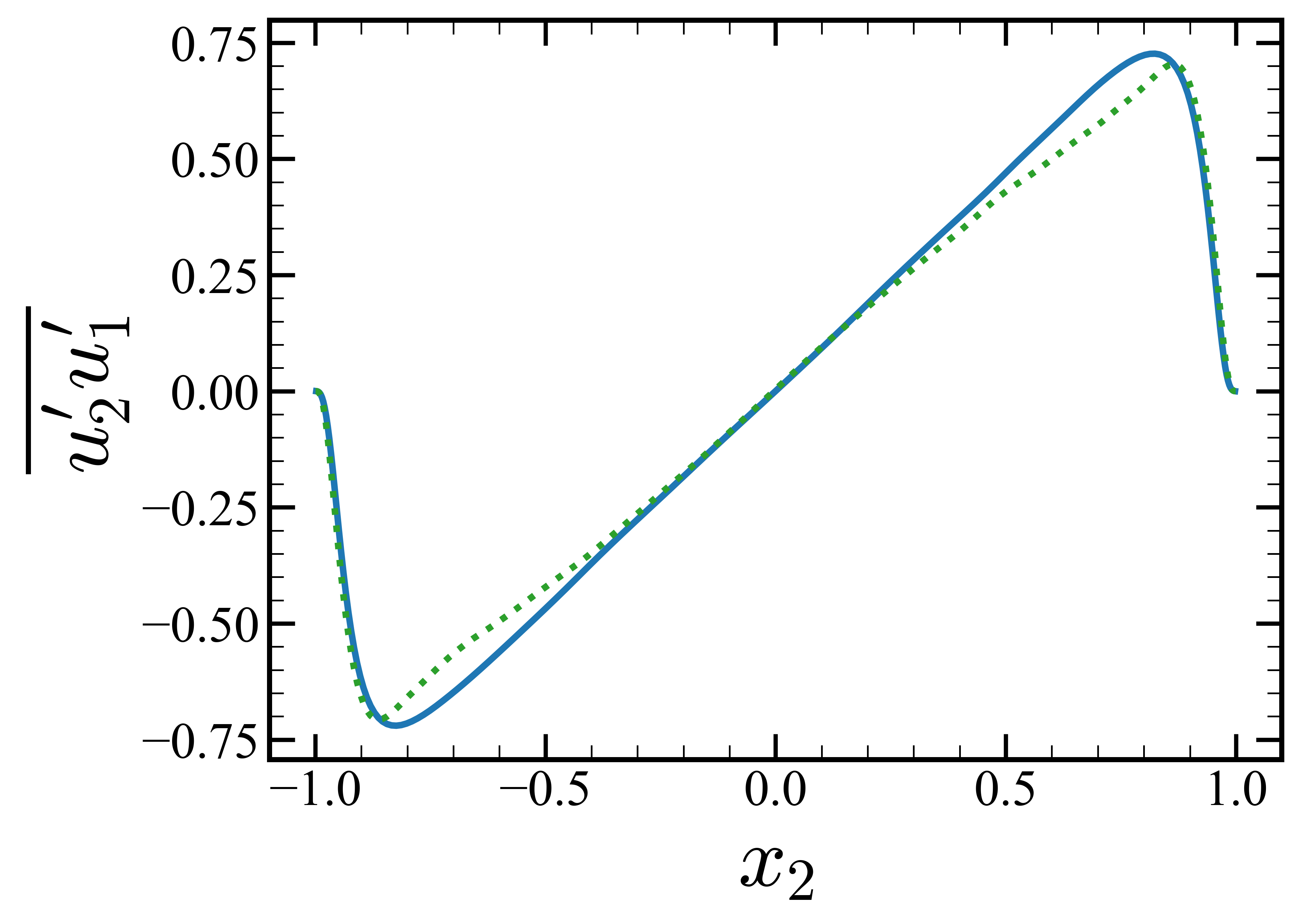}} 
  \subfigure[$D^0_{2221}$]{\includegraphics[width=0.3\linewidth]{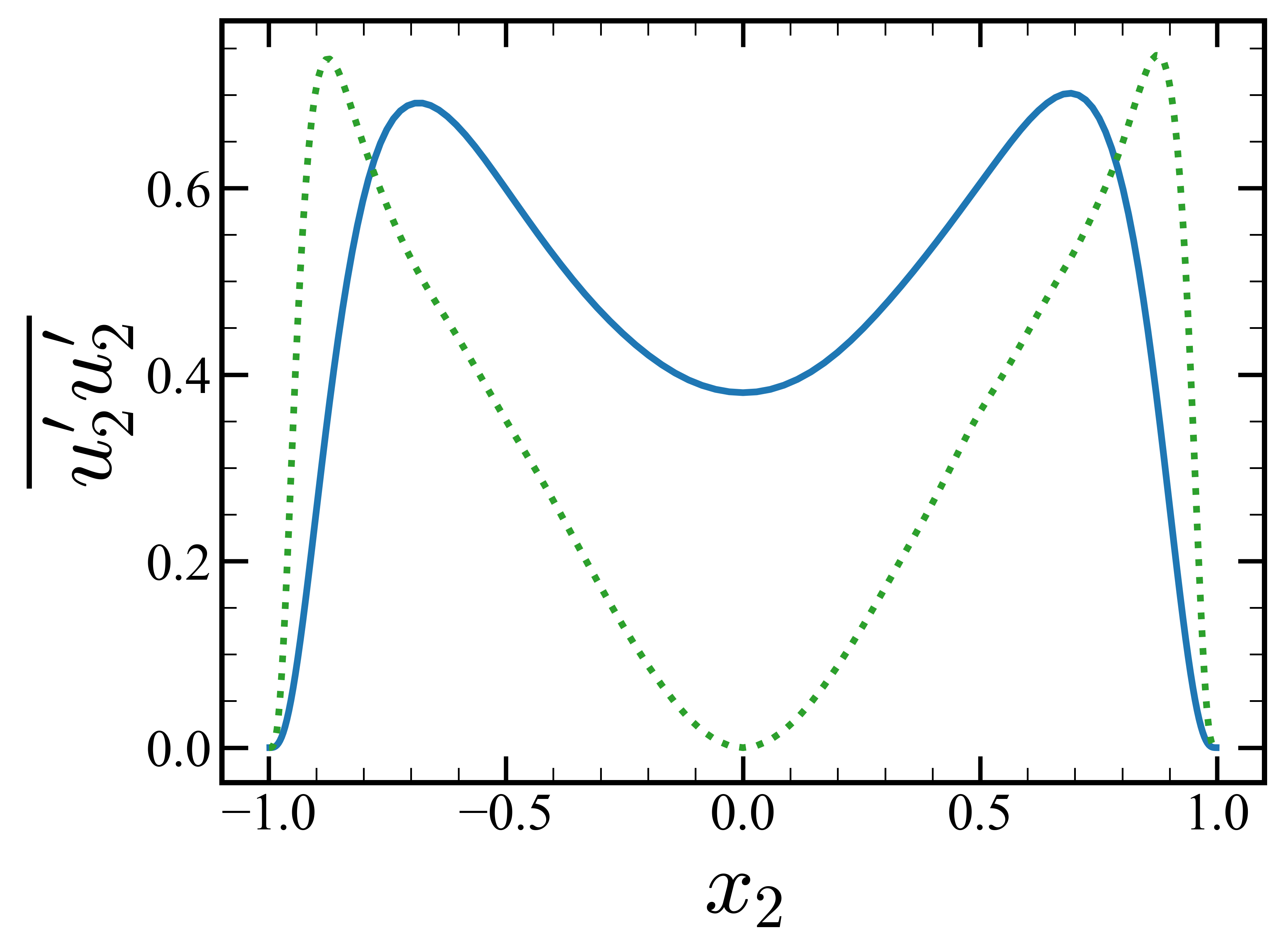}} 
  \subfigure[$D^0_{3321}$]{\includegraphics[width=0.3\linewidth]{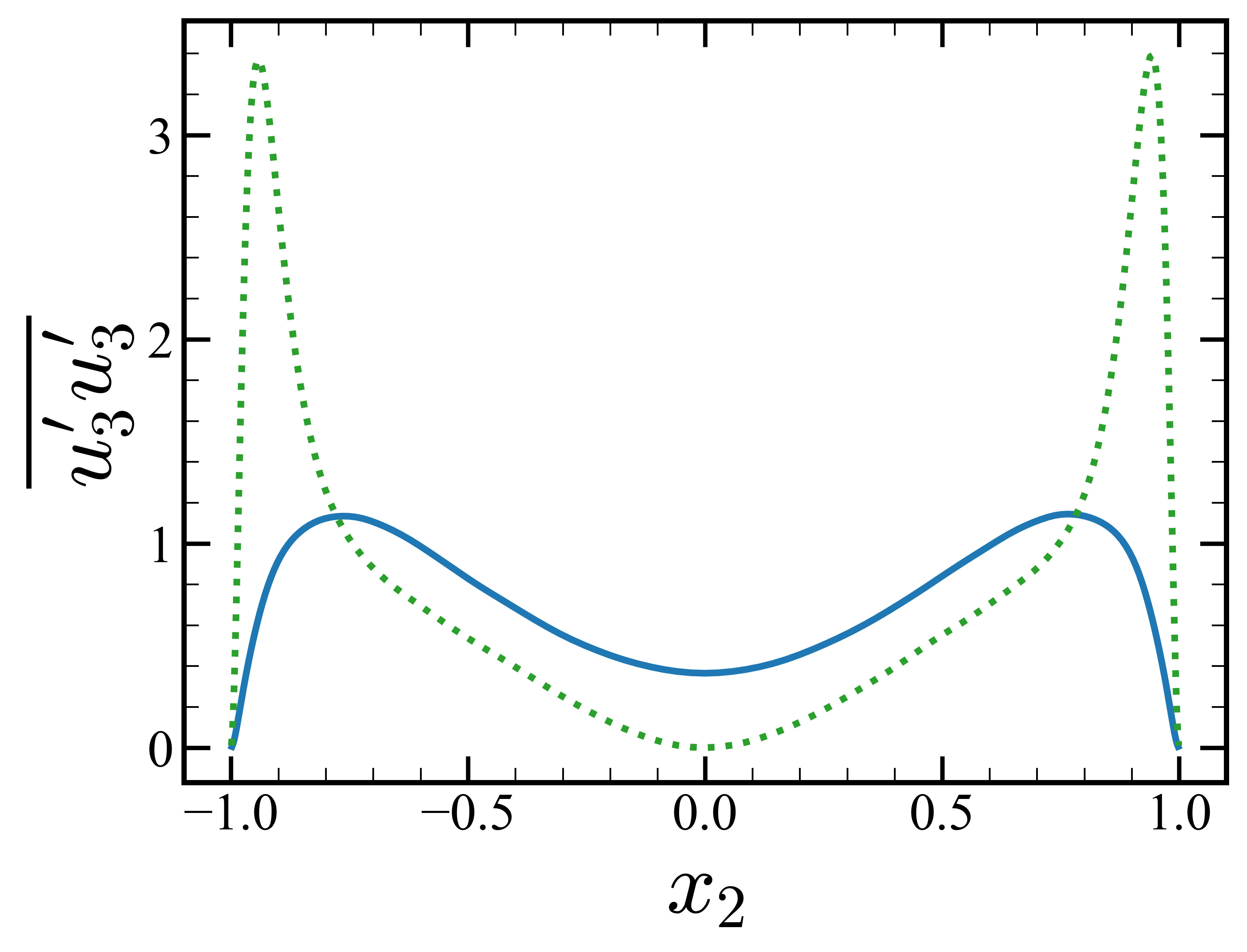}} 
  \caption{Reynolds stresses constructed by the leading-order eddy viscosity tensor associated with (a) $D^0_{1121}$, (b) $D^0_{1221}$, (c) $D^0_{2121}$, (d) $D^0_{2221}$, and (e) $D^0_{3321}$: green dotted line, the reconstructed Reynolds stress by the leading-order eddy viscosity tensor \corr{$\overline{u'_i u'_j}^0$}; blue solid line, the DNS data.}
\label{fig:localresynoldsstress}
\end{figure}

In Section~\ref{standardeddyviscosity}, we computed a RANS solution using $D^0_{2121}$ and compare the solution to that of the DNS to assess appropriateness of the leading-order eddy viscosity for prediction of the mean velocity profile. Another way to make this assessment is to reconstruct Reynolds stress using the computed eddy viscosity tensor and compare it to the Reynolds stress of DNS. This way, we can assess some other components in $D^0_{ijkl}$. The Reynolds stress in the channel flow can be represented in the following way: \corr{$\overline{u'_i u'_j}^0=-D^0_{ij21} {\partial U_1}/{\partial x_2}$}, with the leading-order eddy viscosity tensor $D^0_{ij21}$ computed using MFM and with the mean velocity gradient ${\partial U_1}/{\partial x_2}$ measured from the DNS data. \corr{Here, the superscript zero is added to the Reynolds stress to indicate that this is the leading order reconstruction.}

Figure~\ref{fig:localresynoldsstress} shows the five reconstructed Reynolds stresses associated with the RANS prediction of the channel flow, in comparison with the Reynolds stresses from the DNS data. There are three important observations with the Reynolds stresses that are reconstructed with the leading-order eddy viscosity tensor. 
The first finding is that while the Reynolds stresses reconstructed using the leading-order eddy viscosity show similar qualitative trends and orders of magnitudes to those from DNS, there's still a noticeable difference between the two. This difference is likely due to the leading-order truncation of the eddy viscosity operator. Among various components of the Reynolds stress tensor, only \corr{$\overline{u'_2 u'_1}^0$} is \corr{reasonably constructed. This observation is compatible with the previous observation of reasonable RANS solution for the mean flow, since only this Reynolds stress component is involved in mean momentum mixing for this flow.}

The next important observation is that constructed Reynolds stresses from the leading-order eddy viscosity are not symmetric.
\corr{When the eddy viscosity operator is modeled through a truncated Kramers-Moyal expansion, we introduce an approximation that inherently involves a loss of symmetry. This asymmetry arises from employing the Generalized Momentum Transport (GMT) equation to examine momentum transport, leading to the observation that $\overline{u'_2 v'_1}^0$ is not equal to $\overline{u'_1 v'_2}^0$. However, considering that the Navier-Stokes equations serve as the attractor solution to the GMT framework---as established by \citet{Mani2021}--utilizing exact closure operators ensures that $\overline{u'_i v'_j}$ aligns with $\overline{u'_i u'_j}$, thereby preserving the symmetry of the tensor.} As we shall see, inclusion of the full nonlocal eddy viscosity will eliminate this error. However, the fact that the leading-order \corr{$\overline{u'_2 u'_1}^0$} match the DNS, substantially better than \corr{$\overline{u'_1 u'_2}^0$} indicates that the former Reynolds stress is more local while the latter has substantial nonlocal sensitivity to the mean velocity gradient. While we do not have an intuitive explanation for this observation, we note the coincidence that the former Reynolds stress, represents flux of an active mean momentum component in the direction where its gradients are active. The only way that the latter Reynolds stress could be generated in this setting is through pressure coupling, whose fluctuations are known to nonlocally depend on velocity fluctuations.

Lastly, we observe that the leading-order eddy viscosity cannot reproduce the nonzero Reynolds stresses at the centerline, where the velocity gradient is zero due to the symmetry of the channel flow. Nonlocality needs to be included in eddy viscosity to enable prediction of nonzero Reynolds stresses in regions of zero mean velocity gradient. 

\subsection{Positive Definiteness}
\label{sec:posdef}

It is noted that the Reynolds stress is a positive semi-definite tensor \citep{DuVachat1977, Schumann1977}. We often require eddy viscosity to satisfy the same condition as done in the Boussinesq approximation with $\nu_T \geq 0$ \citep{Speziale1994} for a well-posed closure model. In this section, we discuss whether this condition holds for our leading-order eddy viscosity tensor $D^0_{ijkl}$ as well.

The positive definiteness of the eddy viscosity is closely related to the mean kinetic energy equation, which is the following:
\begin{align}
    \frac{\partial}{\partial t}\left(\frac{U_i U_i}{2}\right)
    + U_j\frac{\partial}{\partial x_j}\left(\frac{U_i U_i}{2}\right) &= \frac{\partial}{\partial x_j}\left(-\frac{P}{\rho}U_j\right)
    + \nu \frac{\partial^2 U_i U_i / 2}{\partial x_j \partial x_j} \\
    &- \nu \frac{\partial U_i}{\partial x_j} \frac{\partial U_i}{\partial x_j}
    - \frac{\partial}{\partial x_j}\left( U_i \overline{u_j^\prime u_i^\prime} \right)
    + \overline{u_j^\prime u_i^\prime} \frac{\partial U_i}{\partial x_j}.
    \label{eq:meanKineticEnergy}
\end{align}

The last term in Equation \ref{eq:meanKineticEnergy} is the negative of turbulent kinetic energy production. It is well-known that this term drains the kinetic energy from the mean flow via interactions of the mean shear and the turbulent fluctuations, and provide energy to the turbulence production. We denote the turbulent kinetic energy production as $P_k = -\overline{u_j^\prime u_i^\prime} \frac{\partial U_i}{\partial x_j}$. In all statistically stationary flows, the volumetric integral of $P_k$ must be non-negative, \corr{otherwise turbulent kinetic energy cannot be sustained}. There are certain cases such as the separation of the shear layer where $P_k$ is locally negative, but even for those cases, the turbulent production is positive for most of the domain \citep{Cimarelli2019} \corr{rendering the total volume integral positive}. The volumetric integral condition for $P_k$ can be expressed using our generalized eddy viscosity expression in Equation~\ref{eq:generalform}.
\begin{align}
\int{P_k}\dd[3]{\vb{x}} &= \int -\overline{u_j^\prime u_i^\prime} \left. \frac{\partial U_i}{\partial x_j}\right\vert_{\vb{x}} \dd[3]{\vb{x}} \\
&= \int\int D_{ijkl}(\vb{x}, \vb{y}) \left. \pdv{U_l}{x_k} \right\vert_{\vb{y}} \left. \frac{\partial U_i}{\partial x_j}\right\vert_{\vb{x}} \dd[3]{\vb{y}} \dd[3]{\vb{x}} \ge 0
\label{eq:production}
\end{align}
\corr{The last relation is the same statement as conditioning the eddy viscosity operator to be positive semi definite. For well-posedness of its RANS mathematical model, any given eddy viscosity field must satisfy this condition for all arbitrary admissible input mean velocity gradients $\partial U_i/\partial x_j$. Otherwise, there will be unstable modes of mean flow that can be energized by the turbulence model, leading to their time exponential blow up.}

\corr{Equation~\ref{eq:production} can be further simplified to a single spatial integral when the eddy viscosity operator is local $D_{ijkl}(\vb{x}, \vb{y})=D^0_{ijkl}(\vb{x})\delta (\vb{y}-\vb{x})$ such as in the case of the leading order eddy viscosity. Substitution of a local model in~\ref{eq:production} results in:} 
\begin{align}
\int{P_k}\dd[3]{\vb{x}} \simeq \int D^0_{ijkl}(\vb{x}) \left. \pdv{U_l}{x_k} \right\vert_{\vb{x}} \left. \frac{\partial U_i}{\partial x_j}\right\vert_{\vb{x}} \dd[3]{\vb{x}}
\label{eq:productionleadingorder}
\end{align}
Since the operator now \corr{involves the local interactions of the mean velocity gradient and since this condition must hold for all fields of $\partial U_i/\partial x_j$,} the positive definiteness must be satisfied for each point. In other words, $D^0_{ijkl} \frac{\partial U_j}{\partial x_i} \frac{\partial U_l}{\partial x_k} \geq 0$ must also be satisfied pointwise for each local $D^0_{ijkl}$ tensor and for all admissible values of mean velocity gradient. Therefore, the quadratic form of the eddy viscosity tensor must be non-negative implying that the eddy viscosity tensor must be positive semi-definite. This result is also similar to the condition considered by \citet{Milani2020} in the context of local scalar transport.

Using our MFM measurement of the eddy viscosity, we can examine whether a local model from the truncated Kramer-Moyal expansion satisfies the positive semi-definite condition. \corr{Since the exact eddy viscosity must satisfy the positive semi-definite condition in~\ref{eq:production}, if this condition is not satisfied for the leading-order eddy viscosity,} it is an indication that the local truncation is not valid and nonlocality is needed for the positive definiteness condition.

To test the positive semi-definiteness of the leading order eddy viscosity tensor, we flatten the eddy viscosity tensor and the velocity gradient. Then, the turbulent production becomes $D^0_{ijkl} \frac{\partial U_j}{\partial x_i} \frac{\partial U_l}{\partial x_k}=\mathbf{z}^T \mathbf{D} \mathbf{z} $ where $\mathbf{z}$ is the flattened velocity gradient $[{\partial U_1} / {\partial x_1}~{\partial U_2}/{\partial x_1}~{\partial U_1}/{\partial x_2}~{\partial U_2}/{\partial x_2}]^T$ and where $\mathbf{D}$ is the following matrix:
\begin{eqnarray*}
  \mathbf{D}
    = 
    \begin{bmatrix} 
    D^0_{1111} & D^0_{1112} & D^0_{1121} & D^0_{1122} \\
    D^0_{1211} & D^0_{1212} & D^0_{1221} & D^0_{1222} \\
    D^0_{2111} & D^0_{2112} & D^0_{2121} & D^0_{2122} \\
    D^0_{2211} & D^0_{2212} & D^0_{2221} & D^0_{2222} \\
    \end{bmatrix}
\end{eqnarray*}

\begin{figure}
  \vspace{0.3cm}
  \centerline{\includegraphics[width=0.5\linewidth]{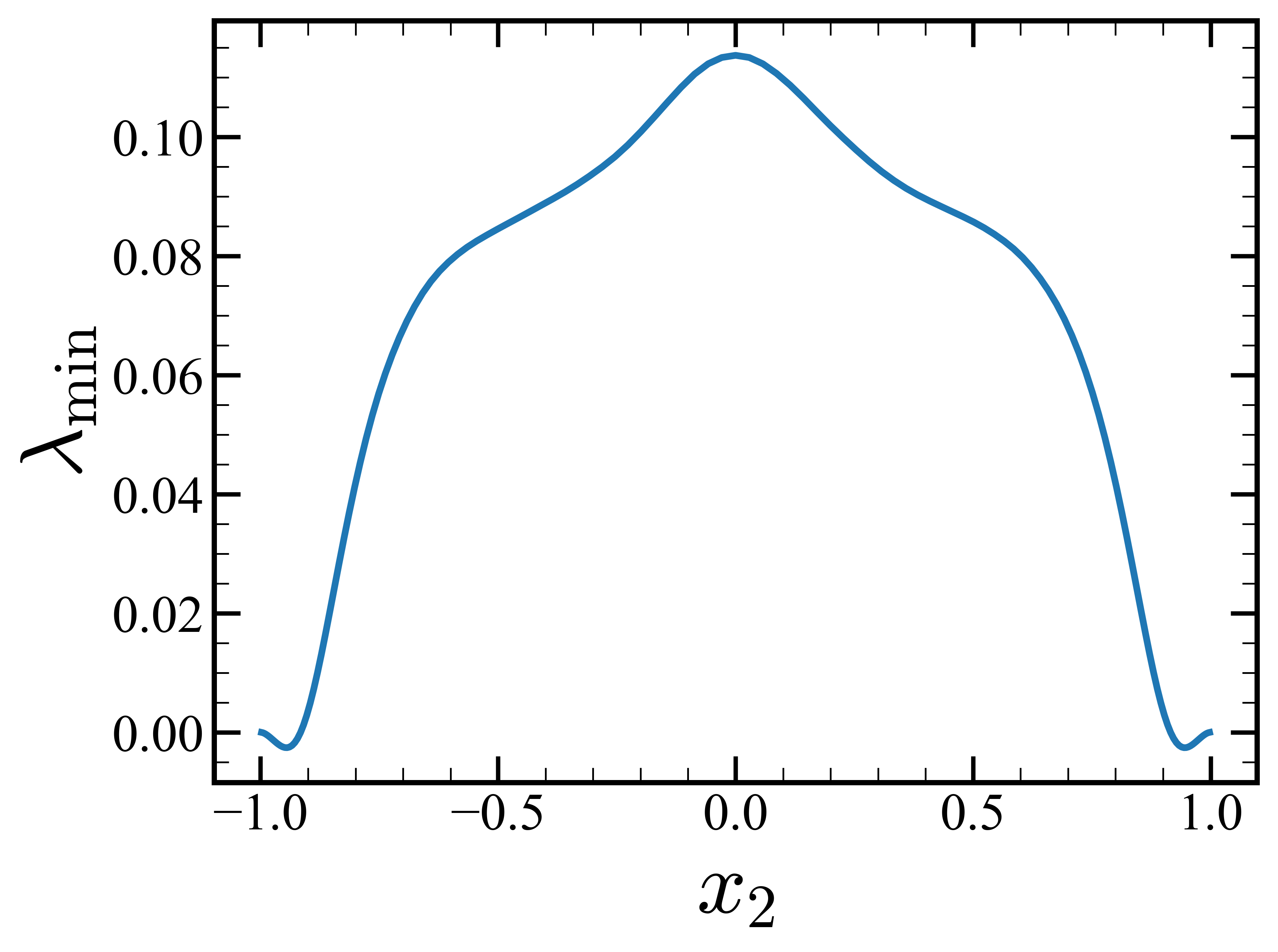}}
  \caption{The minimum eigenvalue of the matrix $\mathbf{C}\left(\mathbf{D}+\mathbf{D}^T\right)\mathbf{C}^T$.}
\label{fig:positiveDefinite}
\end{figure}

It is well known that a symmetric $\mathbf{D}$ is positive semi-definite if and only if \corr{all of its eigenvalues are non-negative} 
However, for our case, $\mathbf{D}$ is non-symmetric and $\mathbf{z}$ is limited to only certain value due to the incompressible condition. Therefore, we modified the quantity of interest. First, instead of the non-symmetric matrix $\mathbf{D}$, we look at the positive definiteness of $\mathbf{D}+\mathbf{D}^T$. If $\mathbf{D}+\mathbf{D}^T$ is positive semi-definite, $\mathbf{z}^T \mathbf{D} \mathbf{z} \geq 0$ also holds \citep{Milani2020}. Secondly, since the flow system is incompressible, $\mathbf{z}$ is limited to certain values satisfying ${\partial U_2} / {\partial x_2} = -{\partial U_1} / {\partial x_1}$. To expand the column vector multiplied to matrix $\mathbf{D}$ to every nonzero real column vector, we must embed the incompressibility condition to the matrix $\mathbf{D}$. We define $\mathbf{z} = \mathbf{C} \mathbf{z^*}$ where $\mathbf{z^*}$ is the reduced flattened velocity gradient $[{\partial U_1} / {\partial x_1}~{\partial U_2}/{\partial x_1}~{\partial U_1}/{\partial x_2}]^T$ and $\mathbf{C}$ is the following matrix:
\begin{eqnarray*}
  \mathbf{C}
    = 
    \begin{bmatrix} 
    1 & 0 & 0 \\
    0 & 1 & 0 \\
    0 & 0 & 1 \\
    -1 & 0 & 0 \\
    \end{bmatrix}
\end{eqnarray*}
Using this definition, $\mathbf{z}^T \mathbf{D} \mathbf{z}$ becomes $\mathbf{z^*}^T \mathbf{C}^T\mathbf{D}\mathbf{C}\mathbf{z^*}$. Combining these two methods, 
we conclude that the eddy viscosity tensor is positive semi-definite when all the eigenvalues of the matrix $\mathbf{C}\left(\mathbf{D}+\mathbf{D}^T\right)\mathbf{C}^T$ is non-negative. We computed the smallest eigenvalue of this matrix at each $x_2$. The resulting plot is shown in Figure~\ref{fig:positiveDefinite}. As shown, except for the thin zones near the wall, we see that the eigenvalues are positive, hence the eddy viscosity tensor is positive definite. 
Near the wall, however, the eigenvalues become negative, indicating that the leading term is not sufficient to capture the positive semi-definiteness of the eddy viscosity. In other words, this negativity occurs due to the local truncation of the eddy viscosity tensor and therefore nonlocality should be incorporated to make the eddy viscosity positive semi-definite. \corr{The same analysis was conducted in the three dimension space context and the conclusion remains the same.}


\section{Nonlocality Analysis}

In Section 3.3, we concluded that aside from $\overline{u'_2 u'_1}$, capturing other components of the Reynolds stress field requires inclusion of nonlocal terms in the eddy viscosity operator. To better understand the nonlocal effect, in this section, we investigate the full eddy viscosity kernel $D_{ijkl}\left(x_2, y_2 \right)$ in Equation \ref{eq:generalformchannel}.

\subsection{Nonlocality}

\begin{figure}
\vspace{0.3cm}
\centering
  \subfigure[$D_{2121}$]{\includegraphics[width=0.3\linewidth]{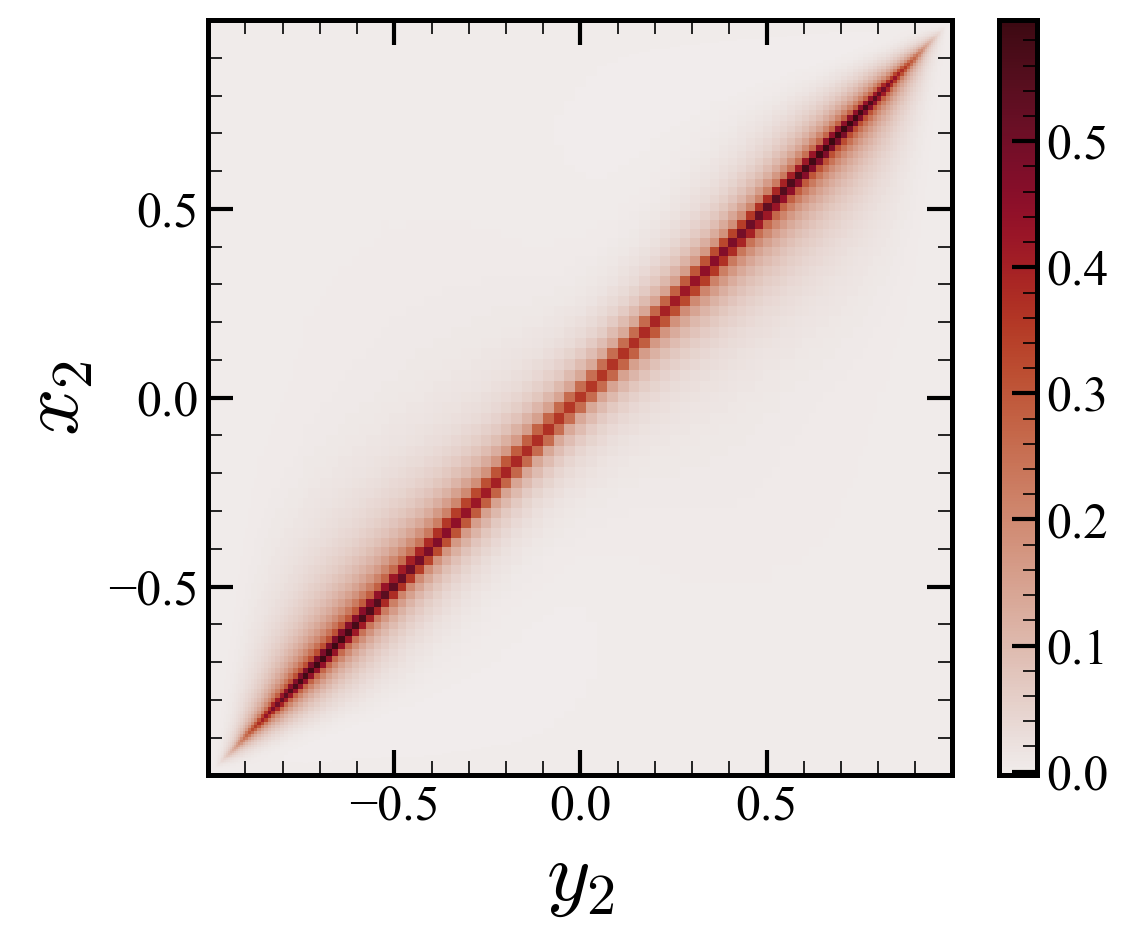}} 
  \subfigure[$D_{2121}$]{\includegraphics[width=0.34\linewidth]{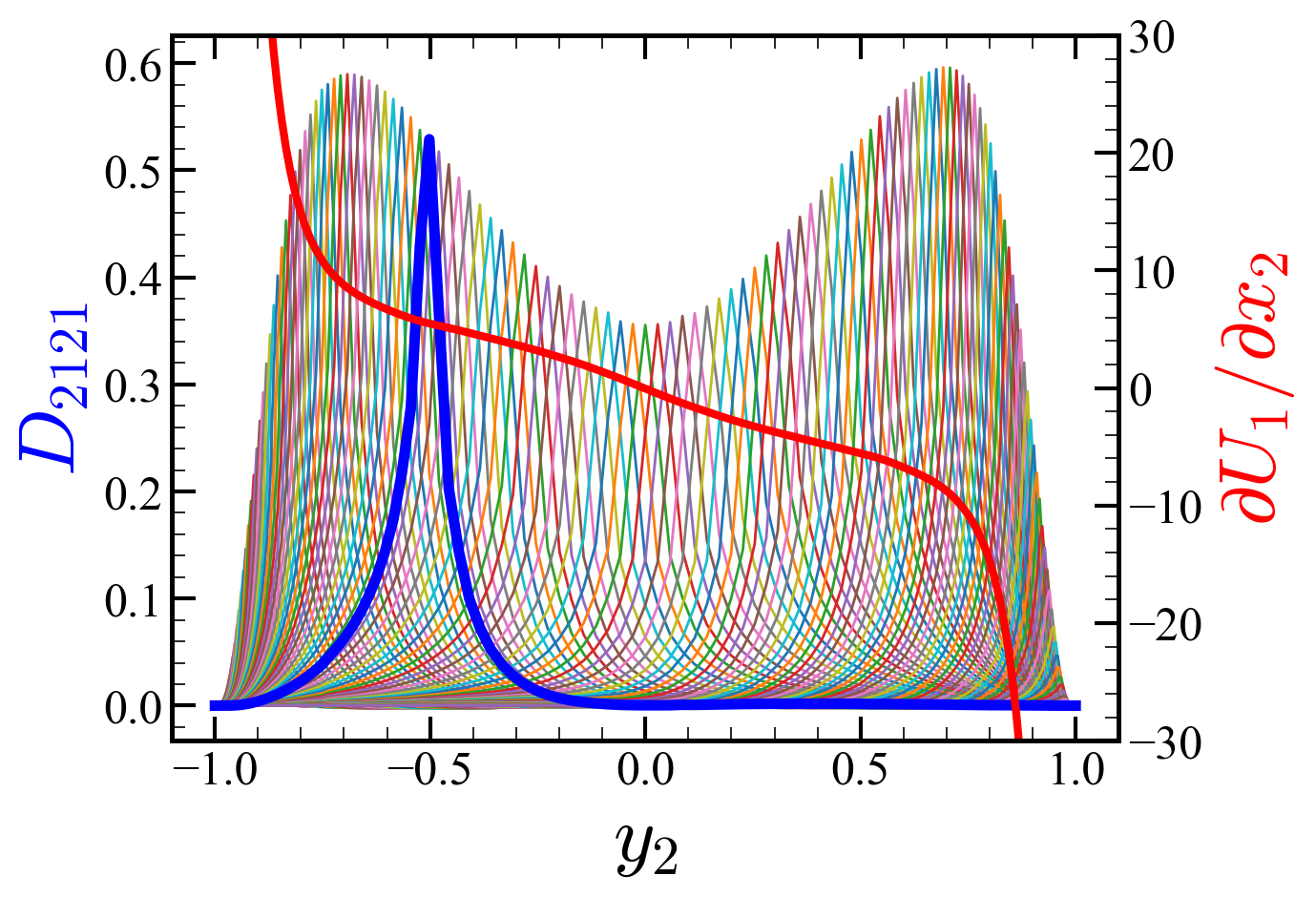}} 
  \subfigure[$\int D_{2121} \mathrm{d}y_2$]{\includegraphics[width=0.3\linewidth]{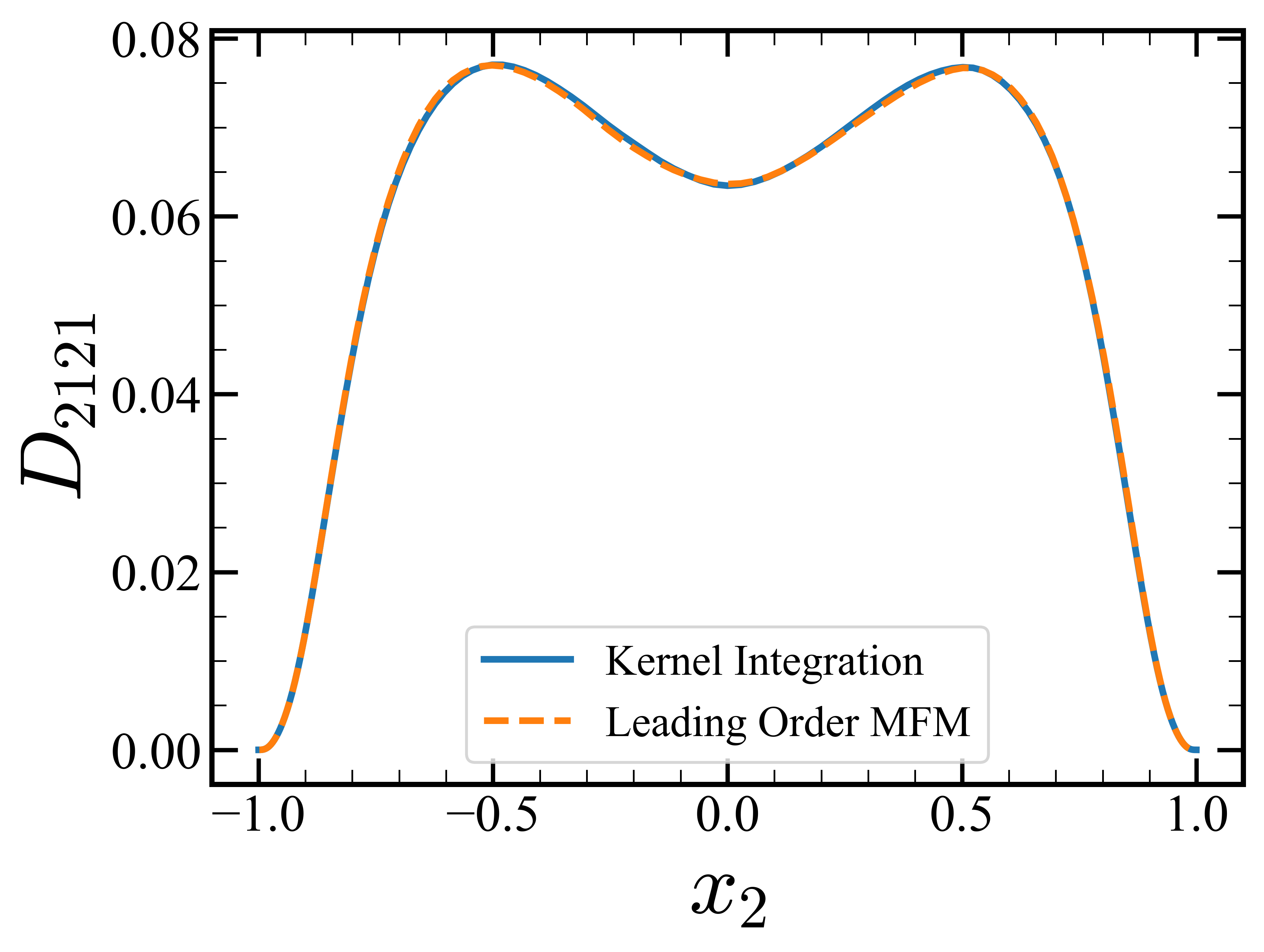}} 
  \caption{Distribution of $D_{2121}$: (a) contour plot of $D_{2121}$, (b) $D_{2121}\left(x_2=x_2^*,y_2\right)$ with various $x_2^*$ where the blue line is at $x_2^*=0.502$ \corr{and the velocity gradient $\partial U_1 / \partial x_2 $ is drawn in red line}, and (c) the blue solid line, $\int D_{2121} \mathrm{d}y_2$ and the orange dashed line,  $D^0_{2121}$.}
\label{fig:D2121}
\end{figure}

Figure~\ref{fig:D2121} shows the full eddy viscosity kernel representation of the $D_{2121}$ component. Each point in Figure~\ref{fig:D2121}(a) represents the effect of the velocity gradient at the location $y_2$ to the Reynolds stress at the location $x_2$. The distribution of $D_{2121}$ is confined to $x_2 \sim y_2$, indicating the approximate locality of this eddy viscosity component. At a given location $x_2$, we can visualize how much contribution the remote velocity gradient at a different location, $y_2$, makes to the Reynolds stress at the location $x_2$.

For instance, in Figure~\ref{fig:D2121}(b), the thick blue line represents $D_{2121}\left(x_2=0.502,y_2\right)$ and the distribution indicates the effects of the velocity gradient nearby. For a purely local eddy viscosity, a delta function around $y_2=0.502$ would b e expected. In the figure, even though the plotted profile is not a Dirac delta function, $D_{2121}\left(x_2=0.502,y_2\right)$ shows concentrated behavior around  $y_2=0.502$. \corr{The blue line peak value is $D_{2121}=0.53$ and the width that the curve drops to one third of its peak value is 0.12.} 
Overall, our narrow banded results indicate that $D_{2121}$ is \corr{relatively} local throughout the domain. \corr{Such locality explains the earlier observation in Section 3 where the leading-order eddy viscosity was shown to construct the shear component of the Reynolds stress within roughly 10\% error, and the mean velocity profile within 1\% error. However, a more quantitatively rigorous assessment would require examination of the mean velocity gradient across the kernel width. The leading-order eddy viscosity relegates the entire sensitivity of Reynolds stresses to the local pointwise value of the mean velocity gradient. If the mean velocity gradient happens to be relatively constant across the kernel width, the pointwise approximation, and hence the local model will be accurate. As shown in Figure~\ref{fig:D2121}(b), the mean velocity gradient has a non-negligible variation across the sample kernel indicated by the blue curve. Therefore, strictly speaking, nonlocal effects in $D_{2121}$ should not be negligible. We conclude that the accurate outcome of the local approximation for $D_{2121}$ is partially owed to the error cancellation due to monotonic variation of the mean velocity gradient in the domain. In other words, the pointwise value of $\partial V_1/\partial x_2$ near the centroid of the kernel, reasonably represents the mean value given that errors from the left side and right side of the kernel partially cancel out each other.}

 Figure~\ref{fig:D2121}(c) compares the results of kernel integration $\int{D_{2121} \mathrm{d}y_2}$ against the leading-order eddy viscosity for the same component $D^0_{2121}$. The definition of $D^0_{2121}$ is the leading-order moment of the eddy viscosity kernel $D_{2121}$. In other word, with correct quantification the integration of the kernel $\int{D_{2121} \mathrm{d}y_2}$ must match $D^0_{2121}$. Figure~\ref{fig:D2121}(c) shows that the two results are collapsing verifying the consistency between our two different MFM measurements.
\begin{figure}
  \vspace{0.3cm}
  \centerline{\includegraphics[width=0.5\linewidth]{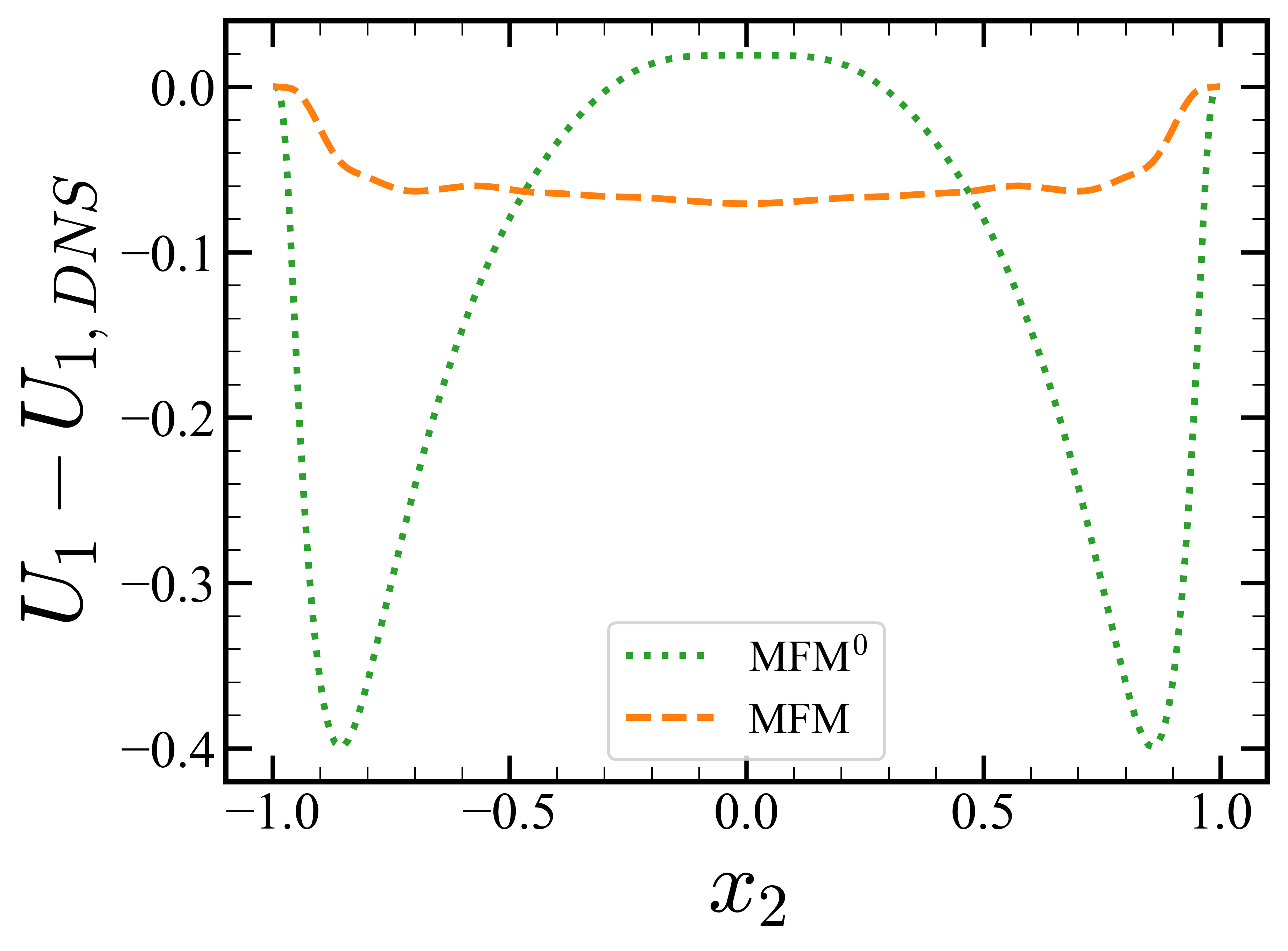}}
  \caption{Error in RANS prediction of the streamwise velocity using eddy viscosity tensor $D^0_{2121}$ (denoted as MFM$^0$) and eddy viscosity kernel $D_{2121}$ (denoted as MFM), $U_{1}-U_{1,\mathrm{DNS}}$; the dotted green line, $U_1$ predicted with $D^0_{2121}$ and the dashed orange line, $U_1$ predicted with $D_{2121}$.}
\label{fig:ranserror}
\end{figure}

In Section 3, we demonstrated that the leading-order eddy viscosity alone can predict an accurate RANS solution for the channel mean velocity with the prediction error around 1\%. This error can be furture reduced by including the nonlocality using the full kernel representation of the eddy viscosity. Figure~\ref{fig:ranserror} shows the two RANS results, one obtained using the leading-order eddy viscosity $D^0_{2121}$ and the other obtained using eddy viscosity kernel $D_{2121}$. Analytically, the full measurement of the kernel is expected to provide the RANS solution that is identical to the averaged DNS result. In our simulation, small errors are due to statistical noise that we expect to resolve with a larger data set. Still, the kernel result is significantly better than the leading-order result, indicating that the RANS solution to the leading order eddy diffusivity model, which is highly local, can be improved using a nonlocal model.

\begin{figure}
\vspace{0.3cm}
\centering
  \subfigure[$D_{1221}$]{\includegraphics[width=0.3\linewidth]{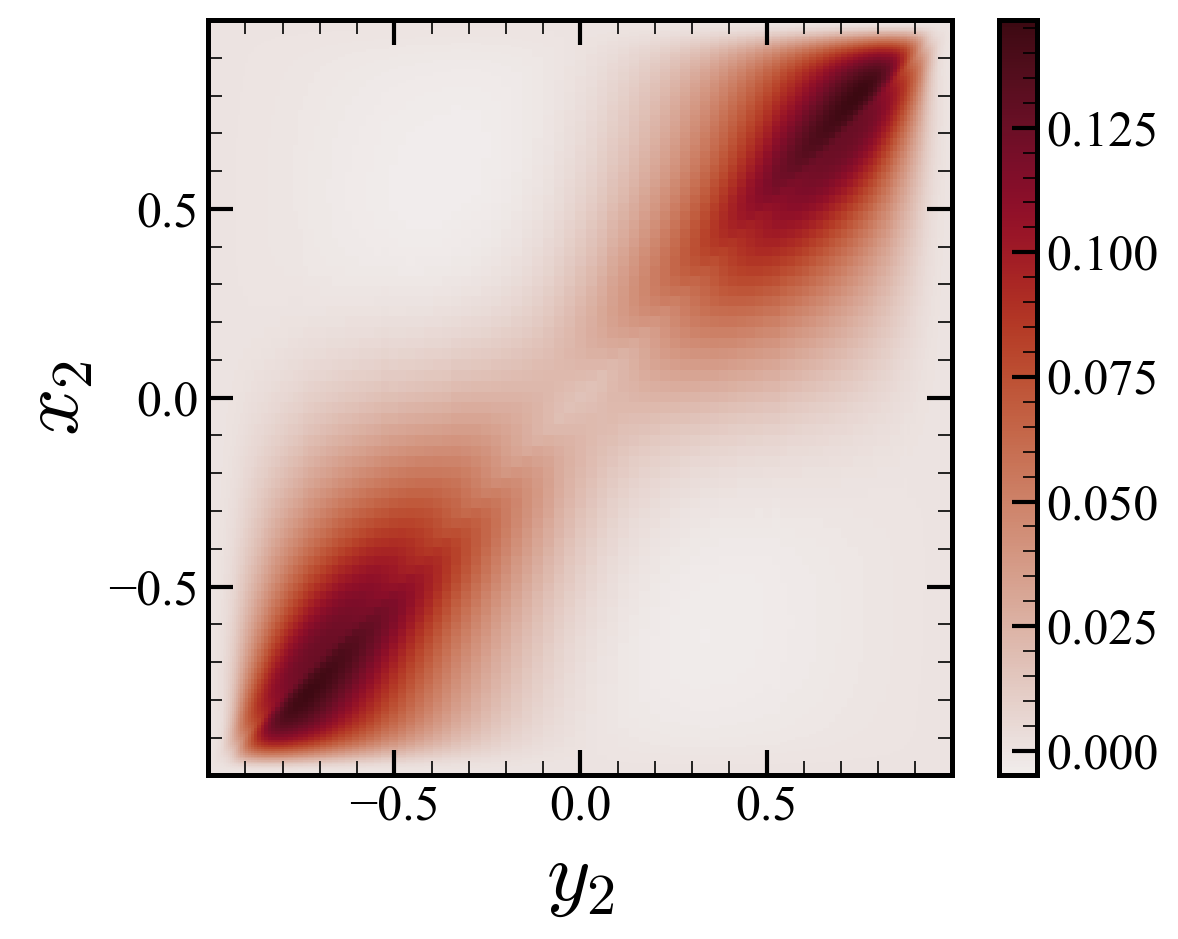}} 
  \subfigure[$D_{1221}$]{\includegraphics[width=0.3\linewidth]{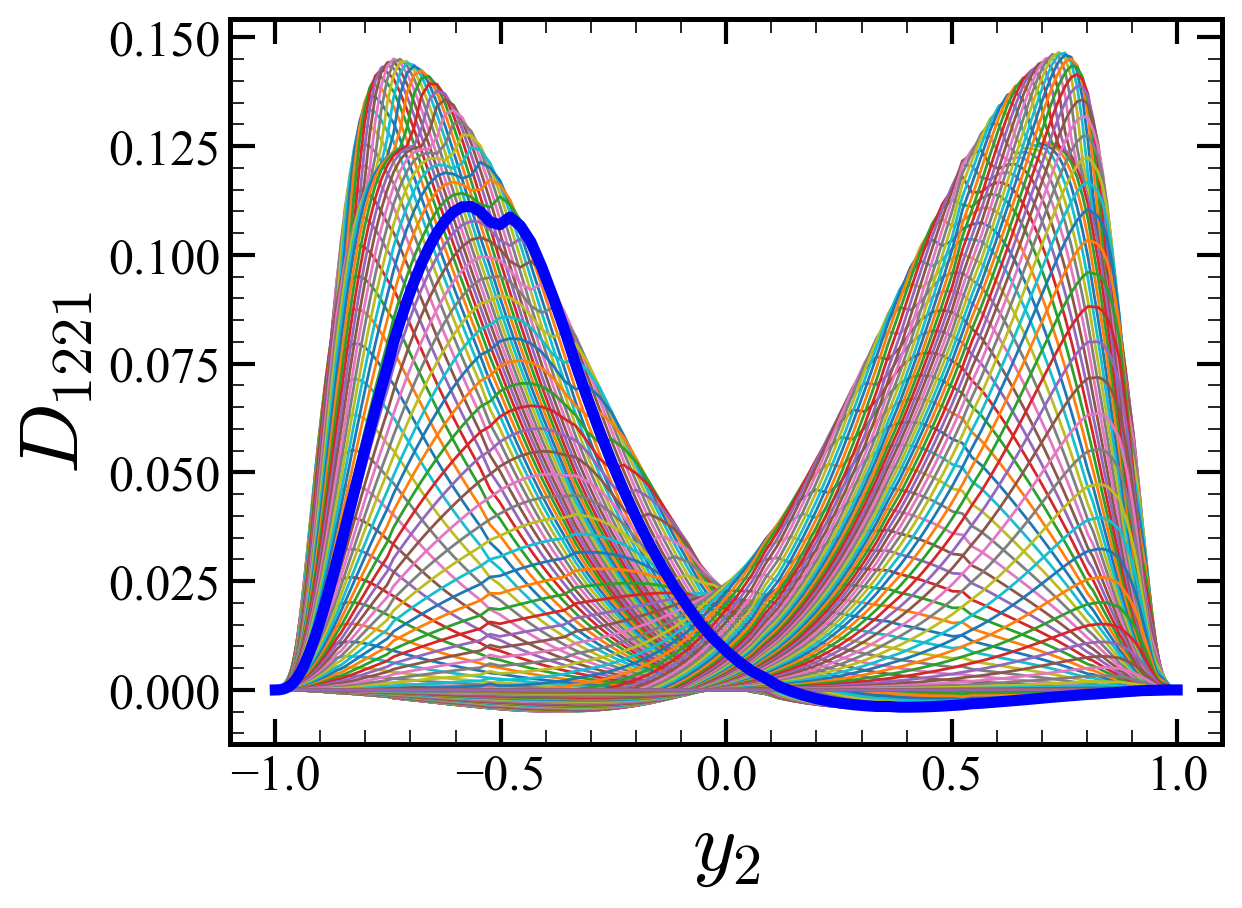}} 
  \subfigure[$\int D_{1221} \mathrm{d}y_2$]{\includegraphics[width=0.3\linewidth]{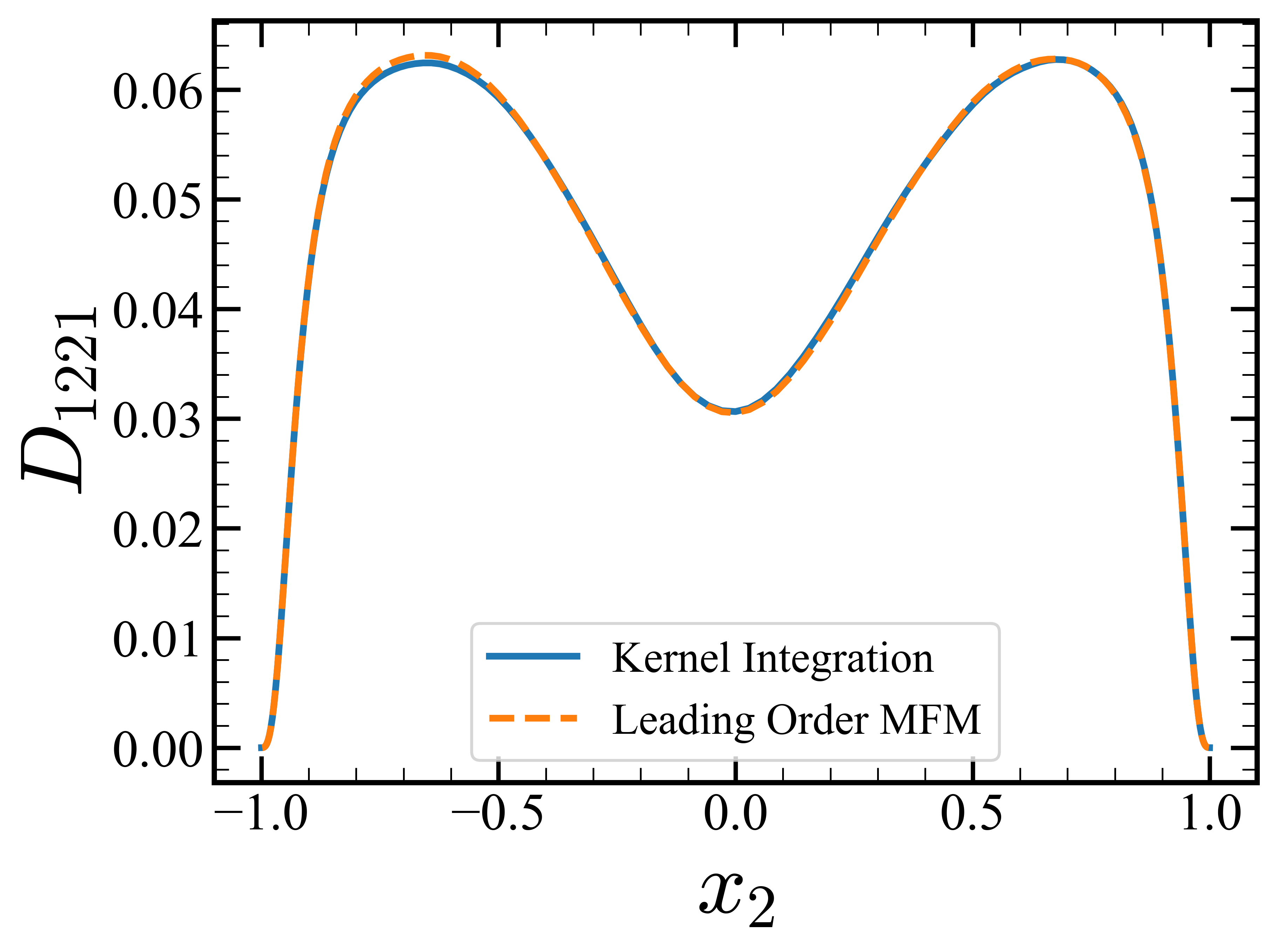}} 
  \caption{Distribution of $D_{1221}$: (a) contour plot of $D_{1221}$, (b) $D_{1221}\left(x_2=x_2^*,y_2\right)$ with various $x_2^*$ where the blue line is at $x_2^*=0.502$, and (c) blue solid line, $\int D_{1221} \mathrm{d}y_2$ and orange dashed line,  $D^0_{1221}$.}
\label{fig:D1221}
\end{figure}

\begin{figure}
\vspace{0.3cm}
\centering
  \subfigure[$D_{1121}$]{\includegraphics[width=0.3\linewidth]{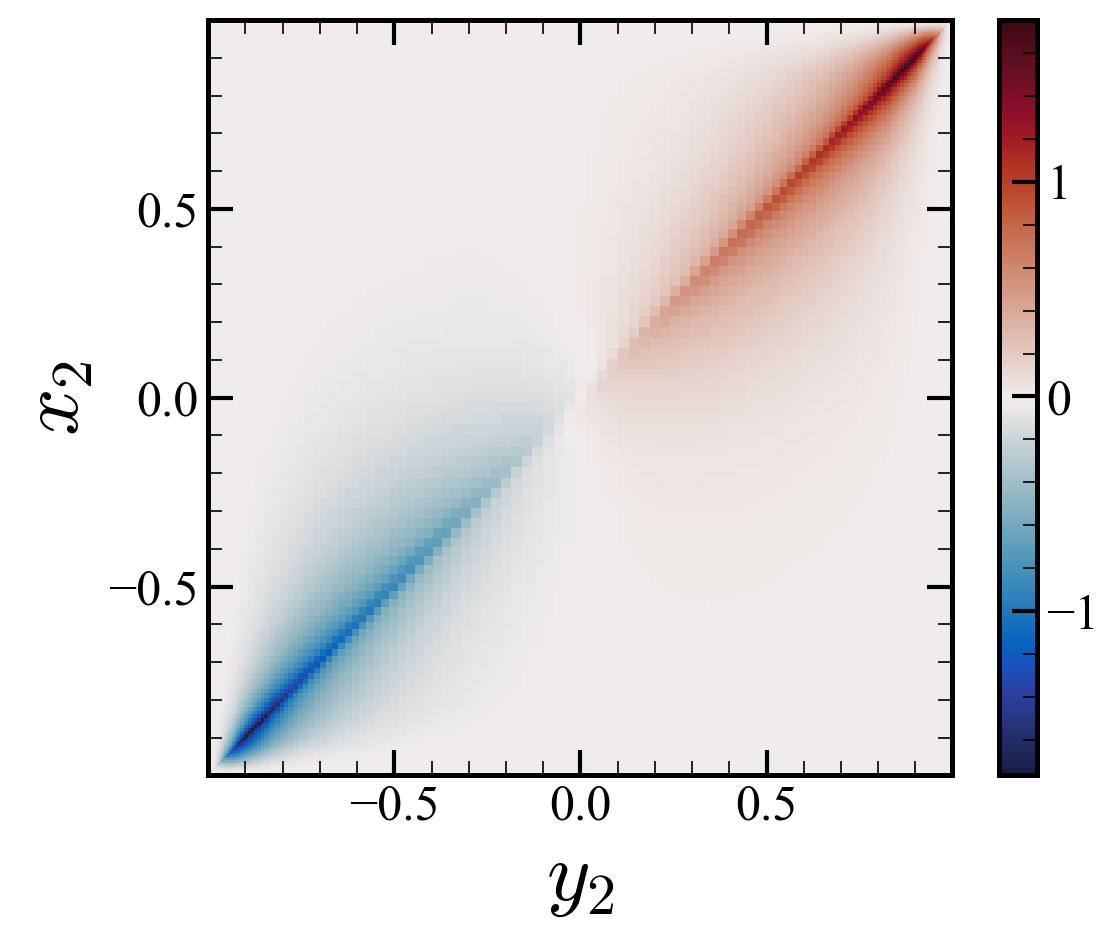}} 
  \subfigure[$D_{2221}$]{\includegraphics[width=0.3\linewidth]{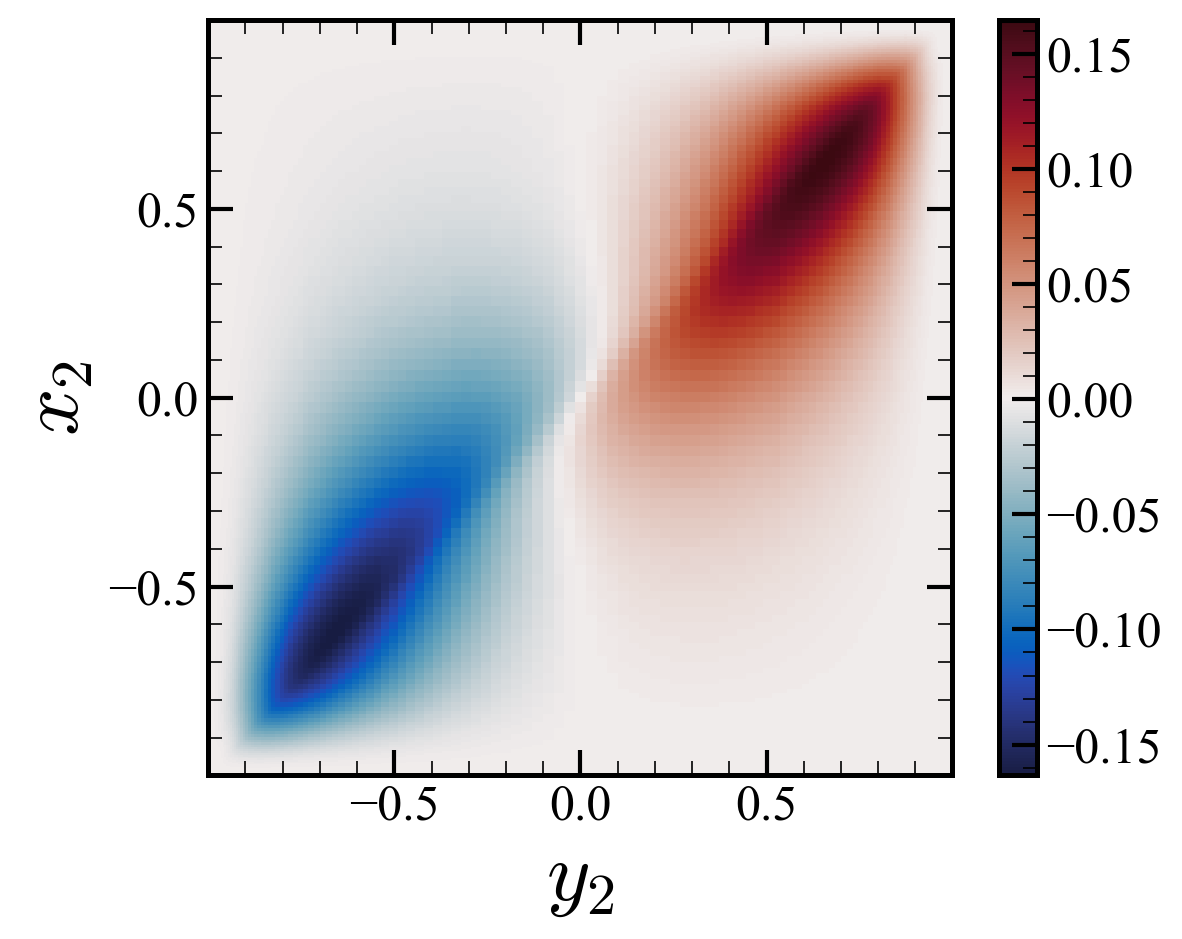}} 
  \subfigure[$D_{3321}$]{\includegraphics[width=0.3\linewidth]{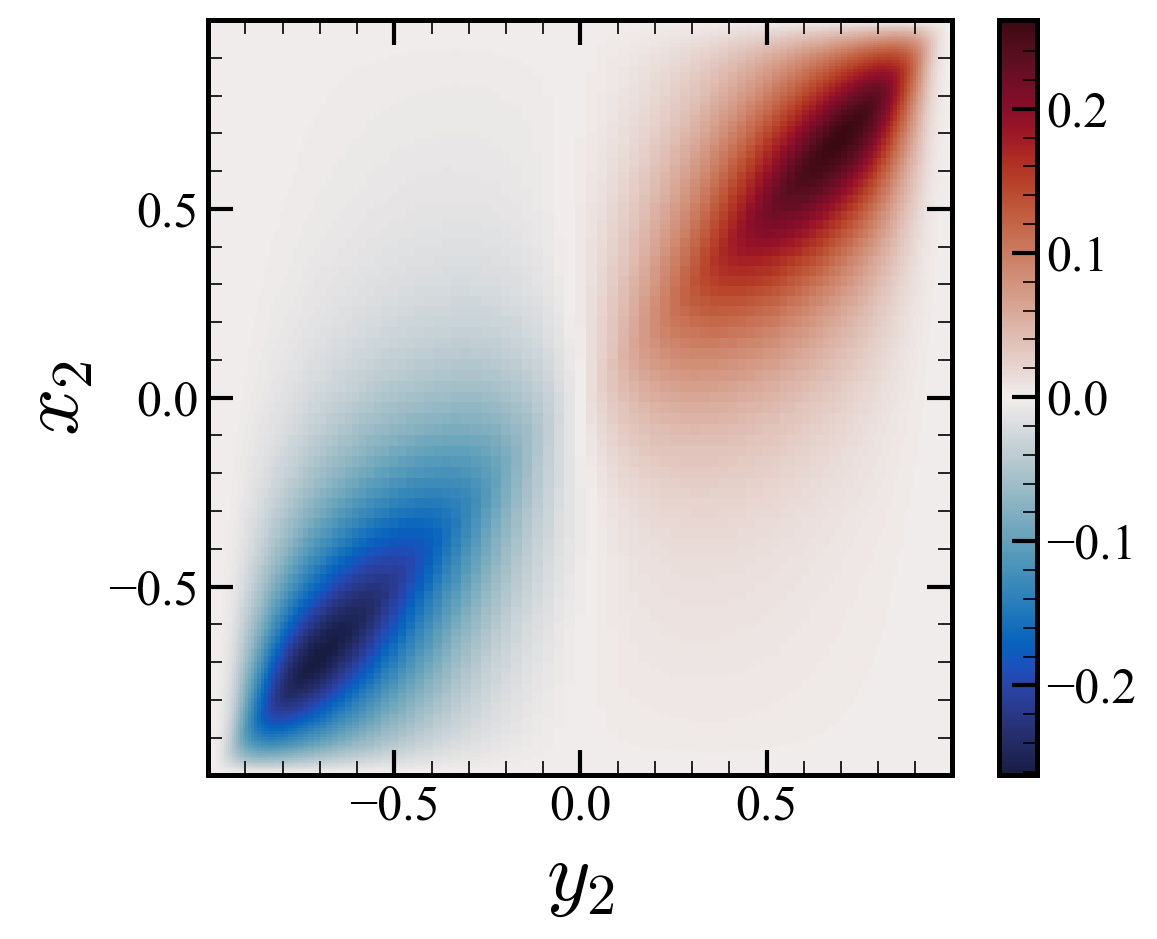}} 
  \caption{Distribution of nonzero $D_{ij21}$ where (a) $D_{1121}$, (b) $D_{2221}$, and (c) $D_{3321}$.}
\label{fig:Dji21}
\end{figure}

Next, we assess nonlocality of other components of $D_{ijkl}$ by examining the corresponding kernels. For example, the eddy viscosity kernel $D_{1221}$ (Figure~\ref{fig:D1221}) is widespread and shows significant nonlocality, invalidating the intrinsic assumption in the Boussinesq approximation. The level of nonlocality is drastic such that the velocity gradient at one half of the channel may affect the Reynolds stress at the other half of the domain. Moreover, the shape of $D_{2121}$ differs from $D_{1221}$, implying the non-universality of the kernel profile across different components of eddy diffusivity. Furthermore, the differences in the kernel shape between $D_{2121}$ and $D_{1221}$, clarifies why a truncated eddy viscosity operator based on the leading term of its Kramer-Moyal expansion can lead to asymmetric Reynolds stresses. Figure~\ref{fig:Dji21} shows the additional three nonzero eddy viscosity kernels $D_{1121}$, $D_{2221}$,  and $D_{3321}$. The rest of the components are zero due to channel symmetry. These three kernels correspond to the trace part of the eddy viscosity kernel and are also highly nonlocal. 

\subsection{Revisit of Reynolds stress}

\begin{figure}
\vspace{0.3cm}
\centering
  \subfigure[$D_{1121}$]{\includegraphics[width=0.3\linewidth]{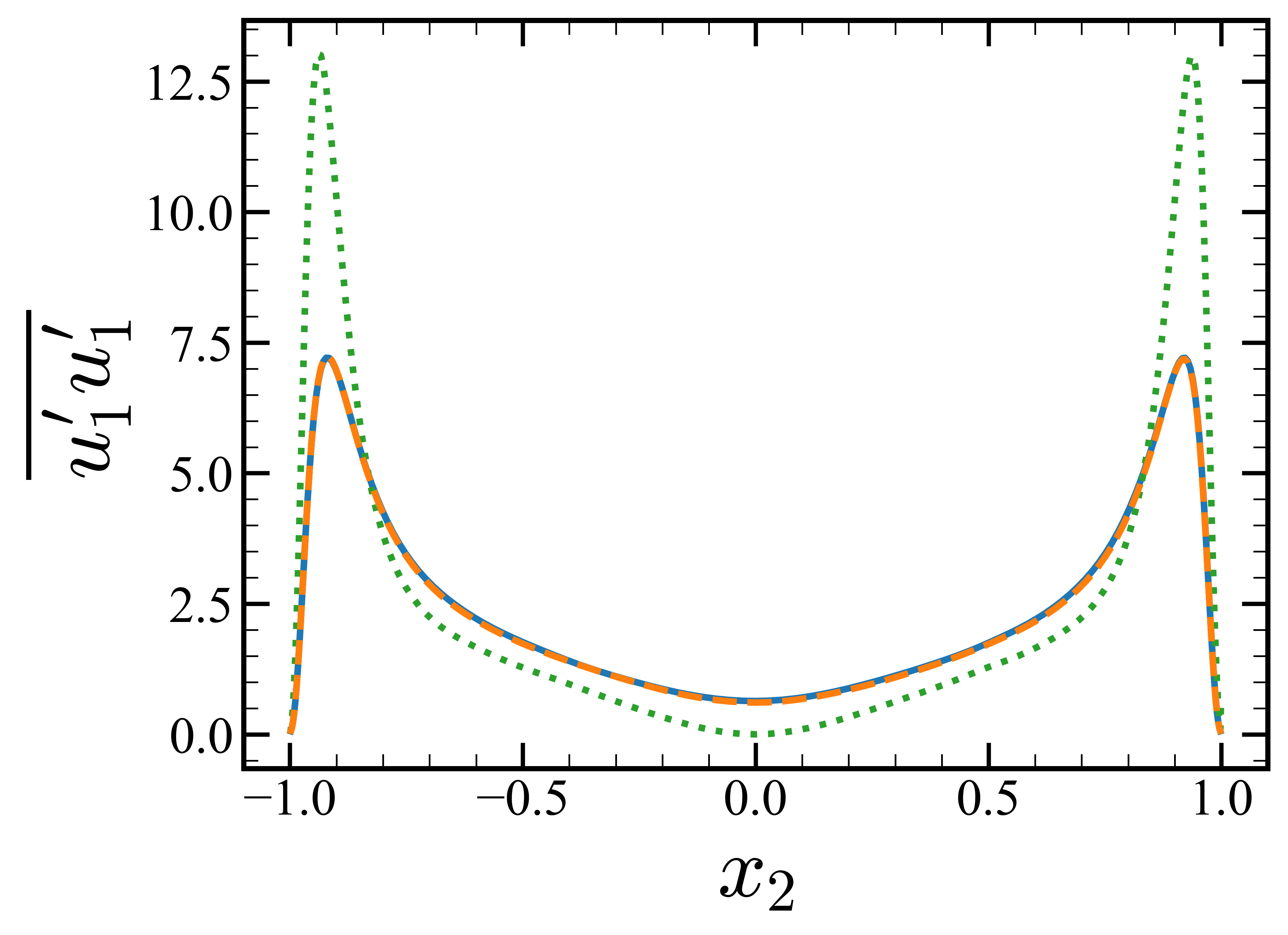}} 
  \subfigure[$D_{1221}$]{\includegraphics[width=0.3\linewidth]{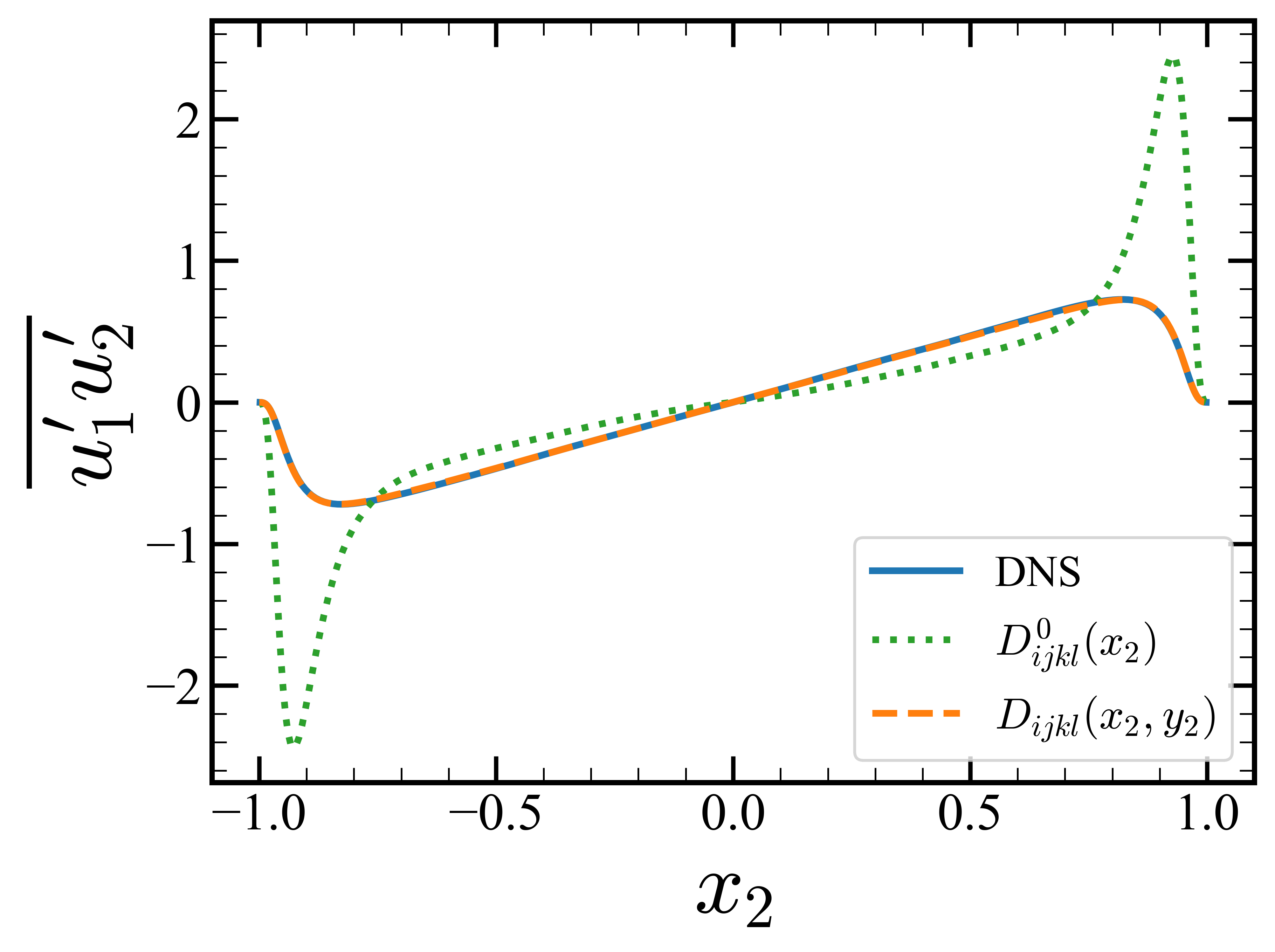}} \\
  \subfigure[$D_{2121}$]{\includegraphics[width=0.3\linewidth]{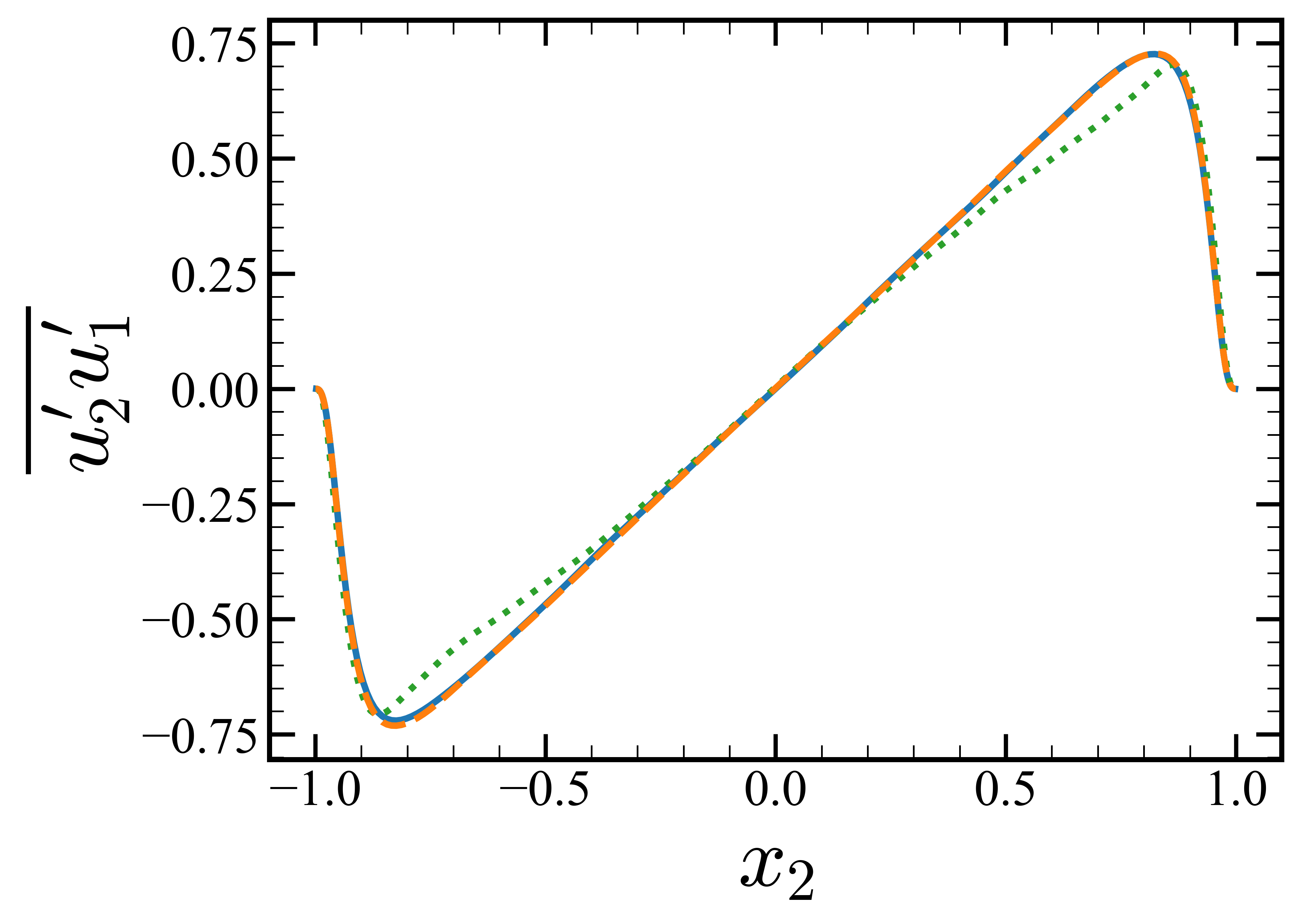}} 
  \subfigure[$D_{2221}$]{\includegraphics[width=0.3\linewidth]{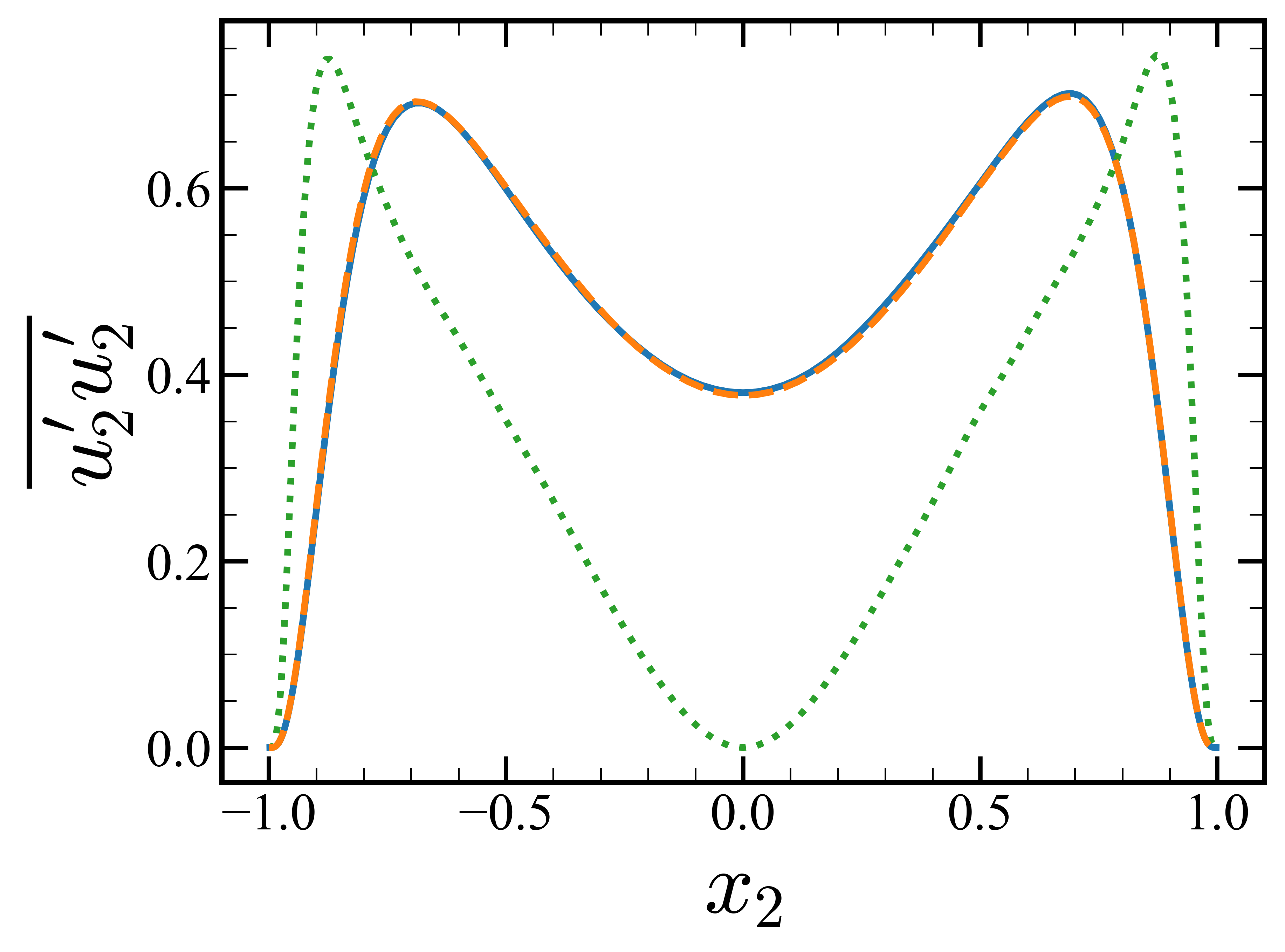}} 
  \subfigure[$D_{3321}$]{\includegraphics[width=0.3\linewidth]{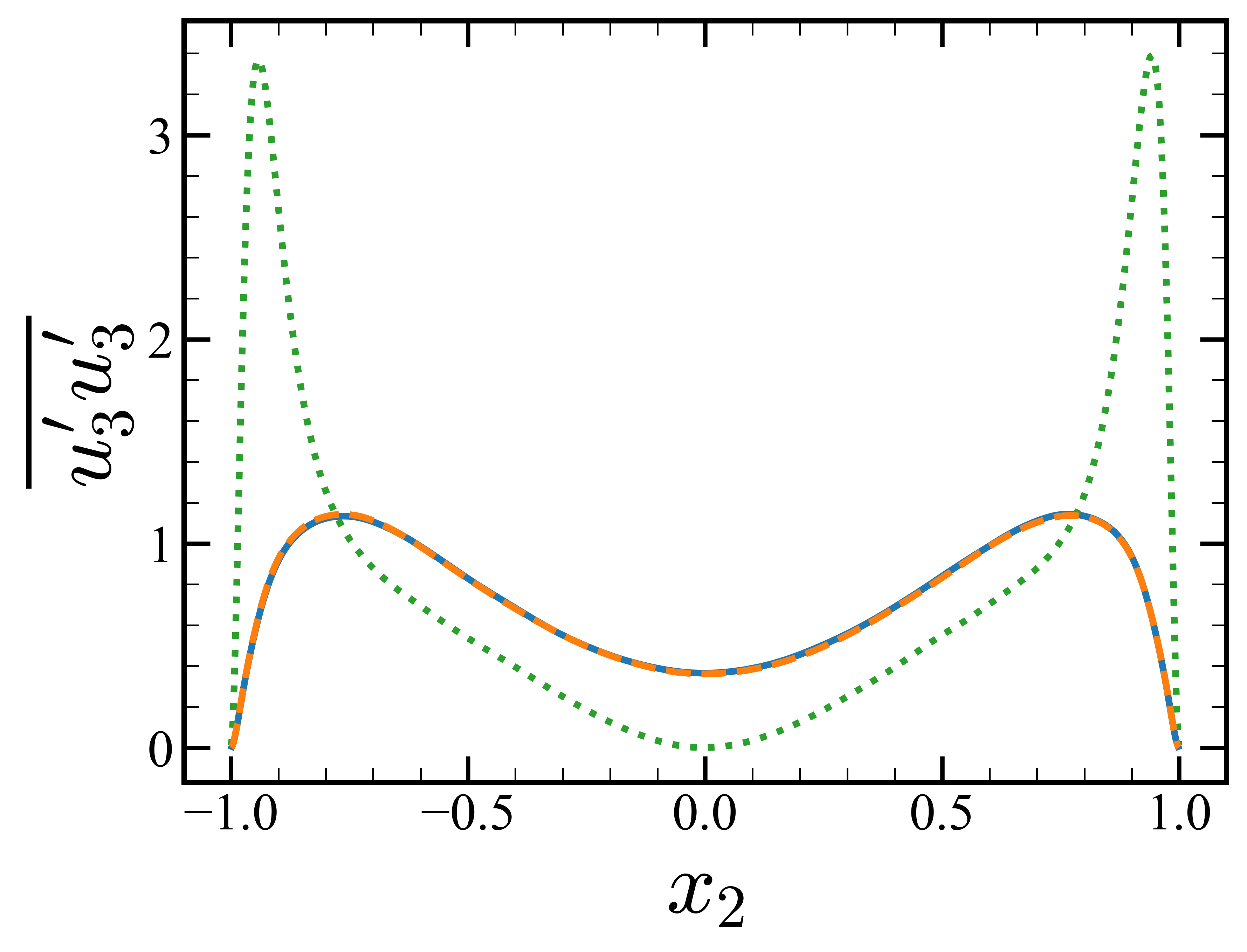}} 
  \caption{Reynolds stresses constructed by the leading-order eddy viscosity tensor and the eddy viscosity kernel associated with (a) $D_{1121}$, (b) $D_{1221}$, (c) $D_{2121}$, (d) $D_{2221}$, and (e) $D_{3321}$: orange dashed line, the reconstructed Reynolds stress by the eddy viscosity kernel; green dotted line, the reconstructed Reynolds stress by the leading-order eddy viscosity tensor \corr{$\overline{u'_i u'_j}^0$}; blue solid line, the DNS data.}
\label{fig:resynoldsstress}
\end{figure}

Lastly, we revisit the Reynolds stress reconstruction with the inclusion of the nonlocal effects. Figure~\ref{fig:resynoldsstress} shows three different ways of constructing the Reynolds stresses. The first shown in the orange dashed line is the reconstructed Reynolds stress by the eddy viscosity kernel from MFM and the mean velocity gradient from DNS. The second shown in green dotted line is the reconstructed Reynolds stress by the leading-order eddy viscosity tensor from MFM and the mean velocity gradient from DNS. The last one is from the mean DNS data shown in blue solid line. The leading-order result and the mean DNS data are shown before in Figure~\ref{fig:localresynoldsstress}.

Unlike the leading-order analysis, the results from the full kernel eddy viscosity matche very well to the DNS data. These plots verify our computational method yielding two findings. First, with full kernels, the Reynolds stresses recover the symmetry that was lost in the leading-order approximation.
Only $D_{2121}$, which is relatively narrow banded, is applicable for the local approximation. Hence, this leads to the symmetry breakage after leading-order approximation. Second, now we can capture the nonzero Reynolds stress at the channel centerline.
Thus, the measured nonlocal eddy viscosity allows prediction of non-zero Reynolds stresses near the centerline, whereas the leading-order approximation fails to do so.

\subsection{Revisit of Positive Definiteness}

In Section 3.5, we discuss the positive definiteness of the local eddy viscosity tensor $D^0_{ijkl}$. Due to the leading-order truncation of the eddy viscosity kernel, $D_{ijkl}$, our result indicated that the local eddy viscosity tensor was not positive definite near the walls. In this section, we introduce the full kernel and examined whether including the nonlocality restores the semi-positive definite condition.

Using the full eddy viscosity kernel expression and applying the fact that only one component of the velocity gradient tensor is nonzero, the turbulent production in Equation~\ref{eq:production} can be written as the following:

\begin{eqnarray*}
    \int P_k \mathrm{d} x_2 = \int \int D_{2121}({x_2}, {y_2})\frac{\partial U_1}{\partial y_2} \frac{\partial U_1}{\partial x_2} \mathrm{ d}y_2 \mathrm{ d}x_2
    = \left[\frac{\partial U_1}{\partial x_2}\right]^T \left[D_{2121}\right] \left[\frac{\partial U_1}{\partial x_2}\right].
\end{eqnarray*}

The far right term represents the discrete form of the expression, where $\left[\frac{\partial U_1}{\partial x_2}\right]$ represents any velocity gradient vector at each point in $x_2$ and $\left[D_{2121}\right]$ represents the discrete matrix value of $D_{2121}({x_2}, {y_2})$. To make the turbulent production non-negative, the matrix $\left[D_{2121}\right]$ needs to be semi-positive definite. Likewise in Section 3.5, we computed eigenvalues of $\left[D_{2121}\right] + \left[D_{2121}\right]^T $ to determine the positive definiteness. The computed eigenvalues range from 0.00 to 7.58, indicating that the eddy viscosity kernel $D_{2121}$ is indeed semi-positive definite, recovering the stability condition that was lost by the leading-order truncation. 

\section{Conclusions}

This study presents a quantification of non-Boussinesq effects in eddy viscosity in a subclass of turbulent wall-bounded flows. The presented analyses is systematically focused on two aspects: anisotropy and nonlocality of momentum mixing. To assess these effects and quantify the deviation from Boussinesq limit, we calculated the eddy viscosity of the turbulent channel flow at $\mathrm{Re}_\tau=180$ using a statistical technique that we recently developed called MFM. Using MFM, we quantified the leading-order eddy viscosity tensor for the analysis of anisotropy and expanded our study to quantify the eddy viscosity tensorial kernel for the analysis of nonlocality.

Our results indicate the following: (1) eddy viscosity is highly anisotropic with some elements orders of magnitude larger than the nominal eddy viscosity; (2) the Reynolds stresses reconstructed from this eddy viscosity depends not only on the mean rate of strain but also on mean rate of rotation; (3) leading-order eddy viscosity, which is obtained by neglecting higher spatial moments of the closure kernel, generates a non-symmetric Reynolds stress tensor; and (4) aside from the shear component of the Reynolds stress, $\overline{u^\prime_2 u^\prime_1}$, \corr{which showed a limited level of nonlocality,} the dependence of other components of Reynolds stress on mean velocity gradient is highly nonlocal at the level where some components of the Reynolds stress are influenced by the velocity gradient on the other half of the channel. 

The exact measurement of the eddy viscosity of the channel flow has different implication for RANS modeling of parallel flows and that of the spatially developing \corr{attached} boundary layers. For parallel flows, only one Reynolds stress component and one velocity gradient are important; hence anisotropy does not influence the prediction of the mean flow as long as $D_{2121}$ is properly modeled. At the same time, not only the anisotropy but also nonlocality may be omitted for the channel flow. \corr{This outcome is in part due to relatively narrower kernel of $D_{2121}(x_2, y_2)$, as shown in our MFM measurement of the eddy viscosity kernel, but also due to coincidental error cancellations that render reasonable estimation of the Reynolds stresses based on a single point quadrature relegating the entire weight of the kernel on the local mean velocity gradient. }

These two findings may explain why the Boussinesq approximation works well for prediction of mean parallel flows. However, our quantification suggests that this conclusion does not hold for normal components of the Reynolds Stress, as well as for spatially developing wall-bounded flows where the non-parallel effects become important. For instance, even a small gradient in the streamwise direction can have a non-neglible effect since $D_{1111}^0$ is very large compared to most of other eddy viscosity components. Our measurements reveal that the eddy viscosity is highly anisotropic and highly nonlocal, when it comes to components other than $D_{2121}$, indicating a clear need to include non-Boussinesq effects in RANS models.

While we focused on full nonlocal analysis in the $x_2$ direction, we did not consider nonlocal spatial effects in other directions and nonlocal temporal effects. Equation \ref{eq:generalformchannel} is a reduced version of Equation \ref{eq:generalform} using leading-order moments in $x_1$, $x_3$, and $t$. These leading-order reductions are justified for channel flow since it is statistically homogeneous in these directions, and are expected to be qualitatively valid for systems with slow variation of turbulence in these directions. While it is possible to quantitatively assess such effects with MFM, we defer analysis of streamwise and spanwise nonlocality in eddy viscosity to a future study. 

\section{Funding}
This work was supported by the Office of Naval Research under grant No. N00013-20-1-2718, and the Boeing Company under grant No. 2017-STU-PA-287. Park was supported by the Stanford Graduate Fellowship and the Kwanjeong Educational foundation.

\section{Declaration of Interests}
The authors report no conflict of interest.

\newpage

\appendix
\section{Estimation of the convergence error}\label{appA}

\begin{figure}
\vspace{0.3cm}
\centering
  \subfigure[Temporal convergence]{\includegraphics[height=0.3\linewidth]{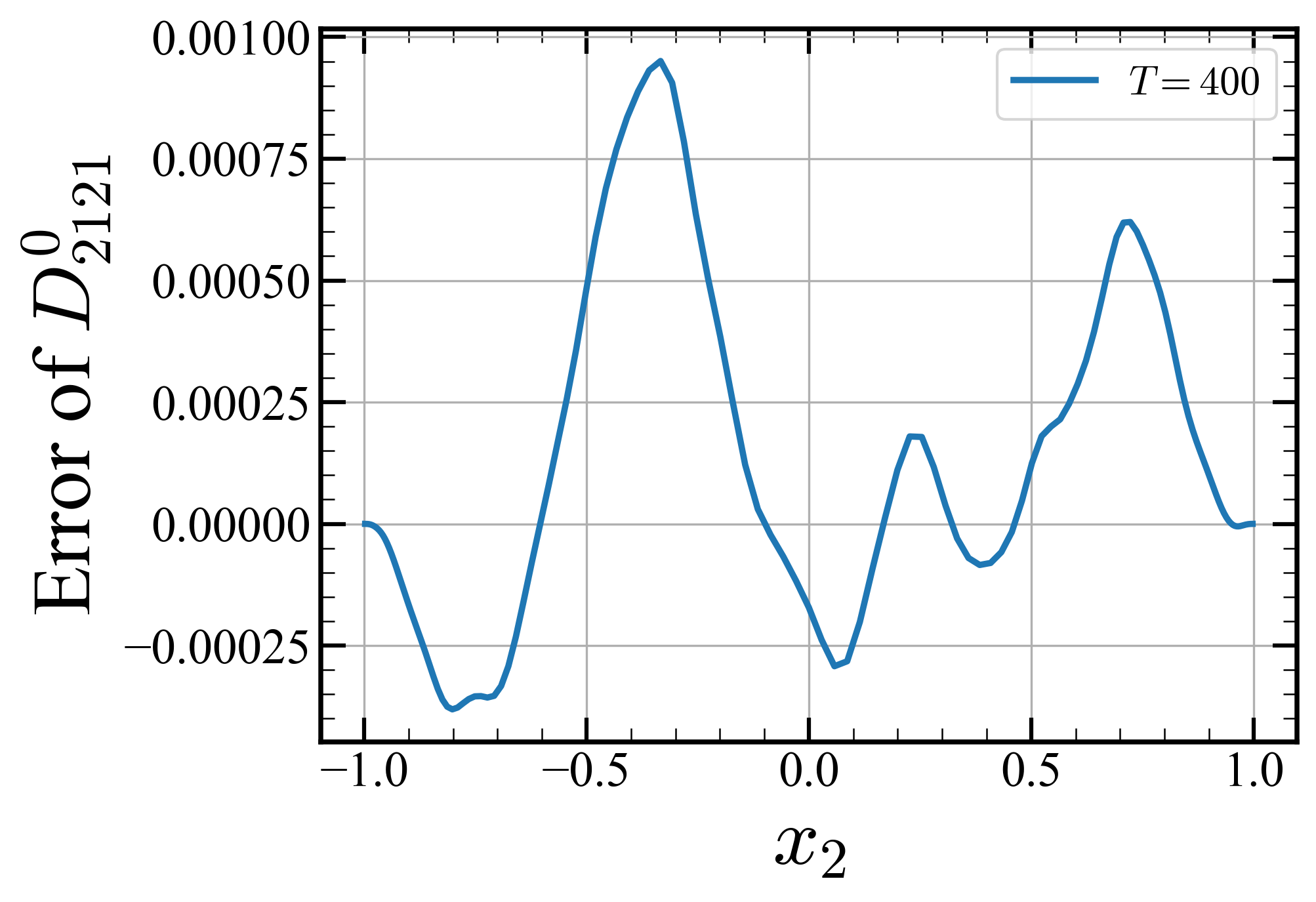}} 
  \subfigure[Domain convergence]{\includegraphics[height=0.3\linewidth]{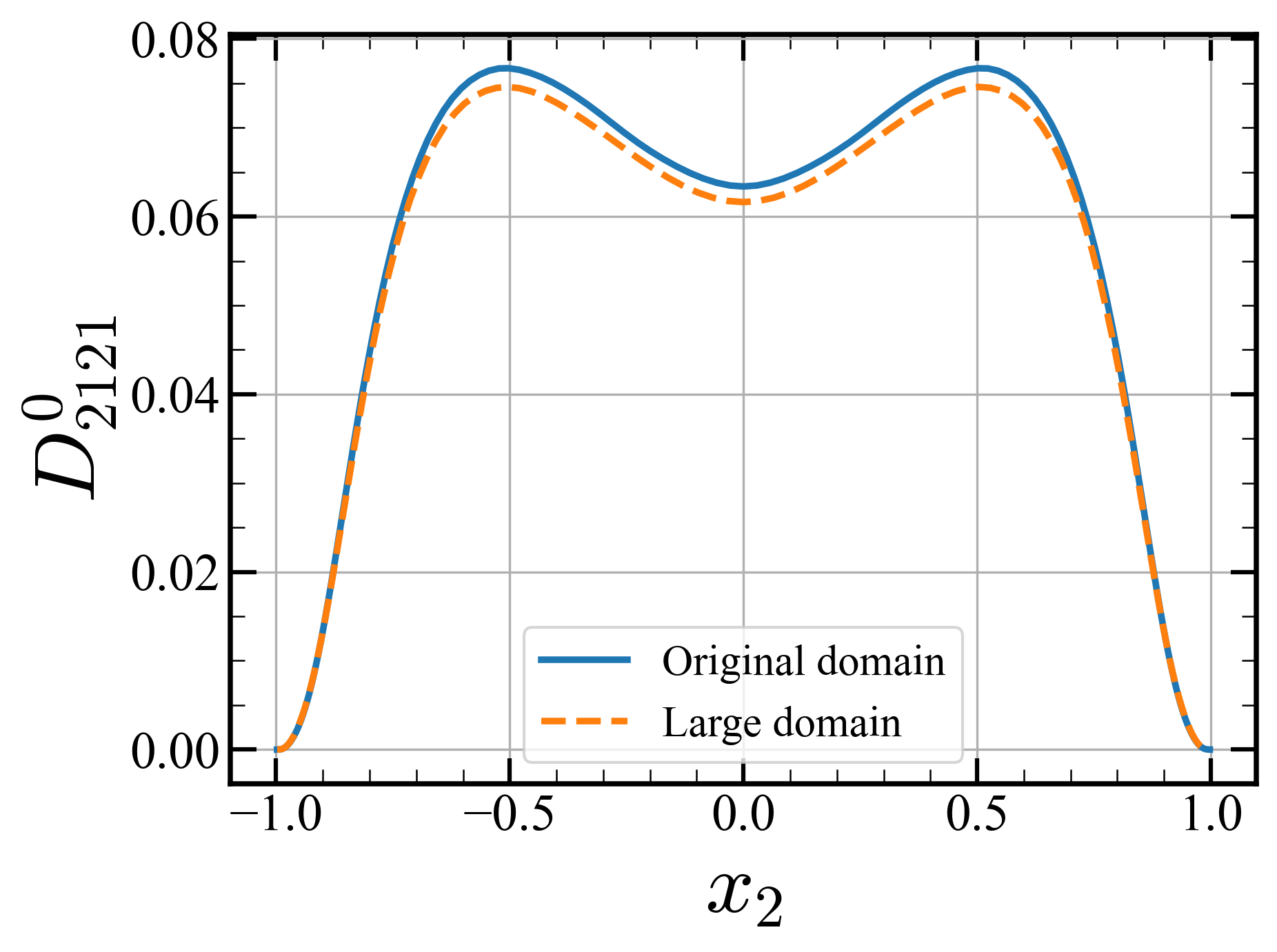}}
  \caption{Convergence studies on $D^0_{2121}$; (a) $D^0_{2121}(T=400)-D^0_{2121}(T=850)$ where $T$ is normalized sampling time period; (b) $D^0_{2121}$ where the blue solid line is from the original domain (Table \ref{table:setup}) and the orange dashed line is from the larger domain where domain length is twice larger in both $x_1$ and $x_3$ directions.}
\label{fig:convergencestudy}
\end{figure}

Figure~\ref{fig:convergencestudy} shows the convergence analysis of MFM results with respect to the spatio-temporal domain size. Figure~\ref{fig:convergencestudy}(a) shows the estimated temporal error due to the finite time horizon of the MFM simulations. In our MFM studies we used a temporal sampling window of $T=850$ in eddy turnover time unit, which is substantially longer than simulation times  typically used in the literature. We estimate the temporal convergence error by comparing $D^0_{2121}$ obtained from a shorter window, $T=400$, with that obtained from the full simulation. Based on the magnitude of the difference, shown in Figure~\ref{fig:convergencestudy}(a), we estimate that the temporal convergence error, is about $1\%$. 

In addition to the sampling time convergence study, we discuss the use of time dependent forcing. MFM restricts the forcing to be in the macroscopic space, the Reynolds-averaged space. For the channel flow, the forcing needs to be only a function of wall-normal direction, i.e. $s_i(x_2)$, and hence, time independent. Therefore, the precise way of conducting the MFM analysis is to estimate the stationary forcing prior to the computation. This is problematic since it is difficult to know the forcing terms before the simulation.
The remedy to this issue is to perform averages over ensembles, instead of using time averages. For a statistically stationary flow, the ensemble-averaged fields tend to time-constant fields as one increases the number of ensembles. Ensemble averages can then be accessed at each time step, in order to estimate $s_i(x_2)$ according to the procedure described in Section 2.1.3. Since channel flow is statistically homogeneous in $x_1$ and $x_3$ directions, instead of creating new simulations, we used these directions for ensemble averaging. We then increased the number of independent ensembles by increasing the domain size in these directions. Figure~\ref{fig:convergencestudy}(b) shows the computed $D^0_{2121}$ in two different domain sizes: one is the original domain size shown in Table \ref{table:setup} and the other is a larger domain which is twice bigger in both $x_1$ and $x_3$. The difference between these two plots are approximately $2\%$. This difference quantitatively represents the error committed by using a weakly time dependent forcings and finite domain size.

\section{Implementation for determining $D^0_{ijkl}$ in a periodic domain}\label{appB}

MFM allows computation of every component in the leading-order eddy viscosity tensor $D^0_{ijkl}$ in Equation \ref{eq:leadingApproxGeneral}. In Section 2.2.2., we briefly explained how $D^0_{ij21}$ is determined via MFM with a forcing that would maintain $V_1=x_2$ and $V_2=V_3=0$. For this case, boundary conditions and the initial condition are easily chosen to be compatible with the MFM instructions; for instance, periodic conditions in $x_1$ and $x_3$ direction and a Dirichlet condition in $x_2$ such as $v_1(x_1,x_2=\pm1,x_3)=\pm1$. The simple generalization of the forcing to other directions is $V_n=x_m$ and $V_{i\neq n}=0$ where $m$ and $n$ are not indices in the index notation, rather a choice of the forcing direction. However, \corr{such forcing is not directly implementable in codes with periodic boundary conditions}. For example, \corr{to compute $D^0_{ij12}$ we need the forcing scenario that sustains $V_2=x_1$ and $V_1=V_3=0$ which} has incompatible boundary conditions with the DNS solver since $V_2=x_1$ is not a periodic field in the streamwise direction. As a remedy, to compute all the components of the eddy viscosity tensor, we modify the GMT to solve for the fluctuating part of the GMT variable $v_i'$ as follows. We start from the GMT equations with forcing of $V_n=x_m$ and $V_{i\neq n}=0$ which allows quantification of $D^0_{ijmn}$ as shown in Equation \ref{eq:leadingOrderMFMPostprocess}. When we subtract the mean of the GMT equation from the GMT equation, the resulting equation becomes
\begin{eqnarray}
    \frac{\partial v_i'}{\partial t} + \frac{\partial }{\partial x_j} \left( u_j v_i' \right) = -\frac{\partial p}{\partial x_i} + \nu\frac{\partial^2 v_i'}{\partial x_j\partial x_j} + s_i -u_m \delta_{in}
    \label{eq:leadingOrderMFM}
\end{eqnarray}
\corr{which is implementable in a periodic solver since it eliminates the need for explicit inclusion of $V_n=x_m$.}
For a given $m$ and $n$, once we numerically solve the equation above, we can determine the nine components of the eddy viscosity tensor by post-processing the results as Equation \ref{eq:leadingOrderMFMPostprocess}. Using different combinations of $m$ and $n$, we reveal all the element in the leading-order eddy viscosity tensor.  
\begin{equation}
    -\overline{u'_i v'_j}(x_2)= \int_{{y_2}}D_{ijkl}({x_2}, {y_2}) \left.\frac{\partial V_l}{\partial x_k}\right\vert_{y_2} \mathrm{ d}y_2= \int_{{y_2}}D_{ijmn}({x_2}, {y_2}) \mathrm{ d}y_2 = D^0_{ijmn}(x_2)
    \label{eq:leadingOrderMFMPostprocess}
\end{equation}

There are multiple advantages of solving for GMT fluctuation equations. The first advantage is that the boundary condition is now compatible with the periodic conditions. Second, all wall boundary conditions for $v'_j$ are easily set with a Dirichet condition of $v_j'=0$. With these two advantages, the solver become more systematic and simple.

 \corr{ Lastly, we note that in RANS solutions $dV_1/dx_1$ can only be present when either $dV_2/dx_2$ or $dV_3/dx_3$ are non-zero. Imagining a 2D flow as a simple example, this implies that $dV_1/dx_1 = - dV_2/dx_2 $. As a result, a macroscopic forcing that honors this constraint can only measure the combined term $D^0_{ij11}-D^0_{ij22}$, but not the individual terms. However, in the procedure described above we have taken advantage of the linearity of the GMT equation in our analysis. In other words, while the solutions to $v'$ fields in equation B1 are obtained by enforcing the divergence free condition, we recognize that these solutions are linear response to the source term, which is the last term in the equation and is controlled by the pre-specified gradient of $V$. This allows quantification of response to independent components of gradient of $V$, and thus prediction of  $D^0_{ij11}$ independent of $D^0_{ij22}$. When the resulting $D^0$ is used in a RANS solver to predict the mean momentum field, one always uses divergence free momentum fields, which bundles back the components of $D^0$ together. Therefore our choice of decomposition by independently activating different components of the mean momentum gradient, would not affect the outcome of Reynolds stress predictions.}

\section{Leading-order eddy viscosity tensor}\label{appC}

\begin{figure}
\centering
  \subfigure[$D_{1211}$]{\includegraphics[width=0.22\linewidth]{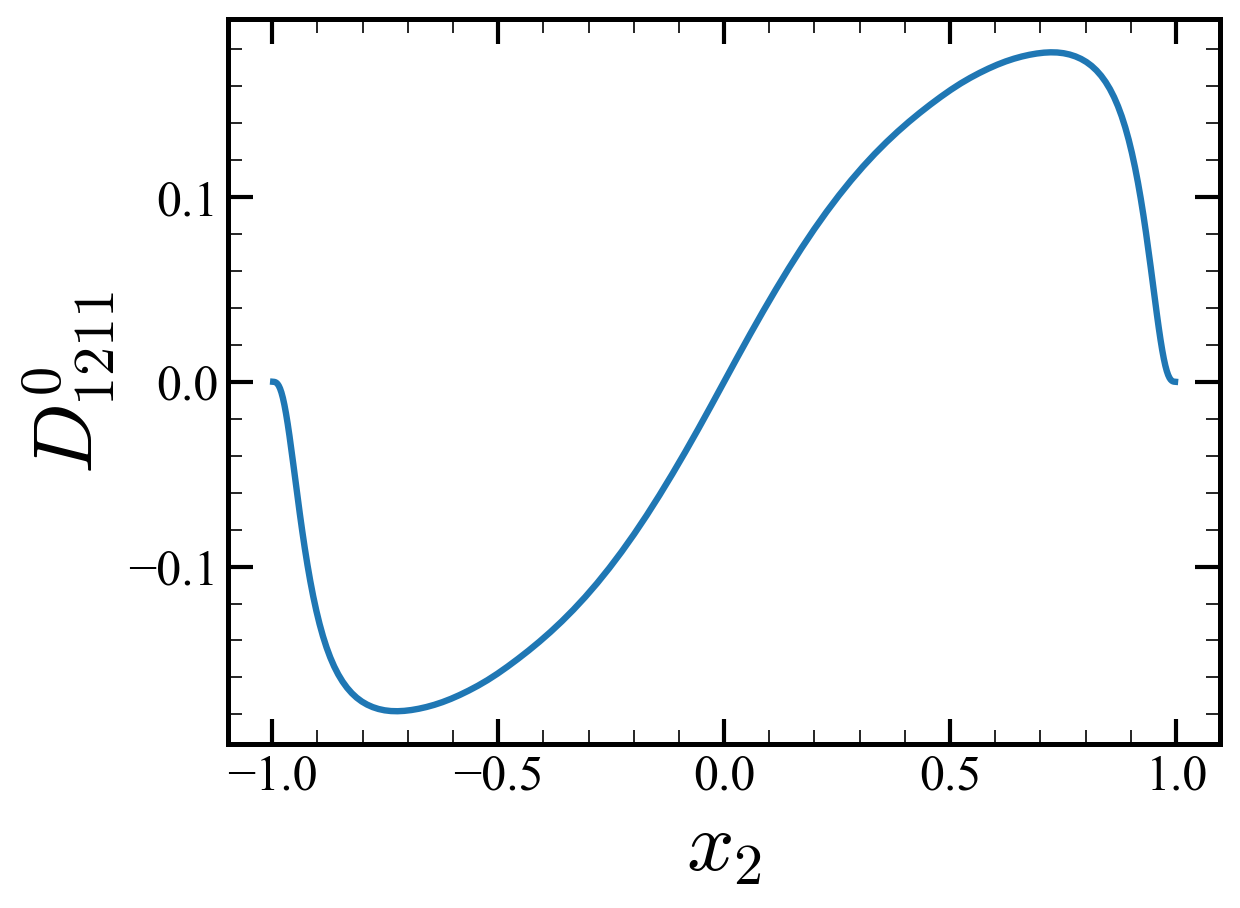}} 
  \subfigure[$D_{2111}$]{\includegraphics[width=0.22\linewidth]{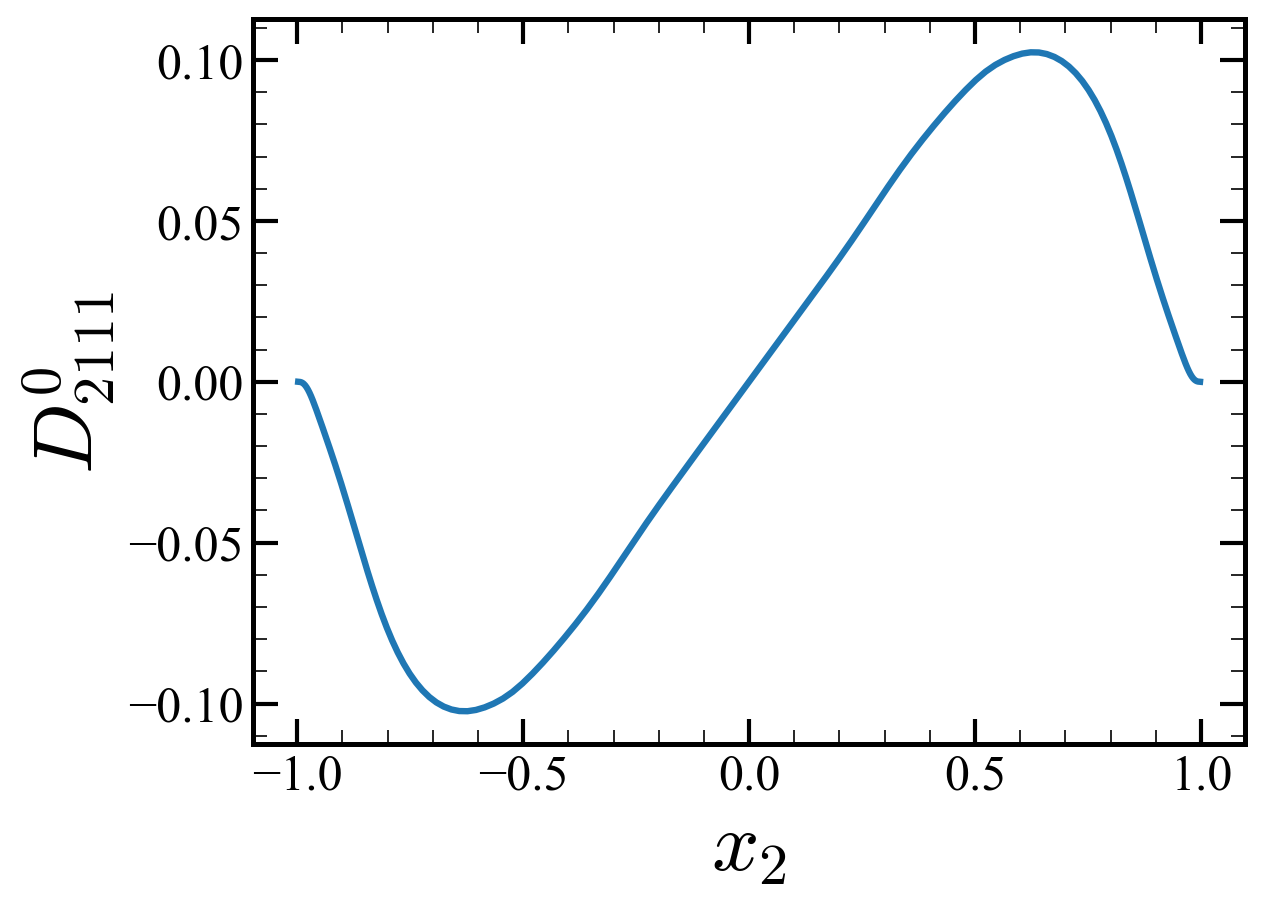}} \\
  \subfigure[$D_{1111}$]{\includegraphics[width=0.22\linewidth]{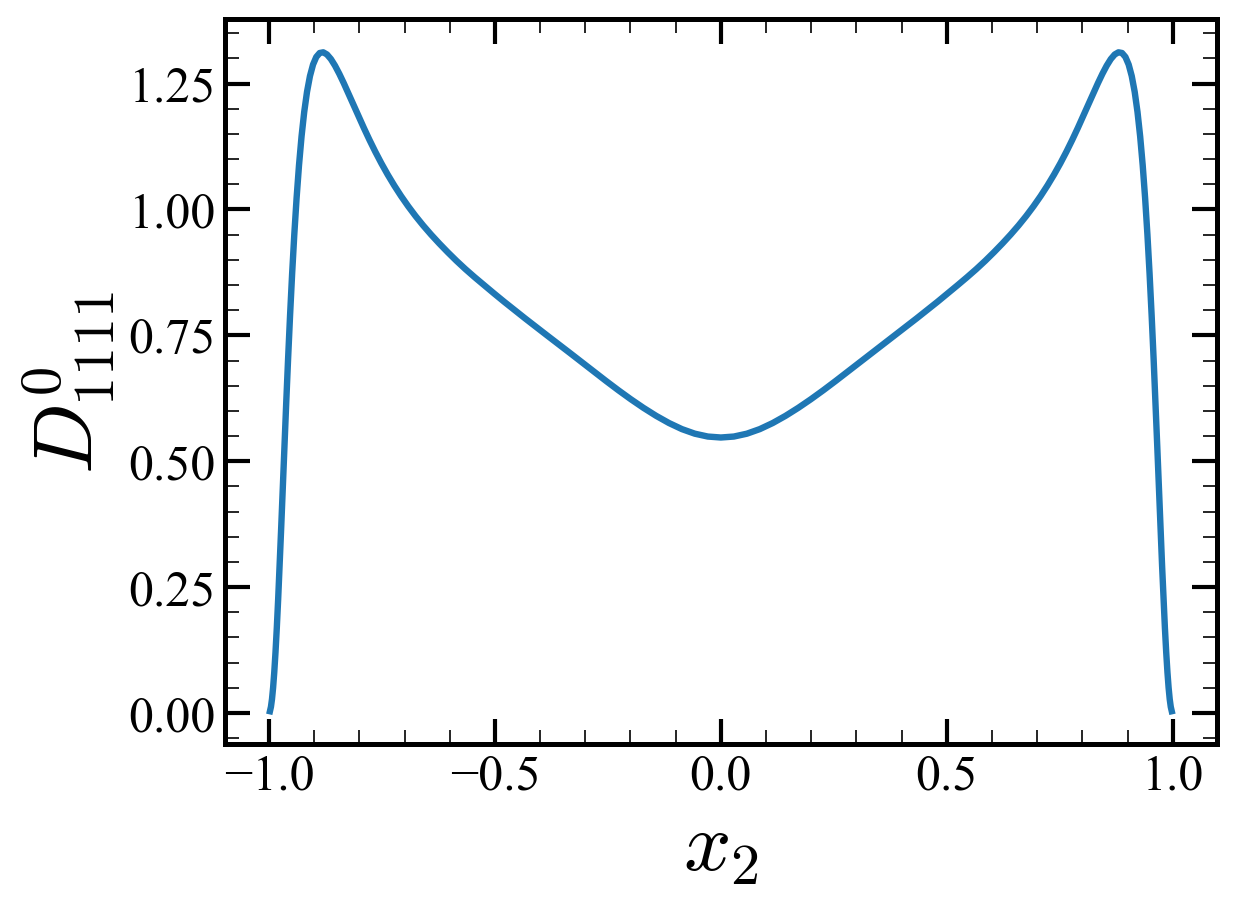}} 
  \subfigure[$D_{2211}$]{\includegraphics[width=0.22\linewidth]{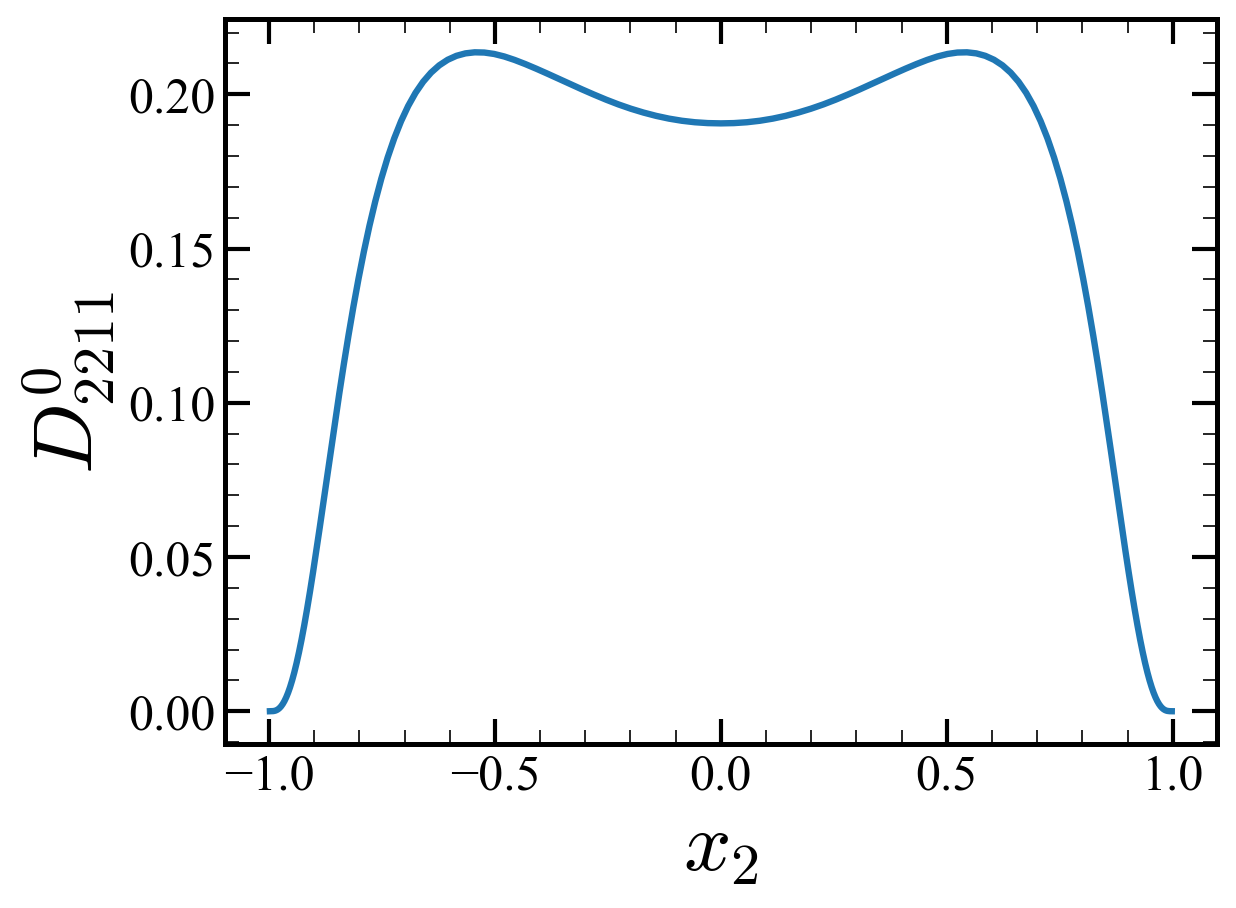}} 
  \subfigure[$D_{3311}$]{\includegraphics[width=0.22\linewidth]{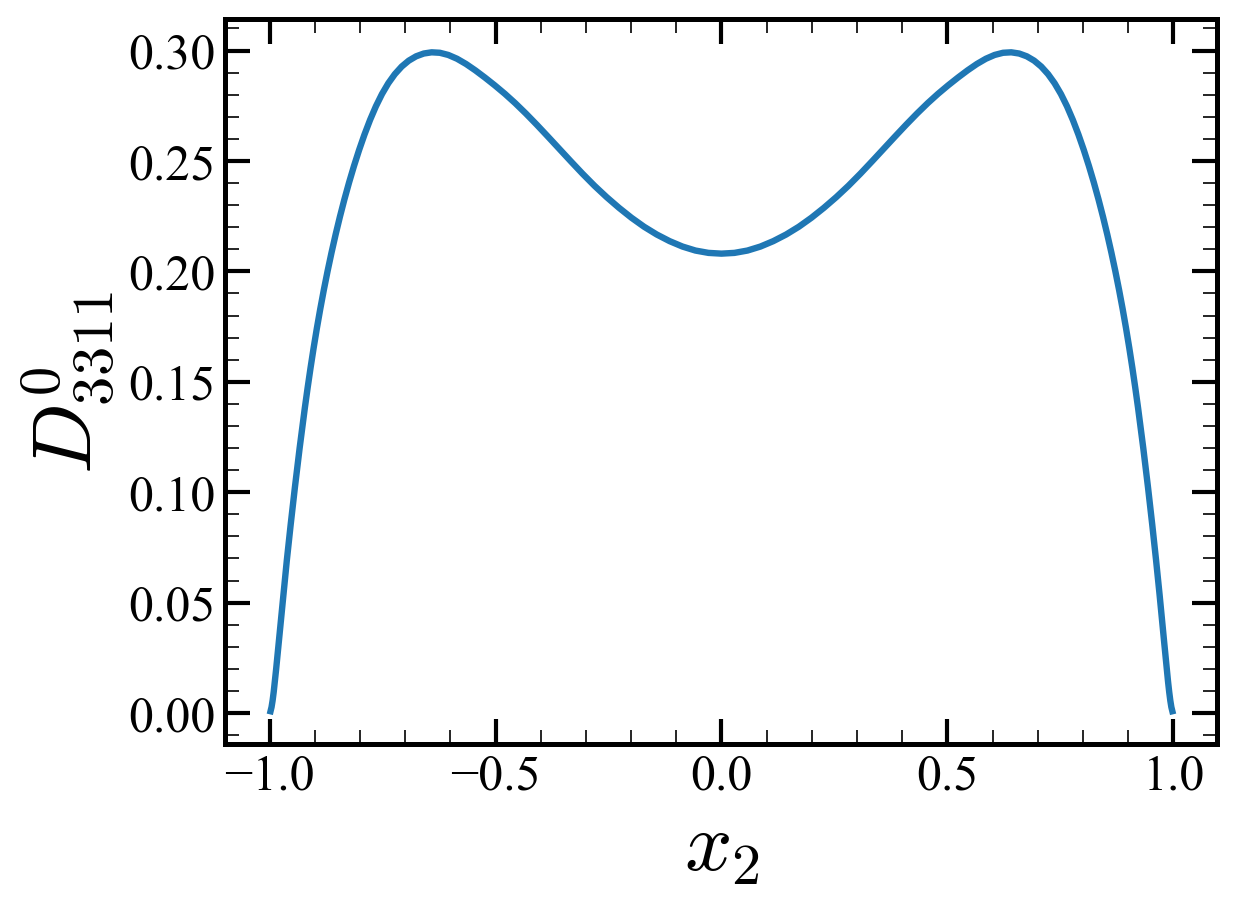}} 
  \caption{Distribution of nonzero $D^0_{ij11}$.}
\label{fig:D0ij11}
\end{figure}

\begin{figure}
\centering
  \subfigure[$D_{1212}$]{\includegraphics[width=0.22\linewidth]{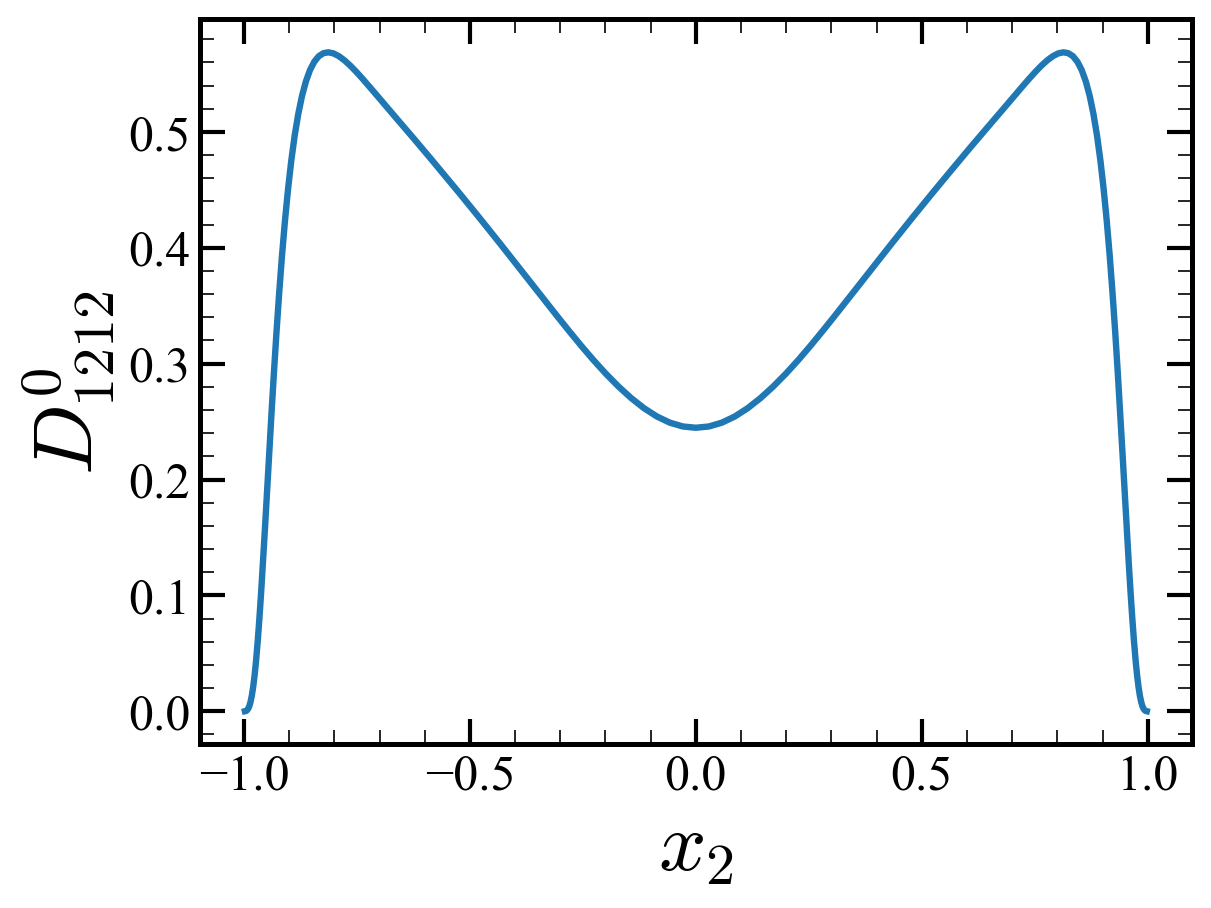}} 
  \subfigure[$D_{2112}$]{\includegraphics[width=0.22\linewidth]{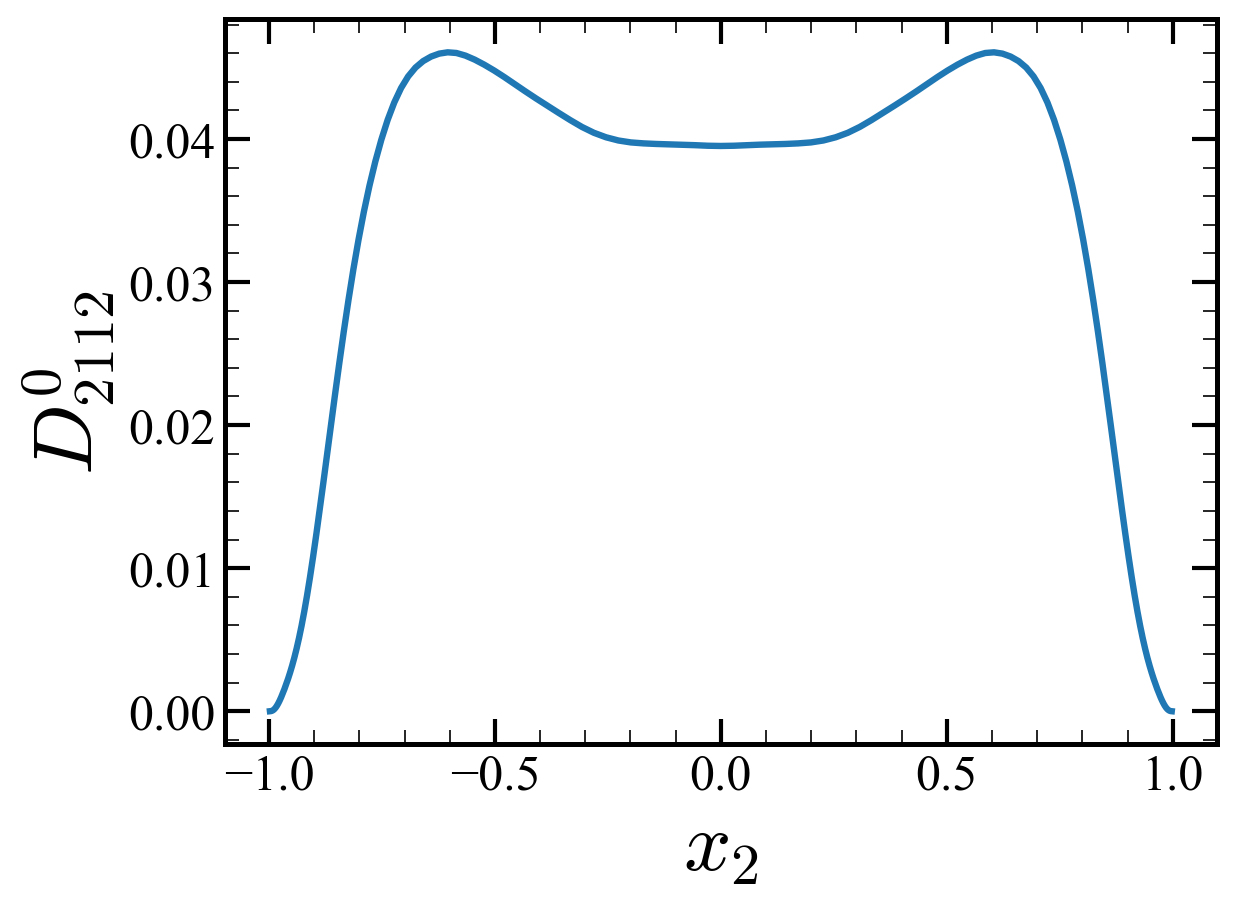}} \\
  \subfigure[$D_{1112}$]{\includegraphics[width=0.22\linewidth]{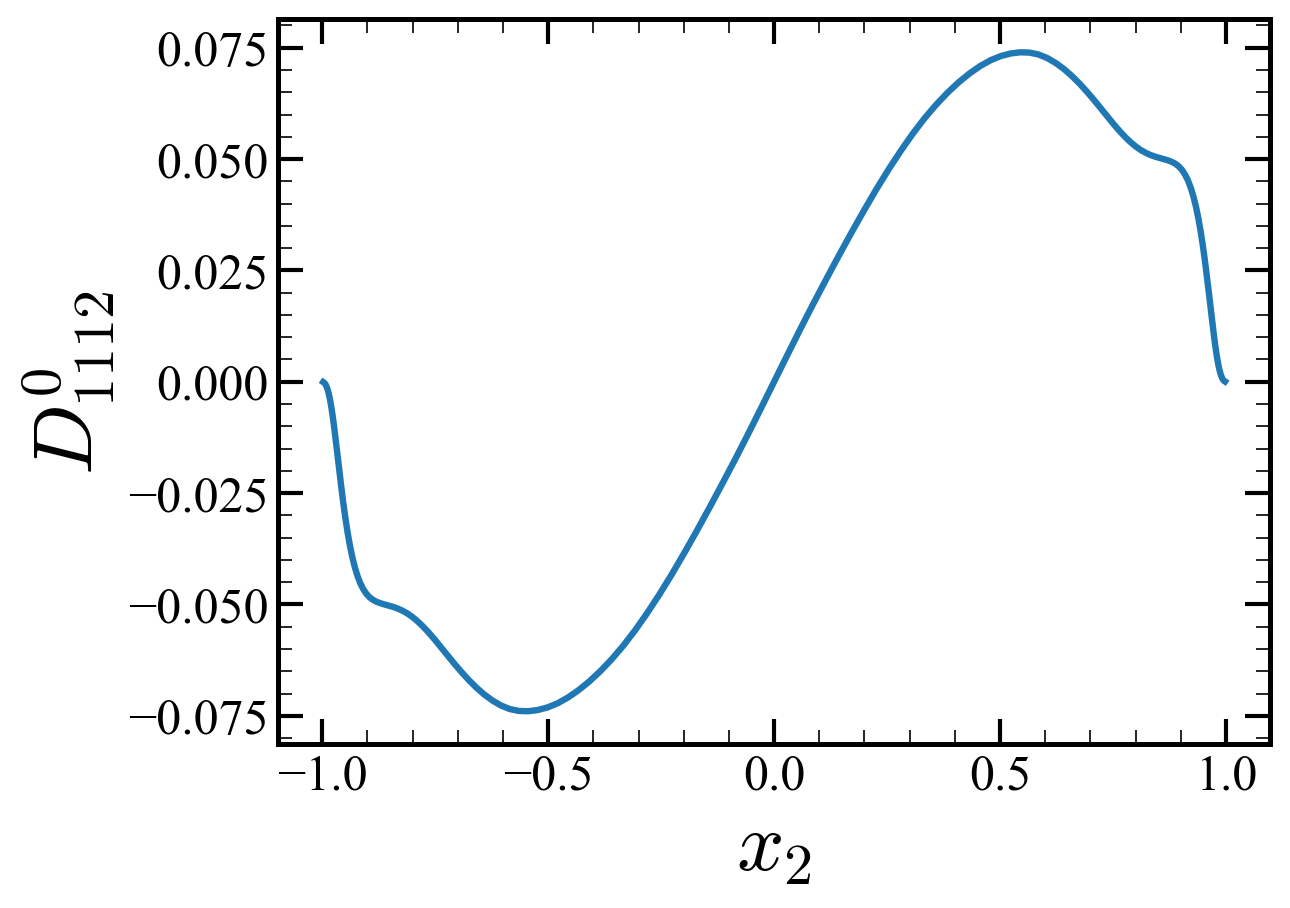}} 
  \subfigure[$D_{2212}$]{\includegraphics[width=0.22\linewidth]{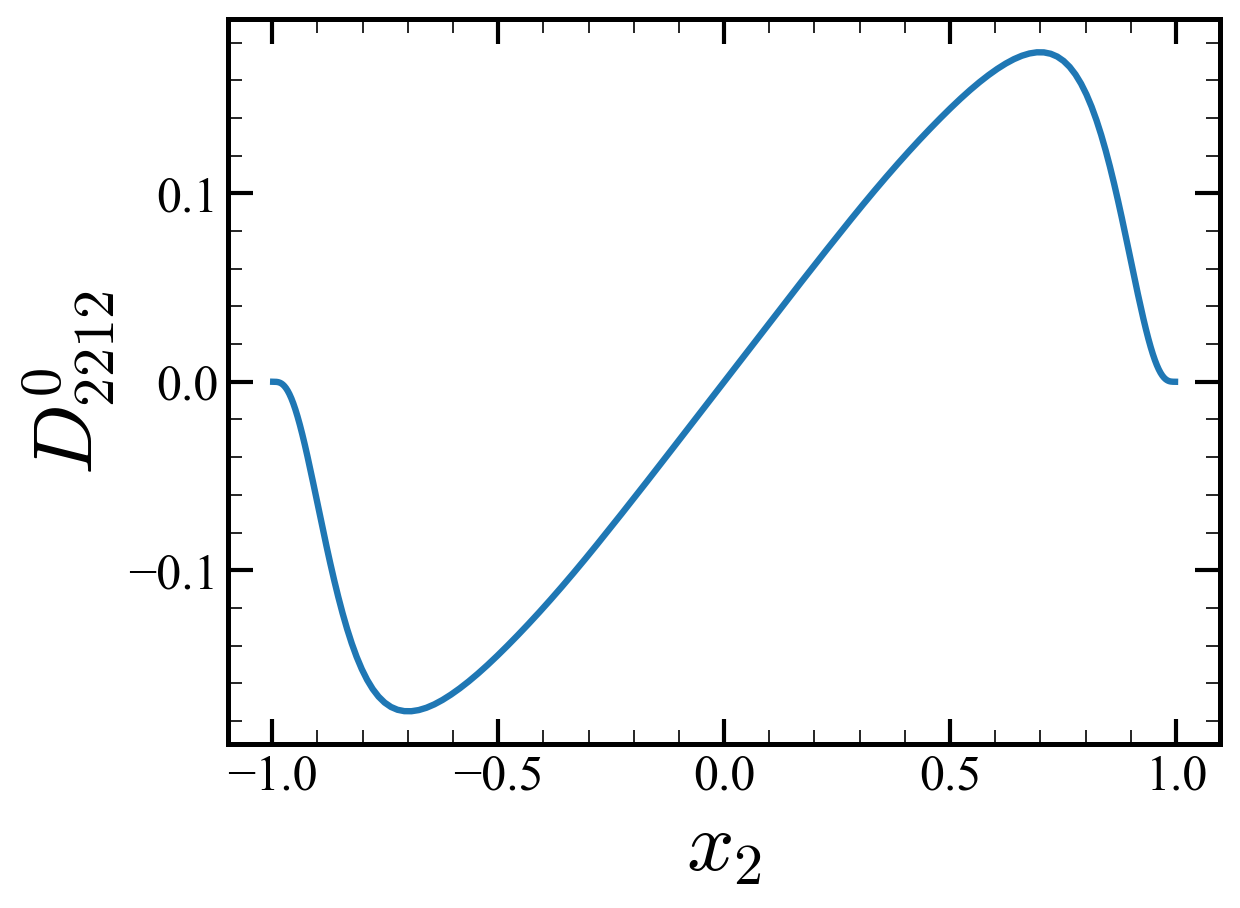}} 
  \subfigure[$D_{3312}$]{\includegraphics[width=0.22\linewidth]{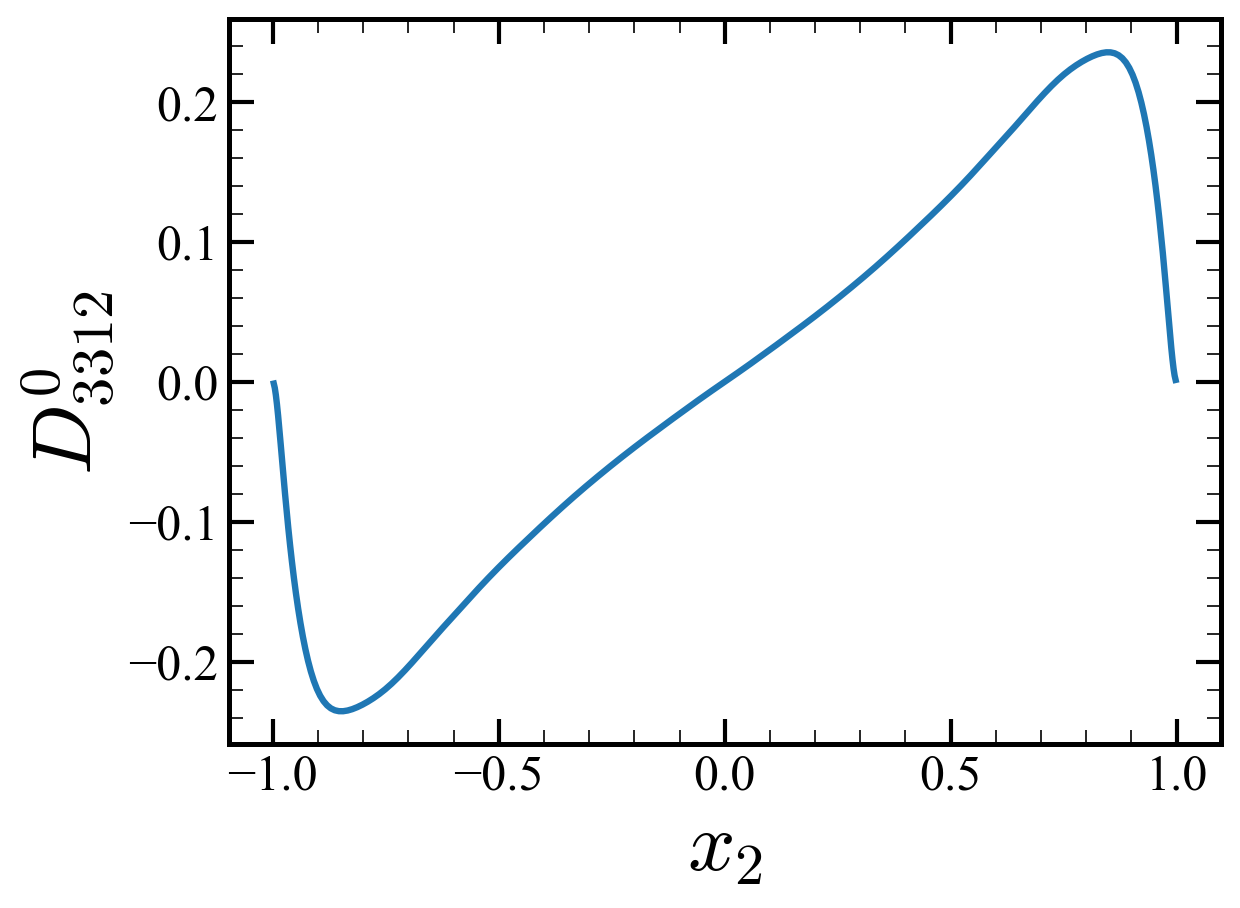}} 
  \caption{Distribution of nonzero $D^0_{ij12}$.}
\label{fig:D0ij12}
\end{figure}

\begin{figure}
\centering
  \subfigure[$D_{1313}$]{\includegraphics[width=0.22\linewidth]{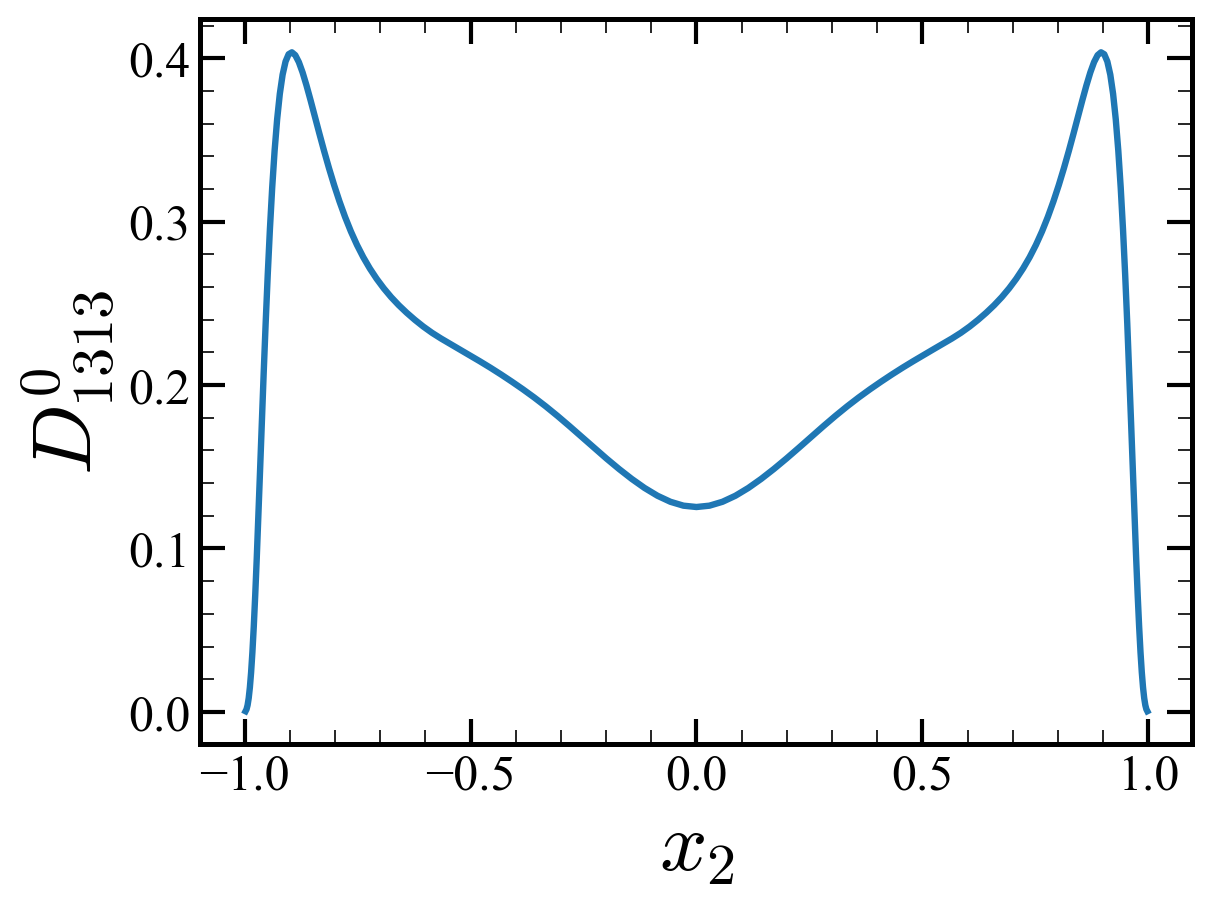}} 
  \subfigure[$D_{2313}$]{\includegraphics[width=0.22\linewidth]{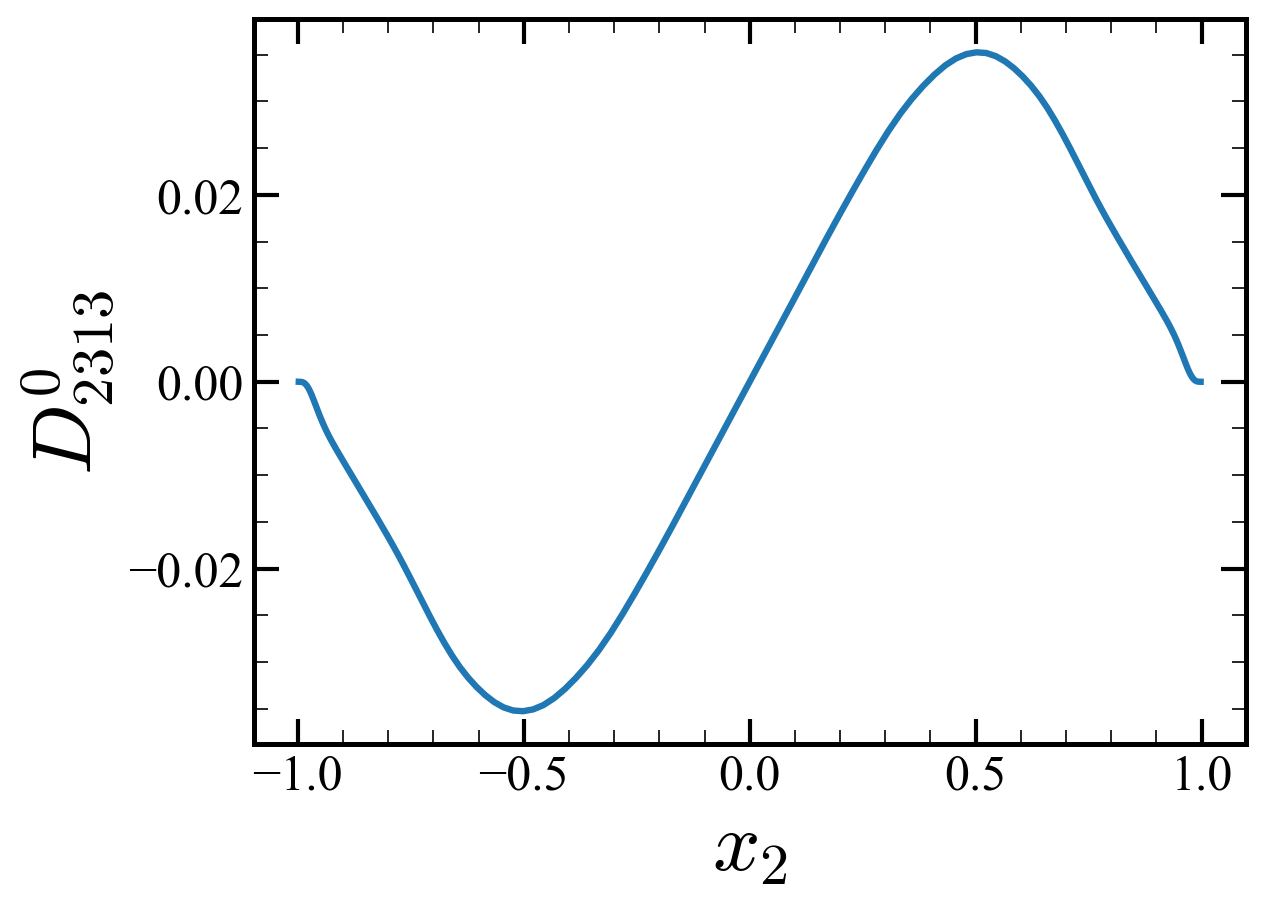}} \\
  \subfigure[$D_{3113}$]{\includegraphics[width=0.22\linewidth]{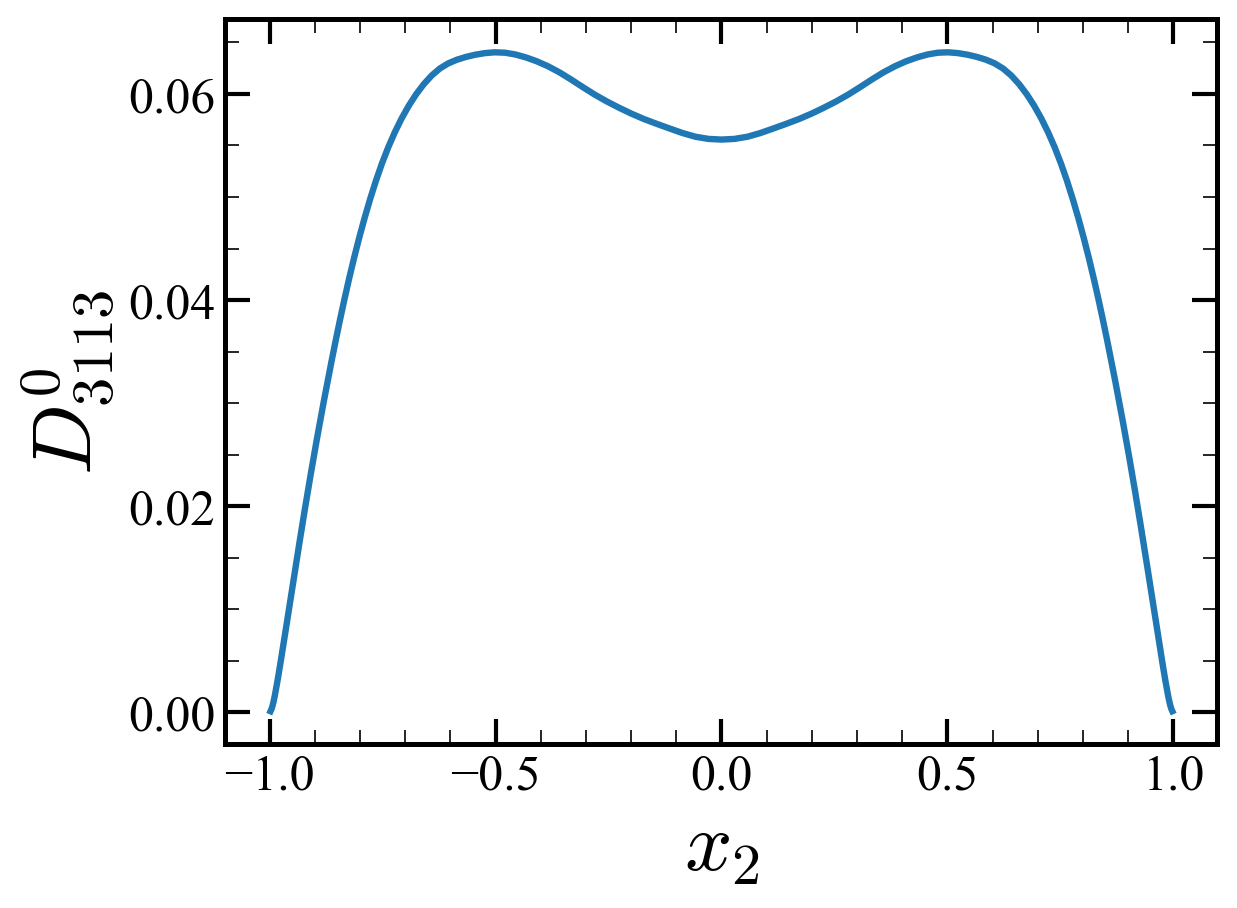}} 
  \subfigure[$D_{3213}$]{\includegraphics[width=0.22\linewidth]{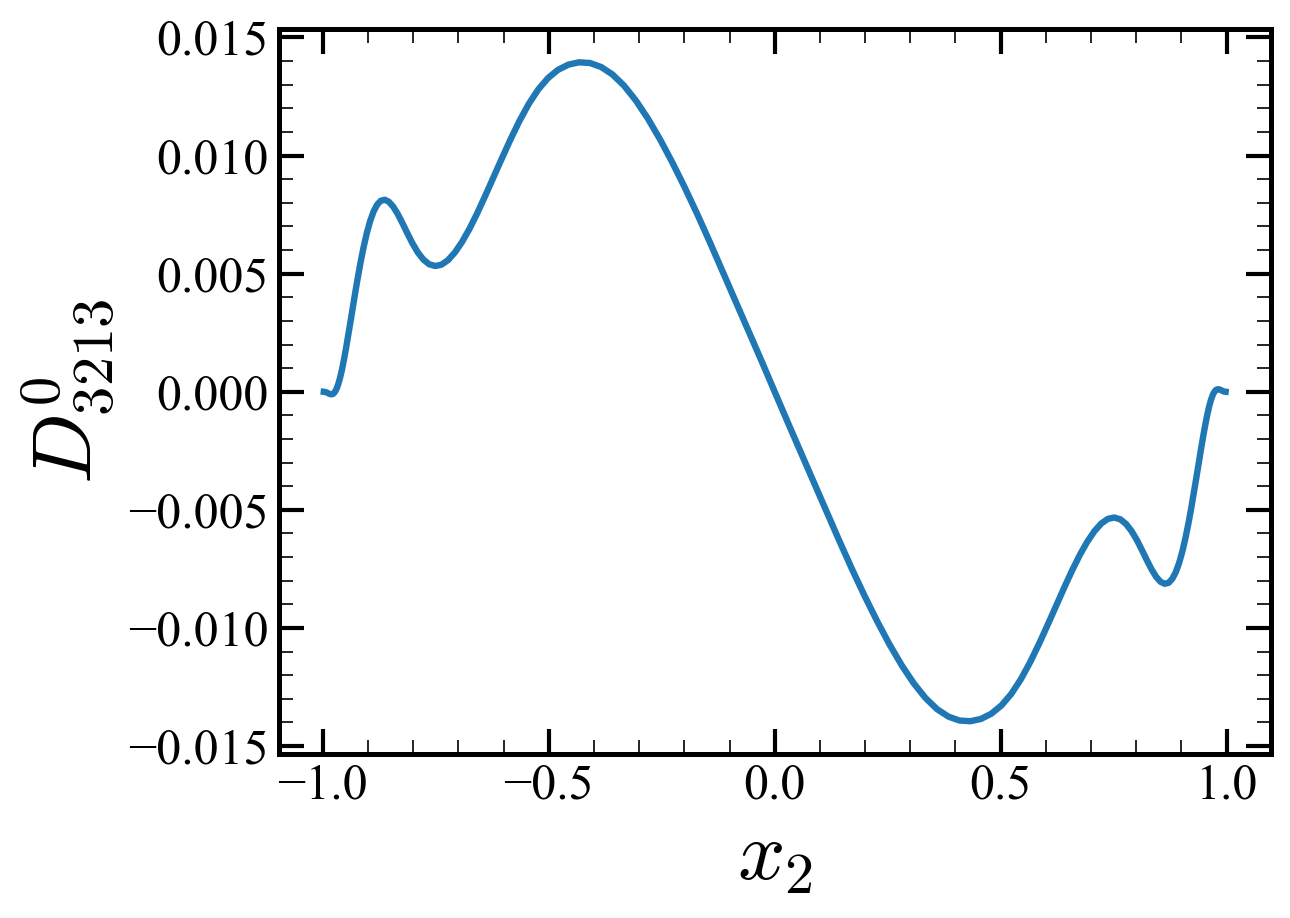}} 
  \caption{Distribution of nonzero $D^0_{ij13}$.}
\label{fig:D0ij13}
\end{figure}

\begin{figure}
\centering
  \subfigure[$D_{1221}$]{\includegraphics[width=0.22\linewidth]{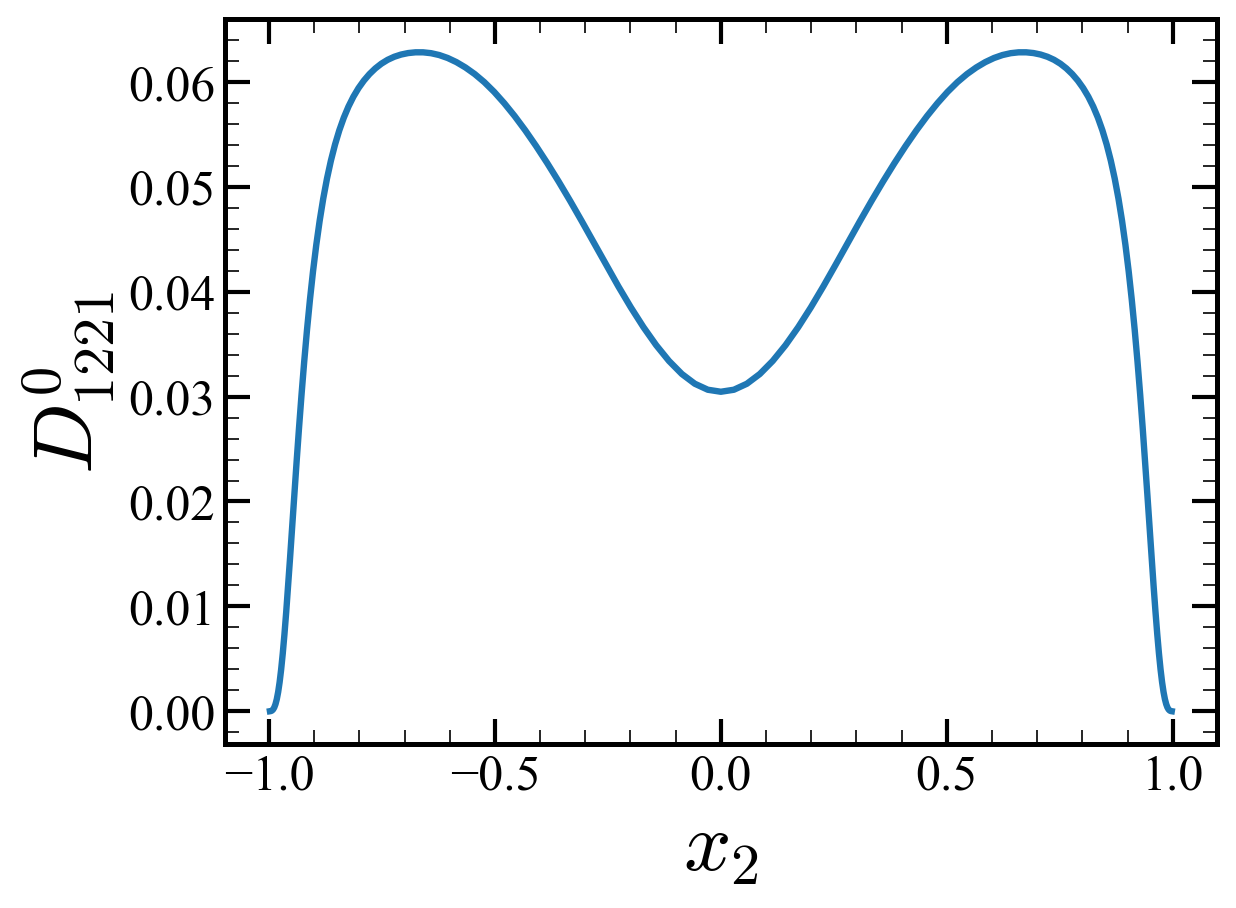}} 
  \subfigure[$D_{2121}$]{\includegraphics[width=0.22\linewidth]{figures/D02121.png}} \\
  \subfigure[$D_{1121}$]{\includegraphics[width=0.22\linewidth]{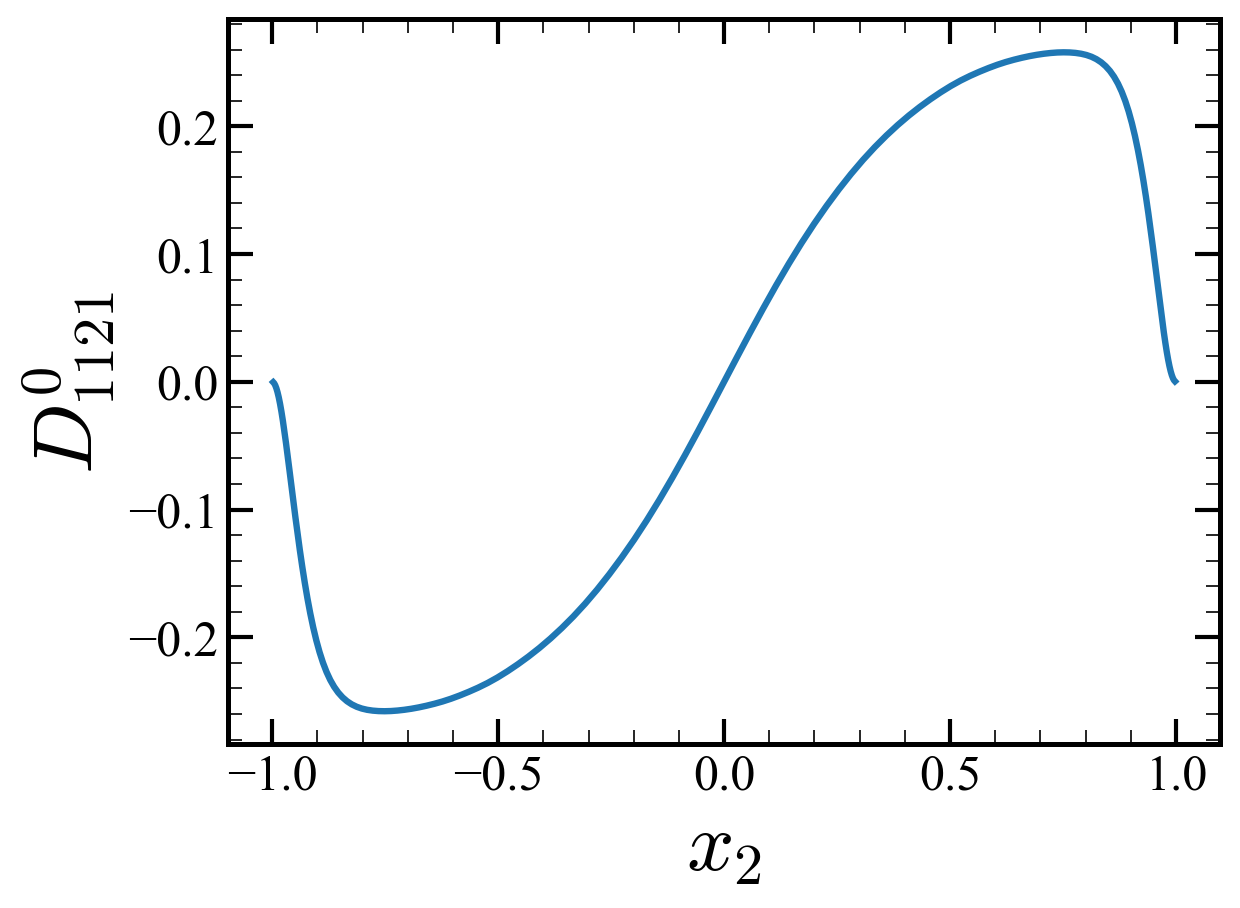}} 
  \subfigure[$D_{2221}$]{\includegraphics[width=0.22\linewidth]{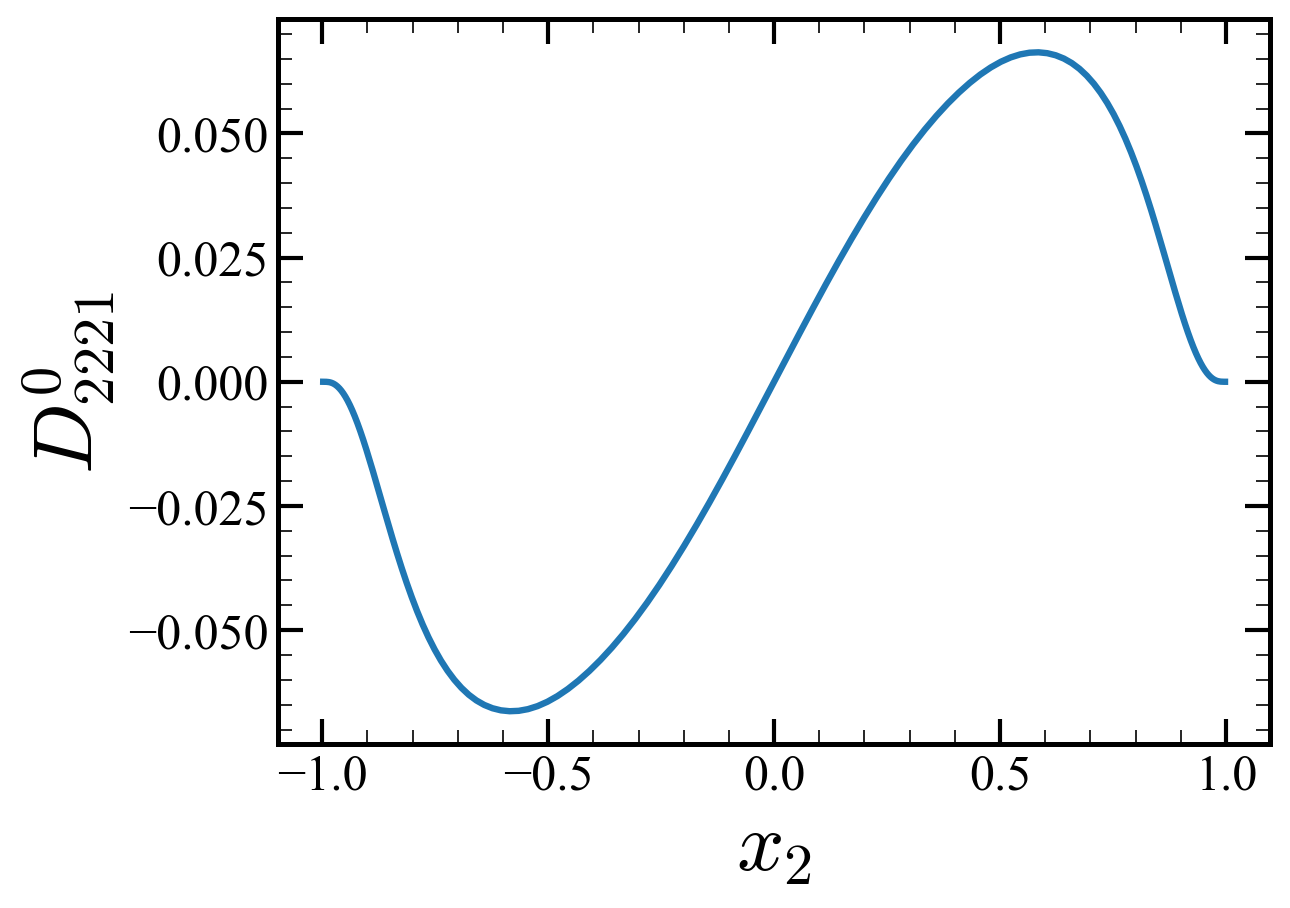}} 
  \subfigure[$D_{3321}$]{\includegraphics[width=0.22\linewidth]{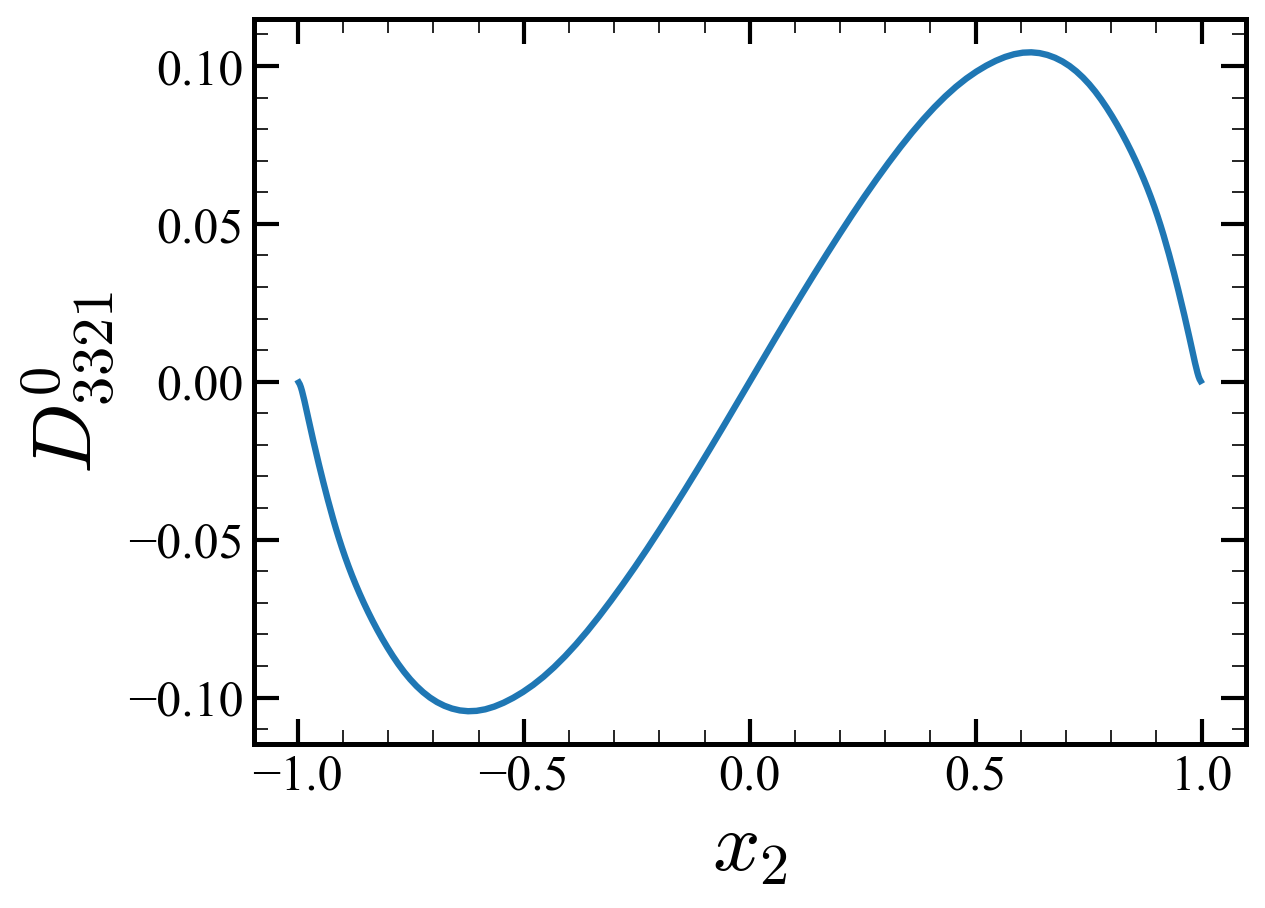}} 
  \caption{Distribution of nonzero $D^0_{ij21}$.}
\label{fig:D0ij21}
\end{figure}

\begin{figure}
\centering
  \subfigure[$D_{1222}$]{\includegraphics[width=0.22\linewidth]{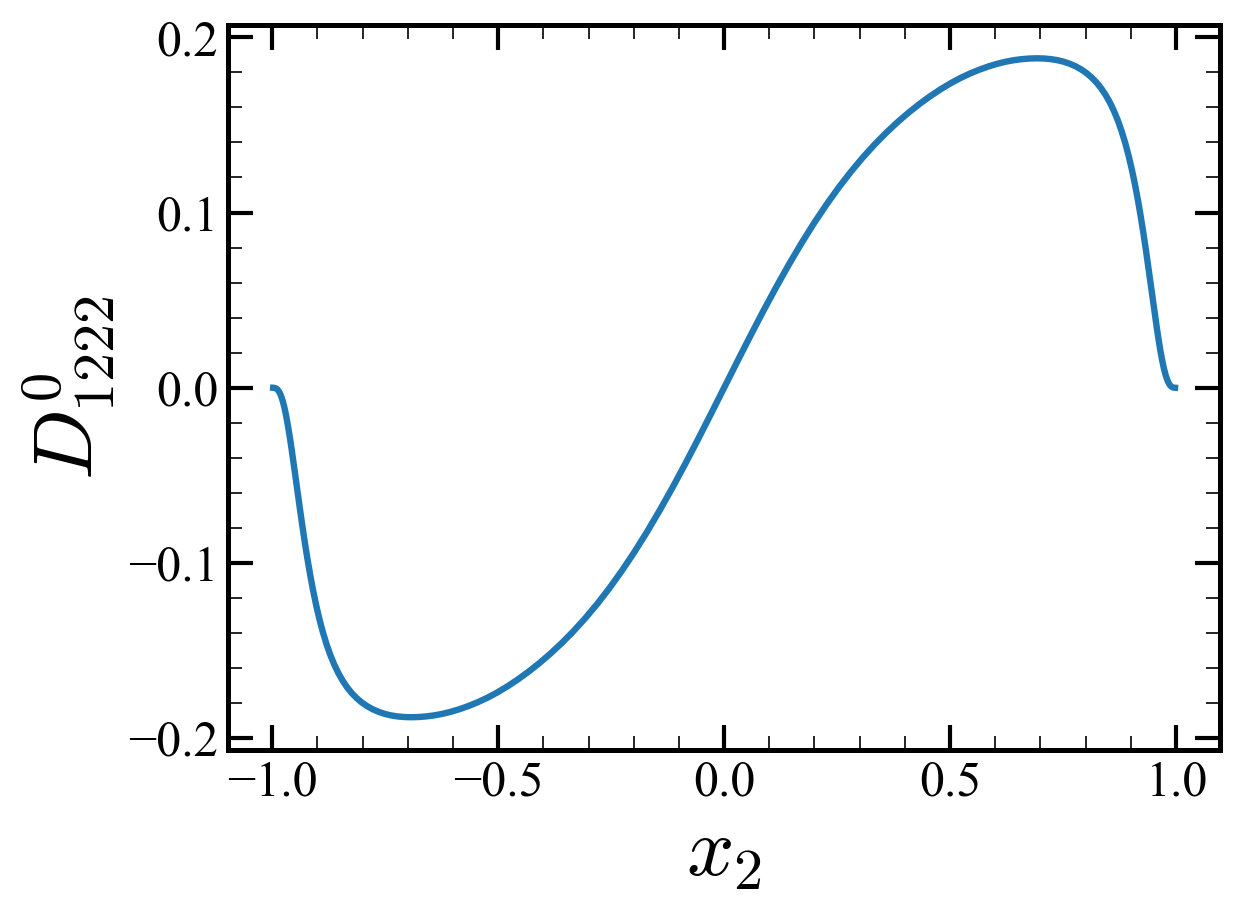}} 
  \subfigure[$D_{2122}$]{\includegraphics[width=0.22\linewidth]{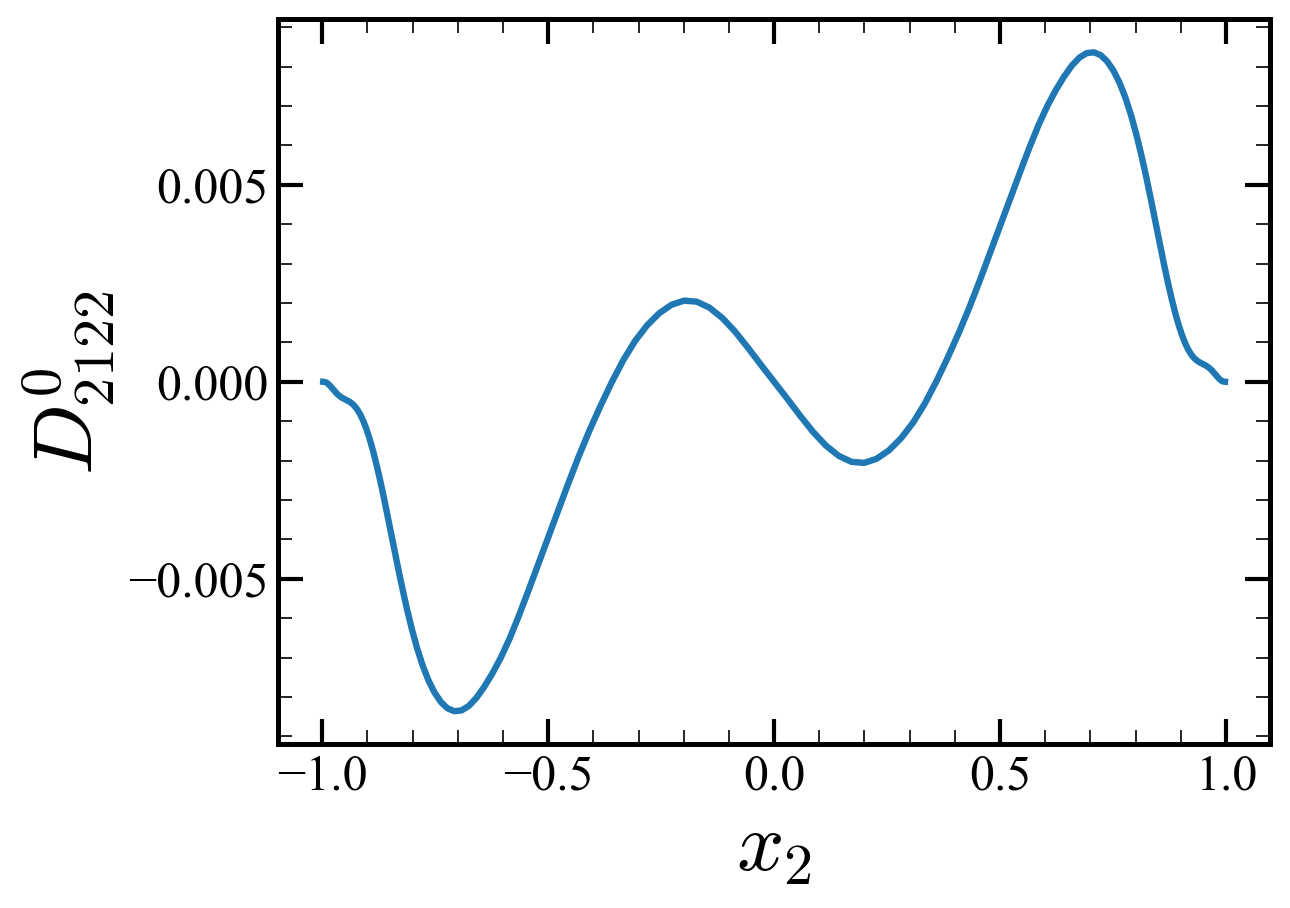}} \\
  \subfigure[$D_{1122}$]{\includegraphics[width=0.22\linewidth]{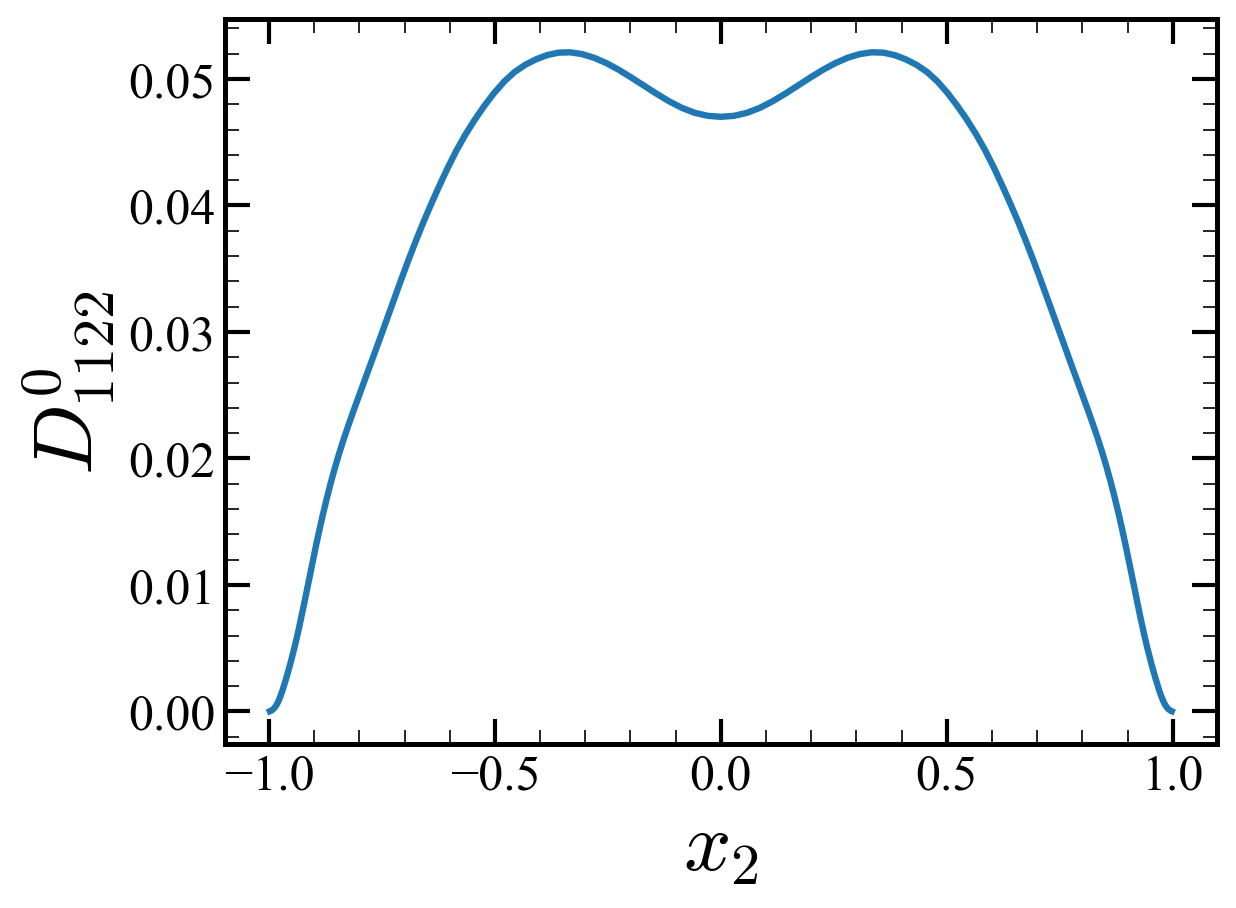}} 
  \subfigure[$D_{2222}$]{\includegraphics[width=0.22\linewidth]{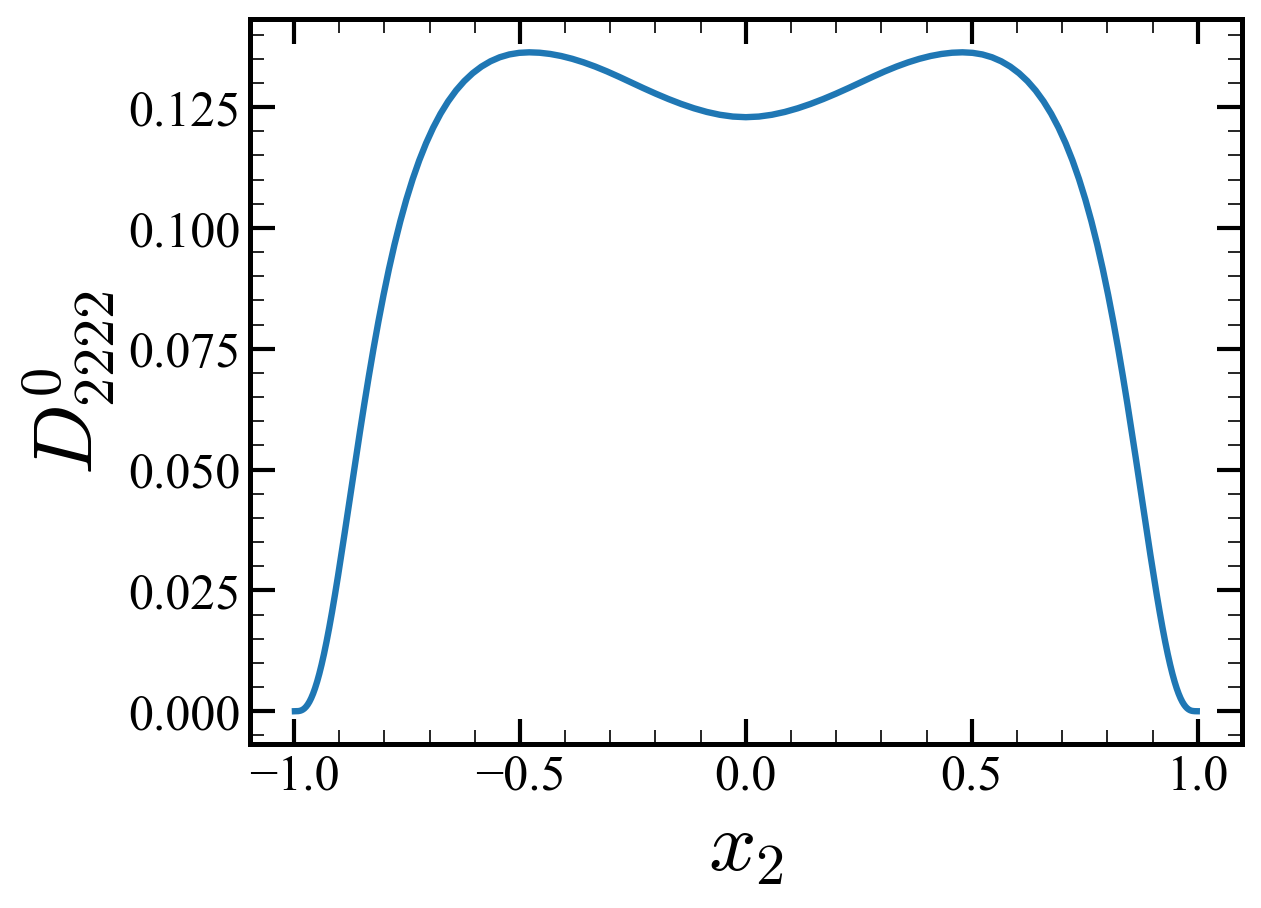}} 
  \subfigure[$D_{3322}$]{\includegraphics[width=0.22\linewidth]{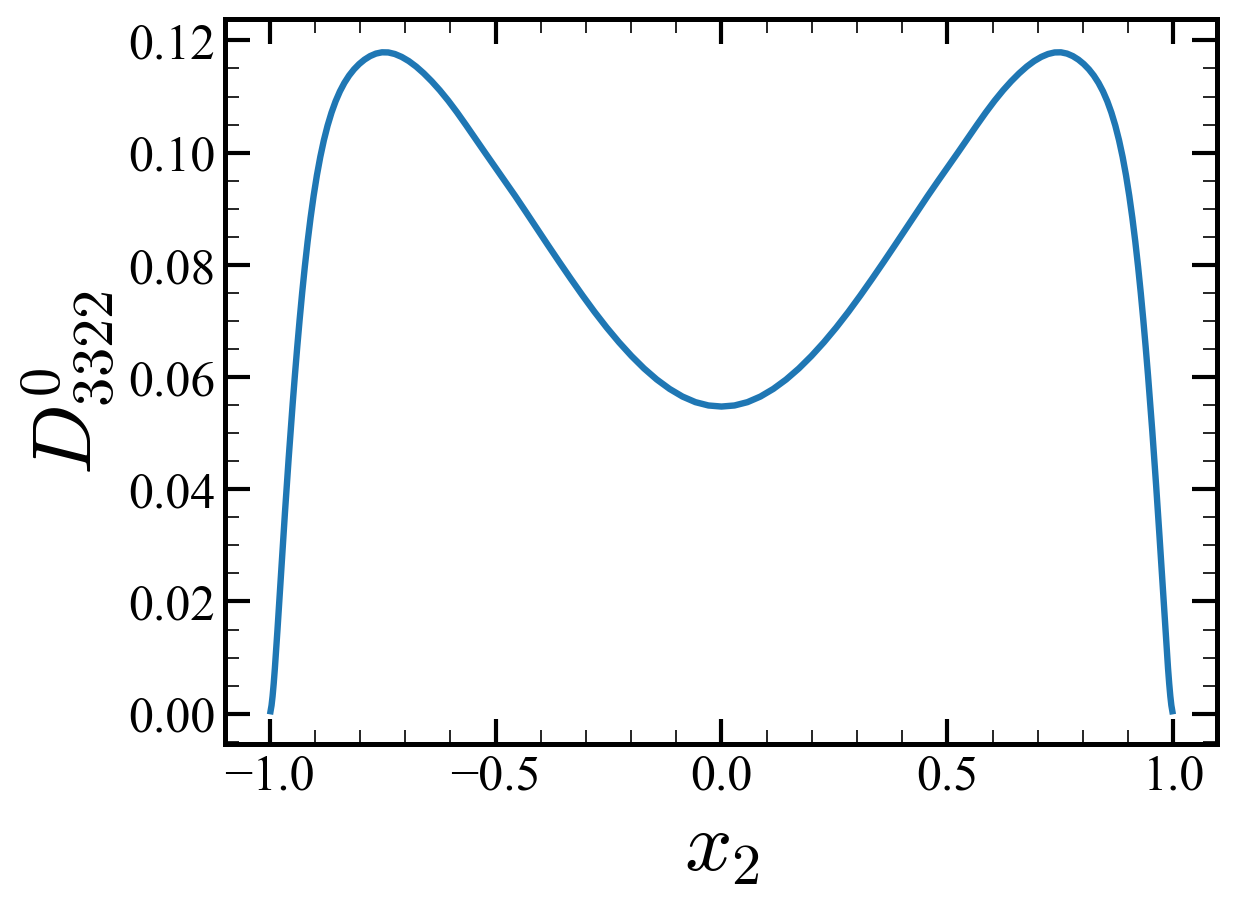}} 
  \caption{Distribution of nonzero $D^0_{ij22}$.}
\label{fig:D0ij22}
\end{figure}

\begin{figure}
\centering
  \subfigure[$D_{1323}$]{\includegraphics[width=0.22\linewidth]{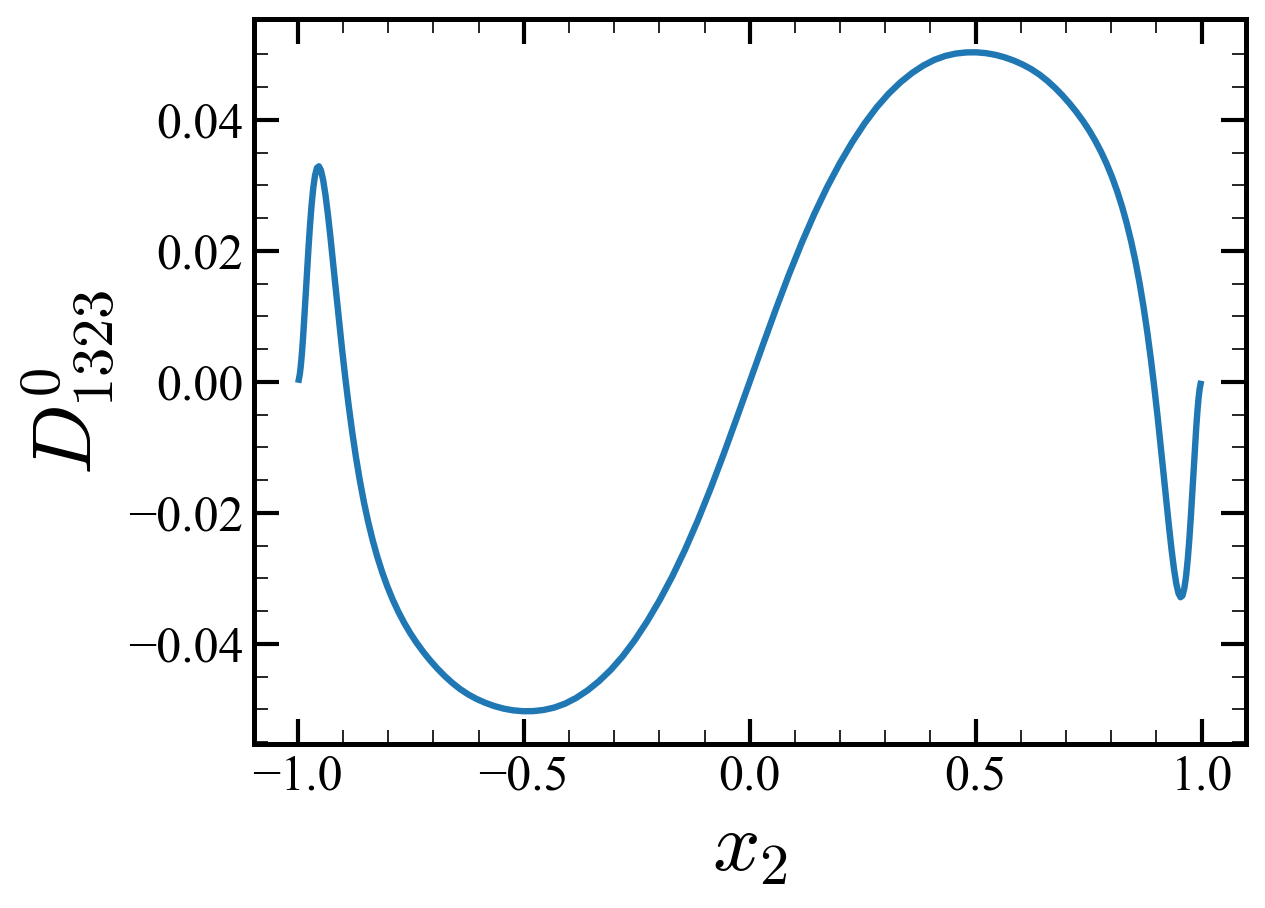}} 
  \subfigure[$D_{2323}$]{\includegraphics[width=0.22\linewidth]{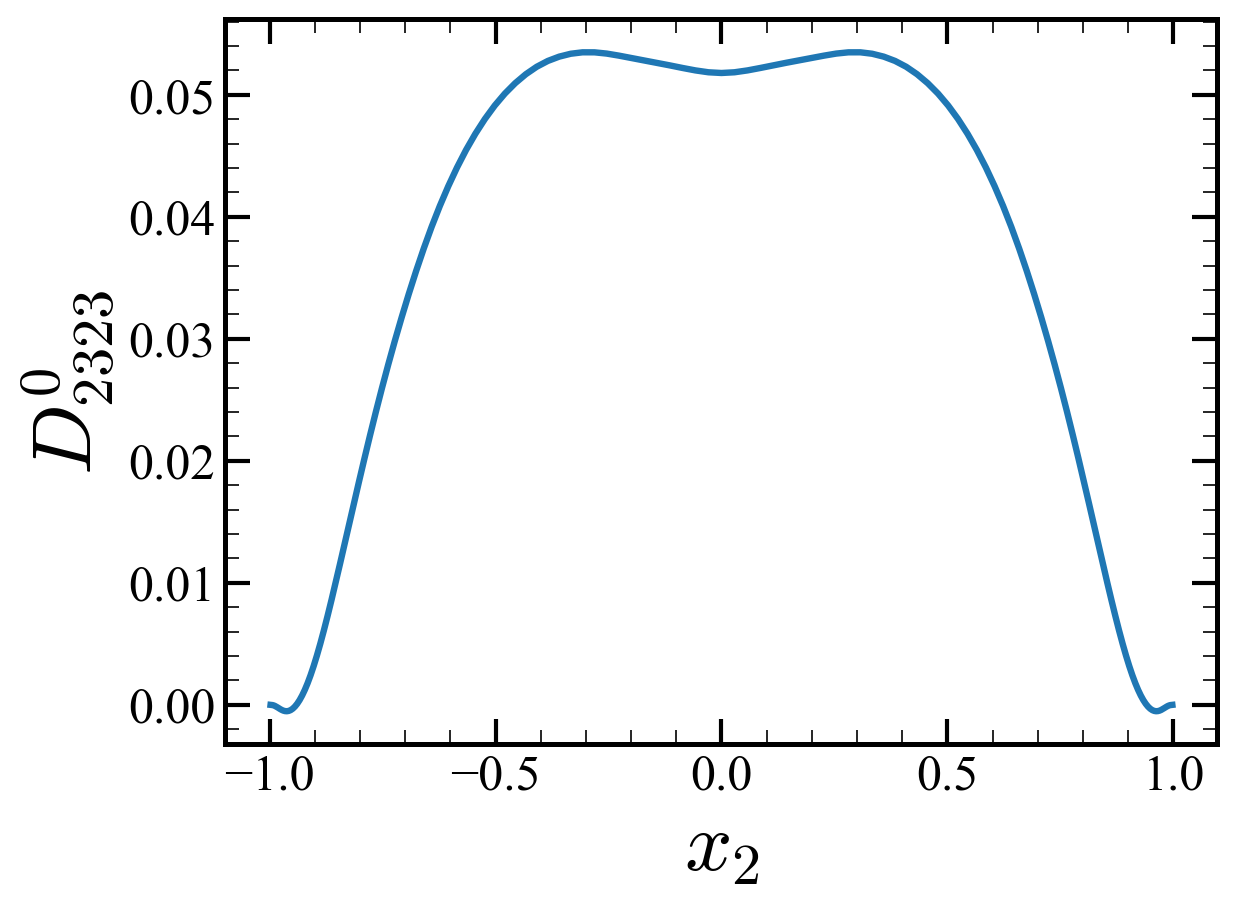}} \\
  \subfigure[$D_{3123}$]{\includegraphics[width=0.22\linewidth]{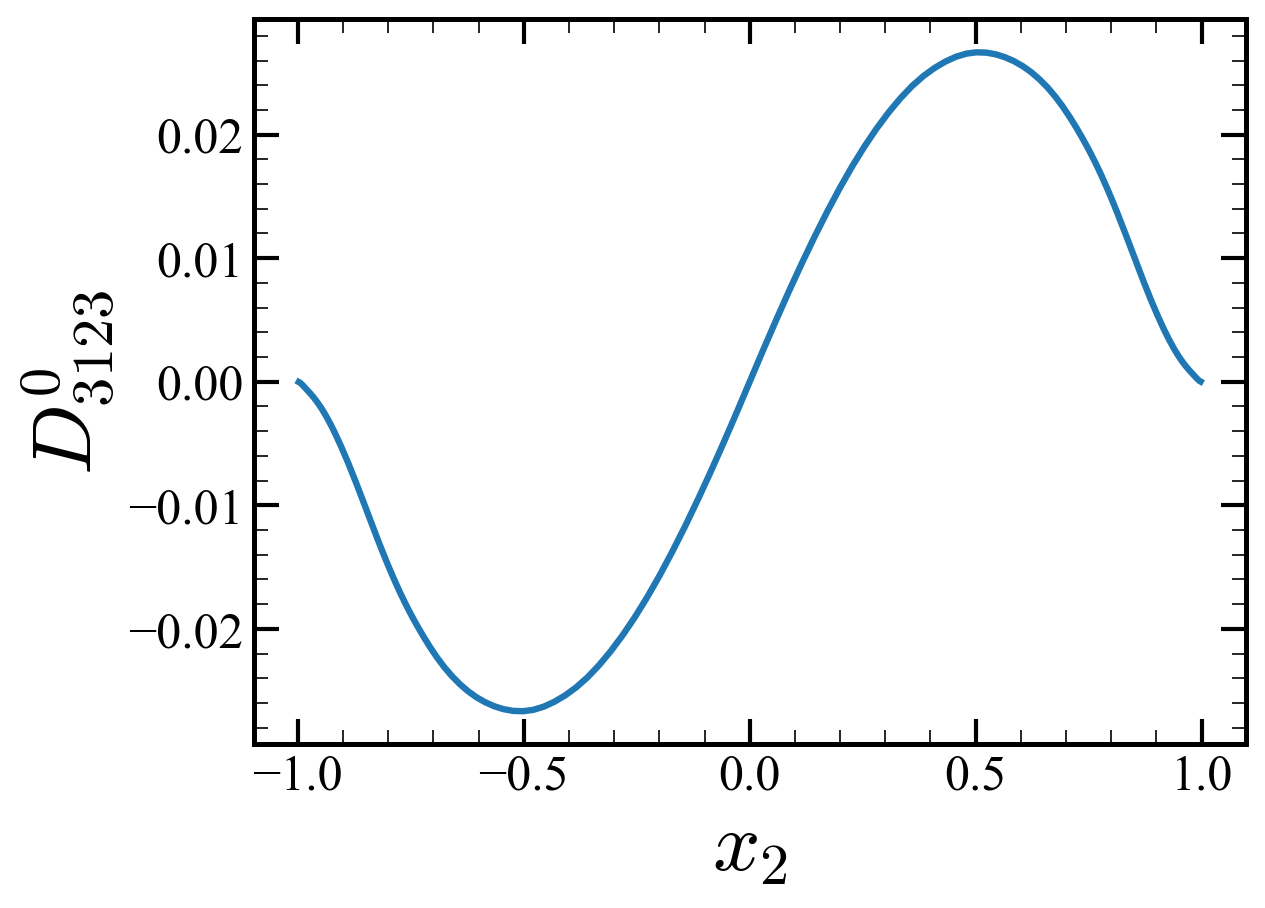}} 
  \subfigure[$D_{3223}$]{\includegraphics[width=0.22\linewidth]{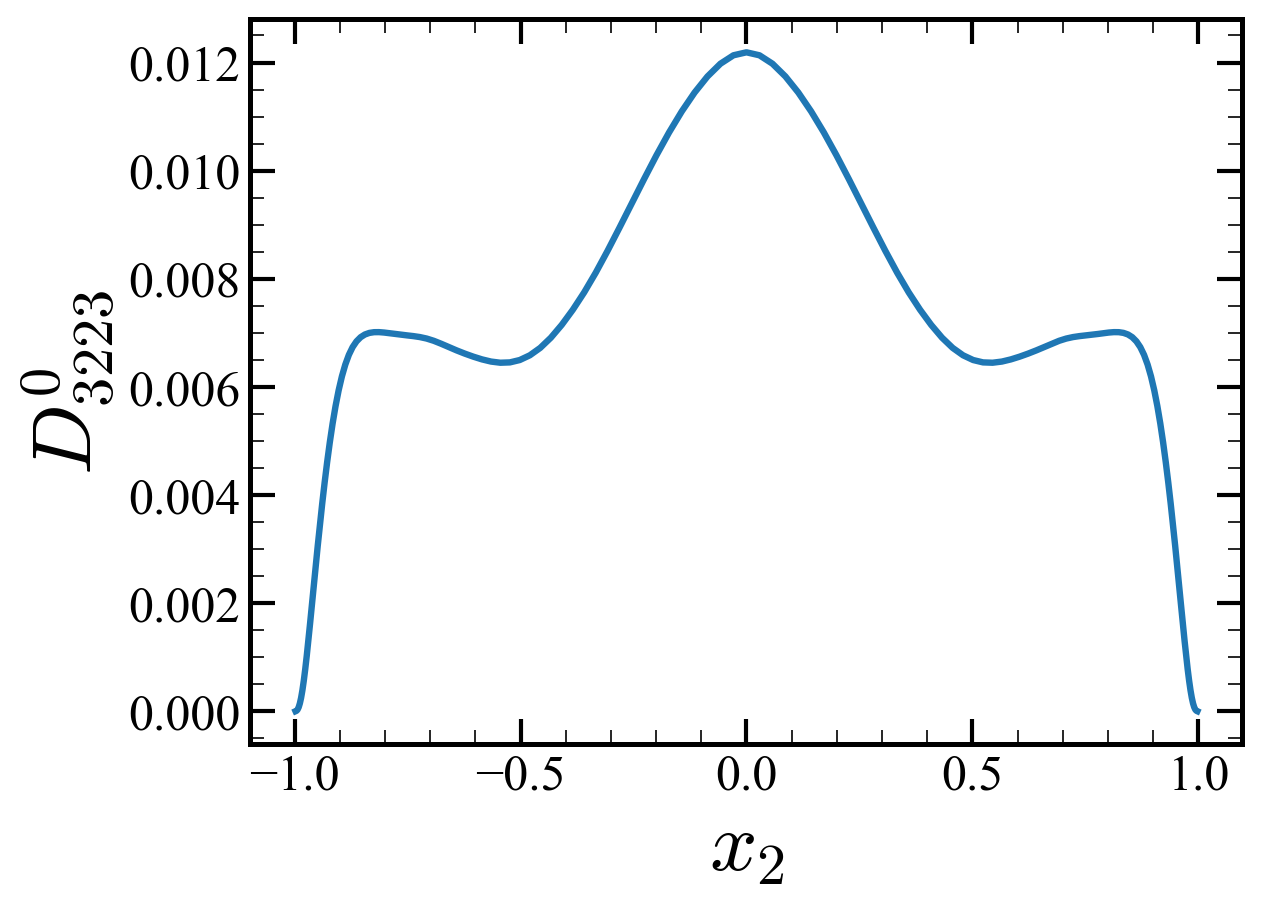}} 
  \caption{Distribution of nonzero $D^0_{ij23}$.}
\label{fig:D0ij23}
\end{figure}

\begin{figure}
\centering
  \subfigure[$D_{1331}$]{\includegraphics[width=0.22\linewidth]{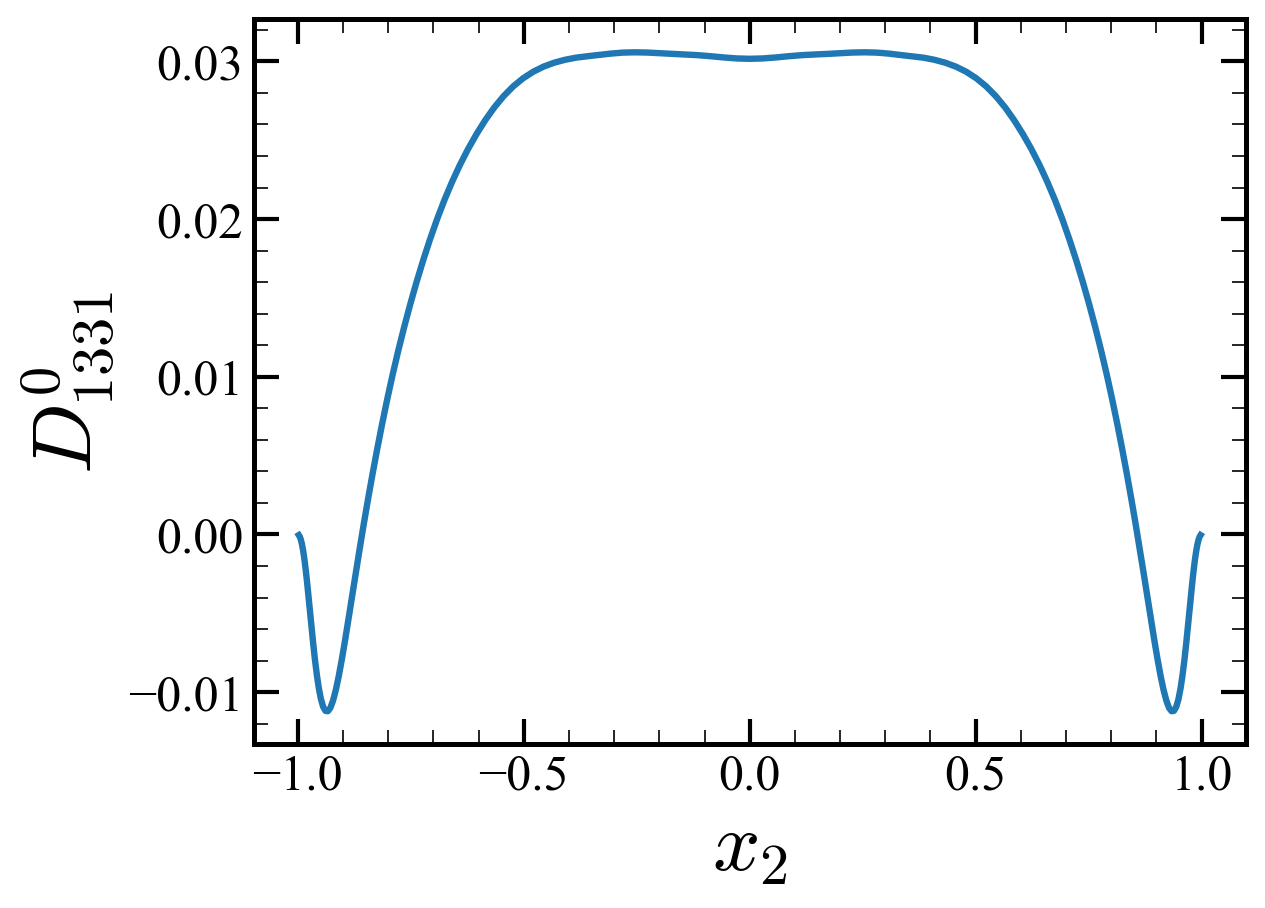}} 
  \subfigure[$D_{2331}$]{\includegraphics[width=0.22\linewidth]{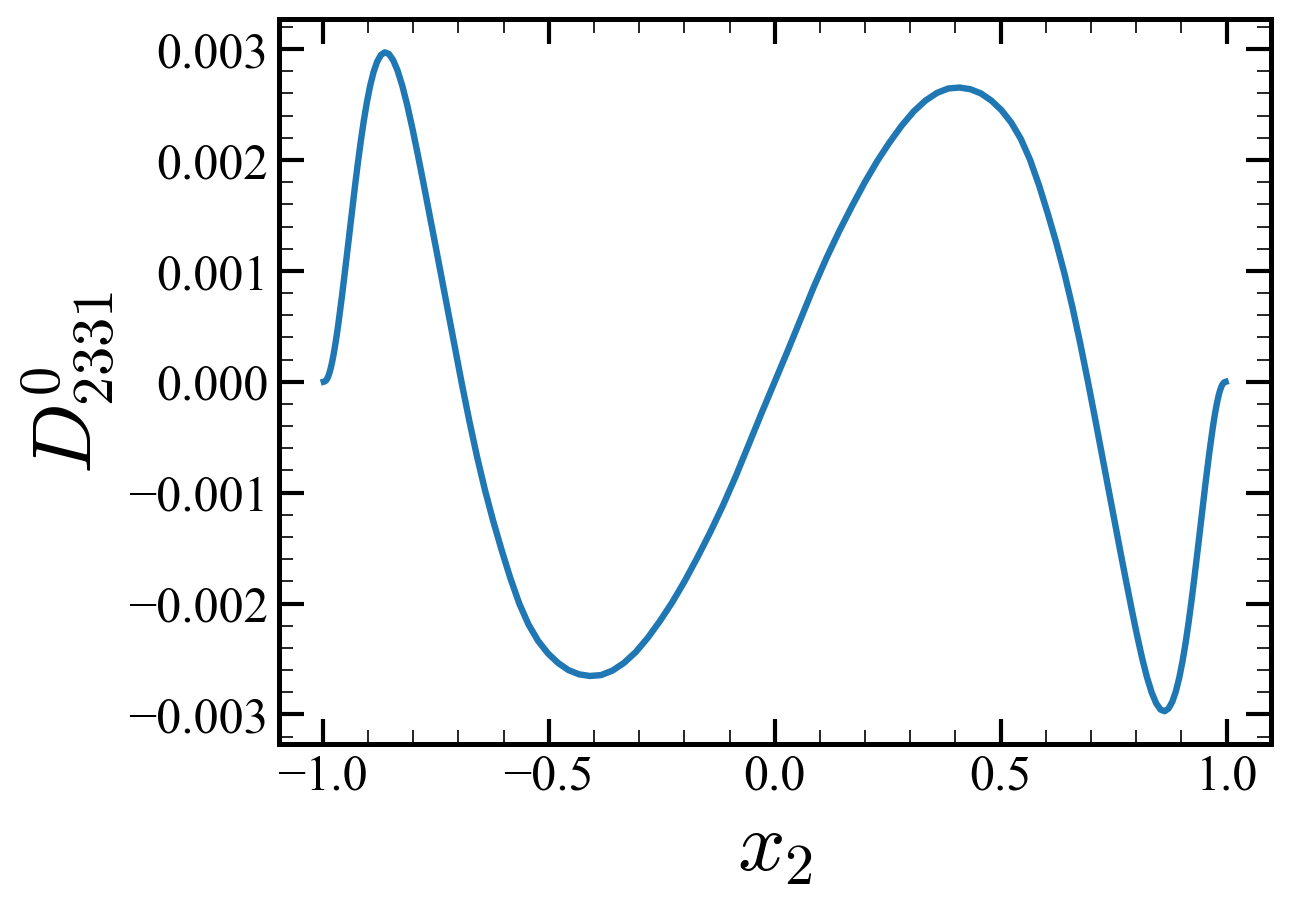}} \\
  \subfigure[$D_{3131}$]{\includegraphics[width=0.22\linewidth]{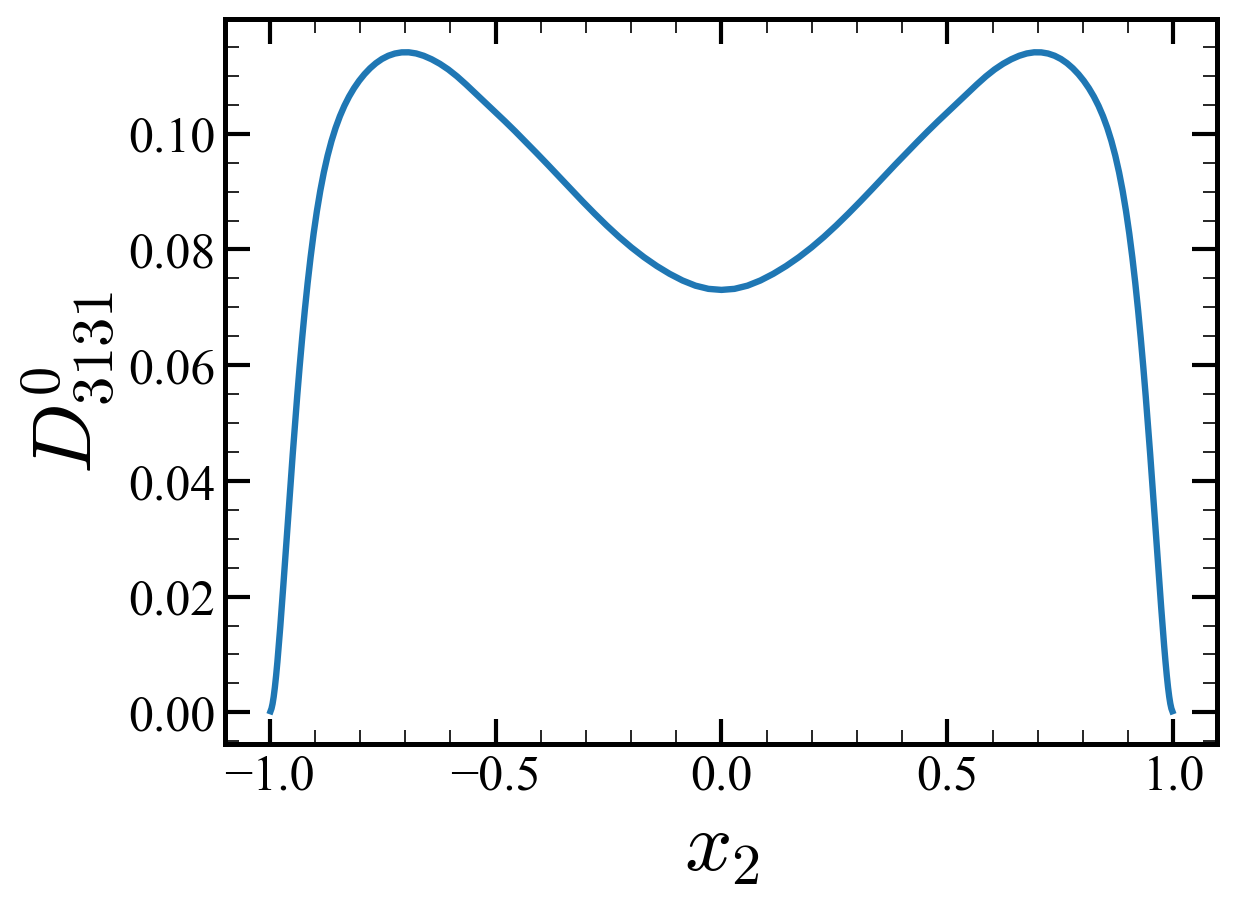}} 
  \subfigure[$D_{3231}$]{\includegraphics[width=0.22\linewidth]{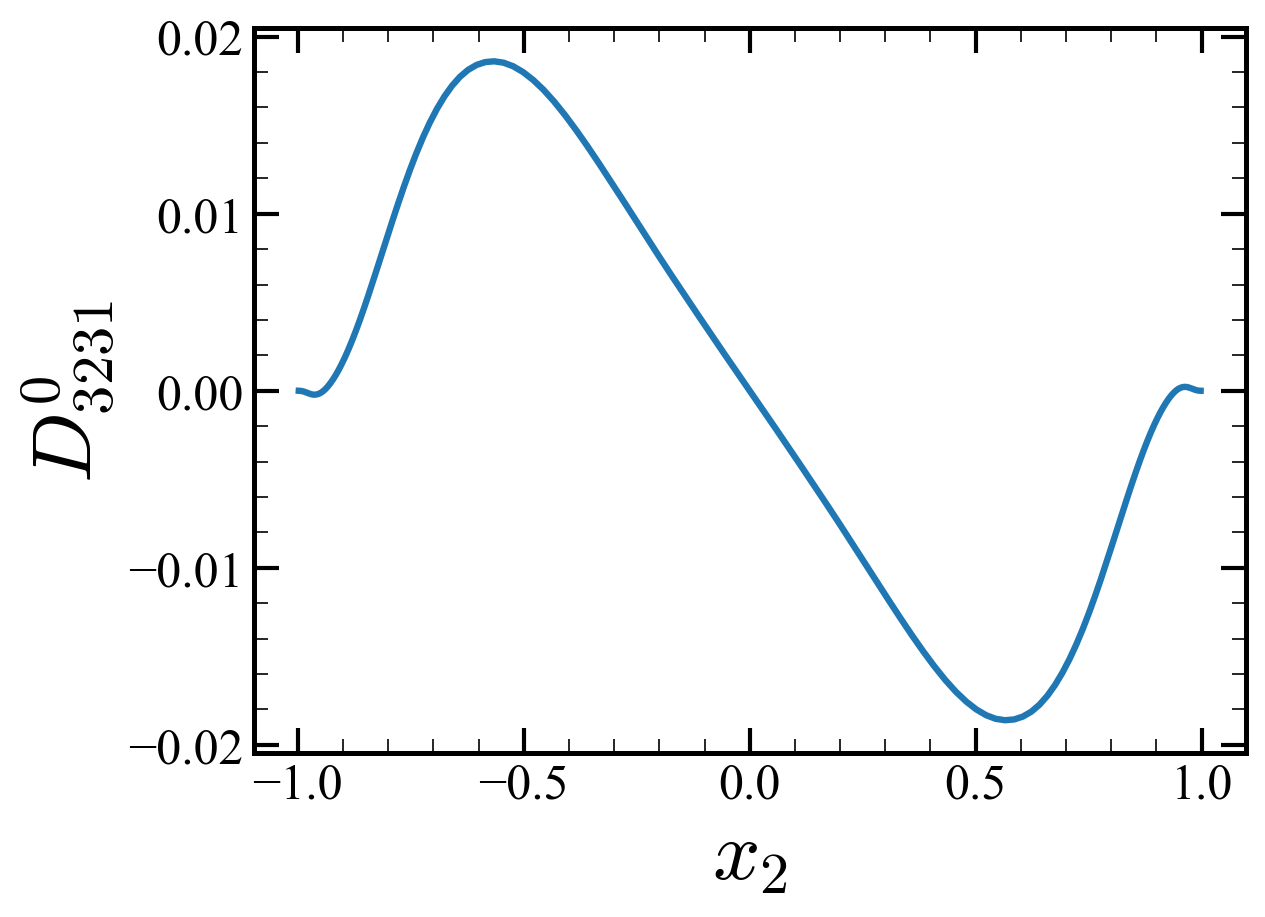}} 
  \caption{Distribution of nonzero $D^0_{ij31}$.}
\label{fig:D0ij31}
\end{figure}

\begin{figure}
\centering
  \subfigure[$D_{1332}$]{\includegraphics[width=0.22\linewidth]{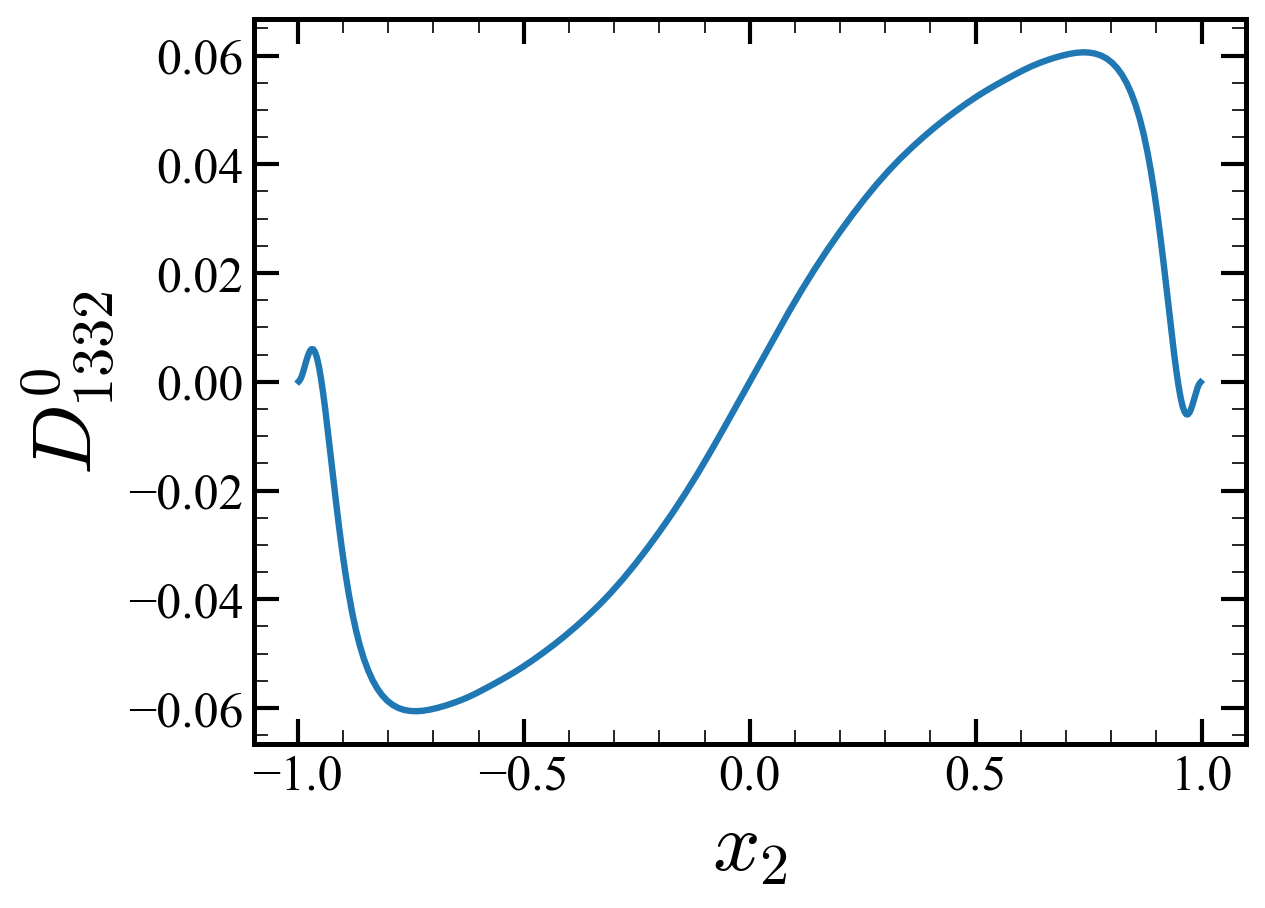}} 
  \subfigure[$D_{2332}$]{\includegraphics[width=0.22\linewidth]{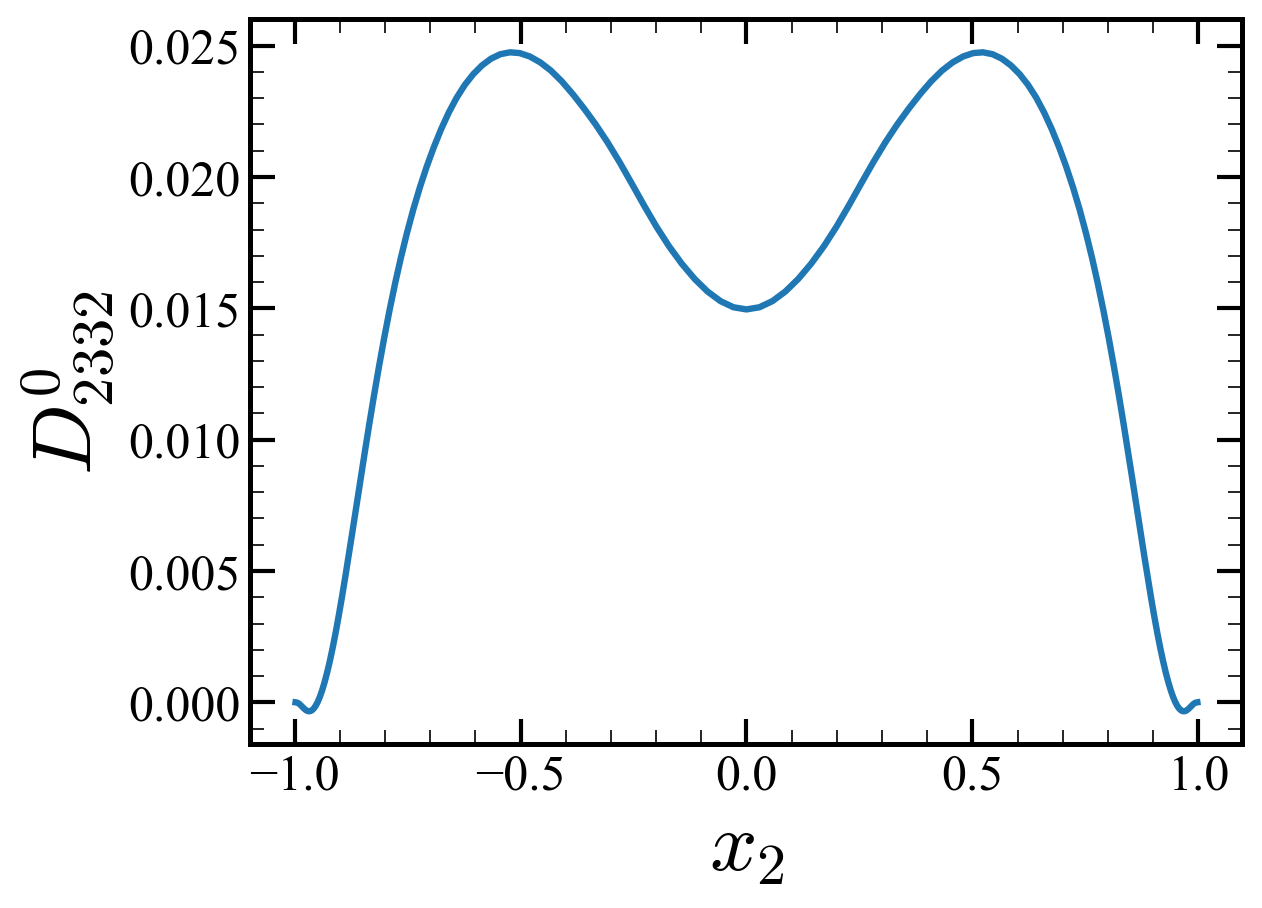}} \\
  \subfigure[$D_{3132}$]{\includegraphics[width=0.22\linewidth]{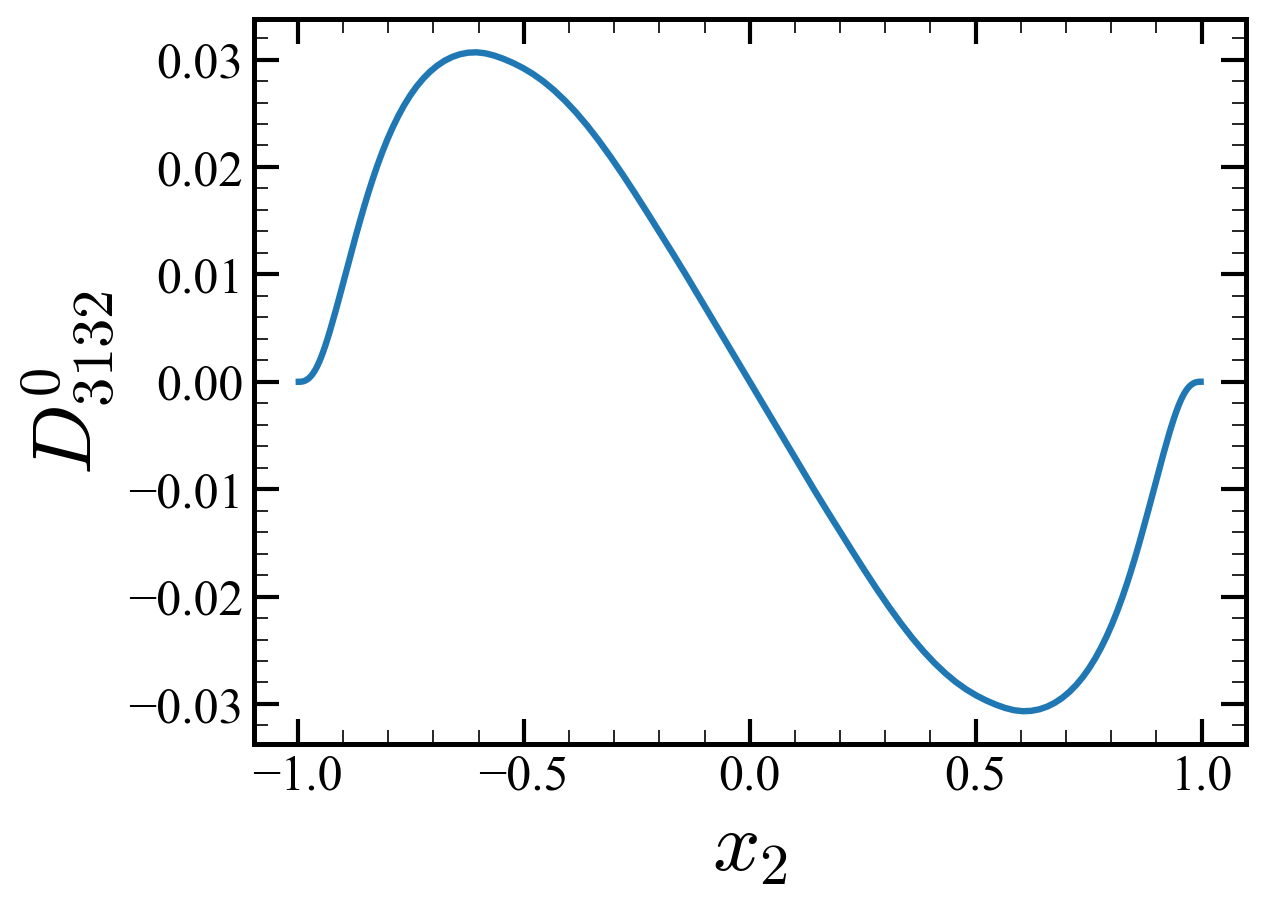}} 
  \subfigure[$D_{3232}$]{\includegraphics[width=0.22\linewidth]{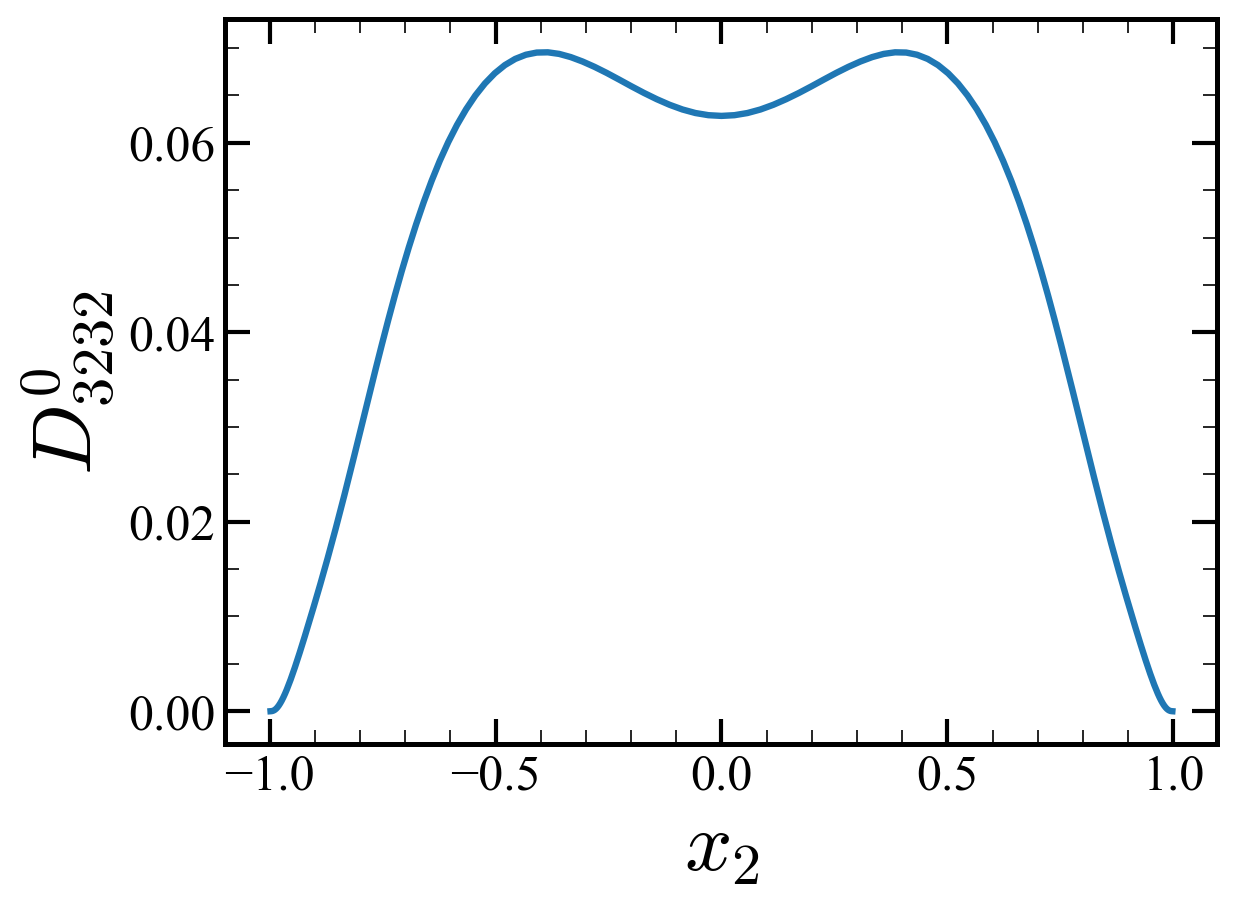}} 
  \caption{Distribution of nonzero $D^0_{ij32}$.}
\label{fig:D0ij32}
\end{figure}

\begin{figure}
\centering
  \subfigure[$D_{1233}$]{\includegraphics[width=0.22\linewidth]{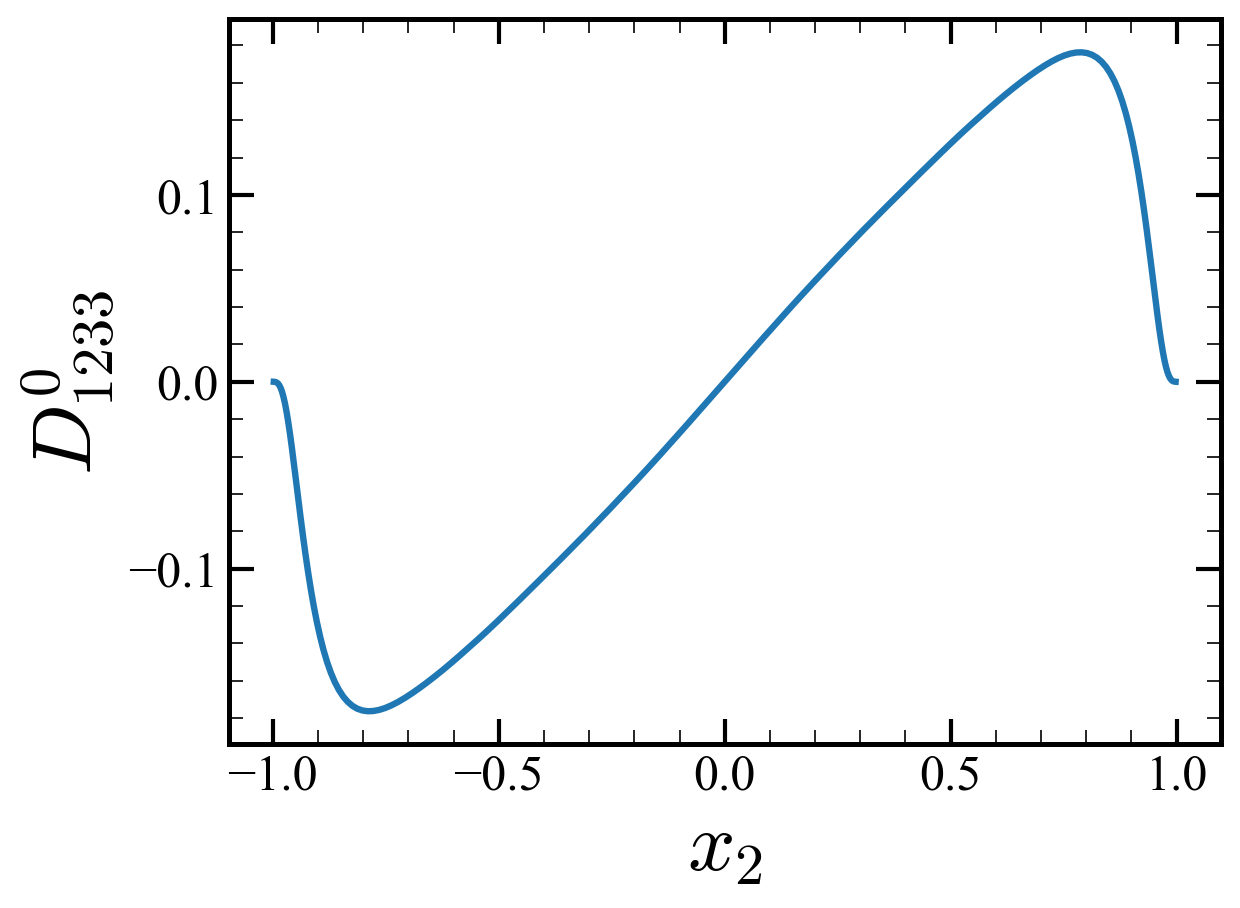}} 
  \subfigure[$D_{2133}$]{\includegraphics[width=0.22\linewidth]{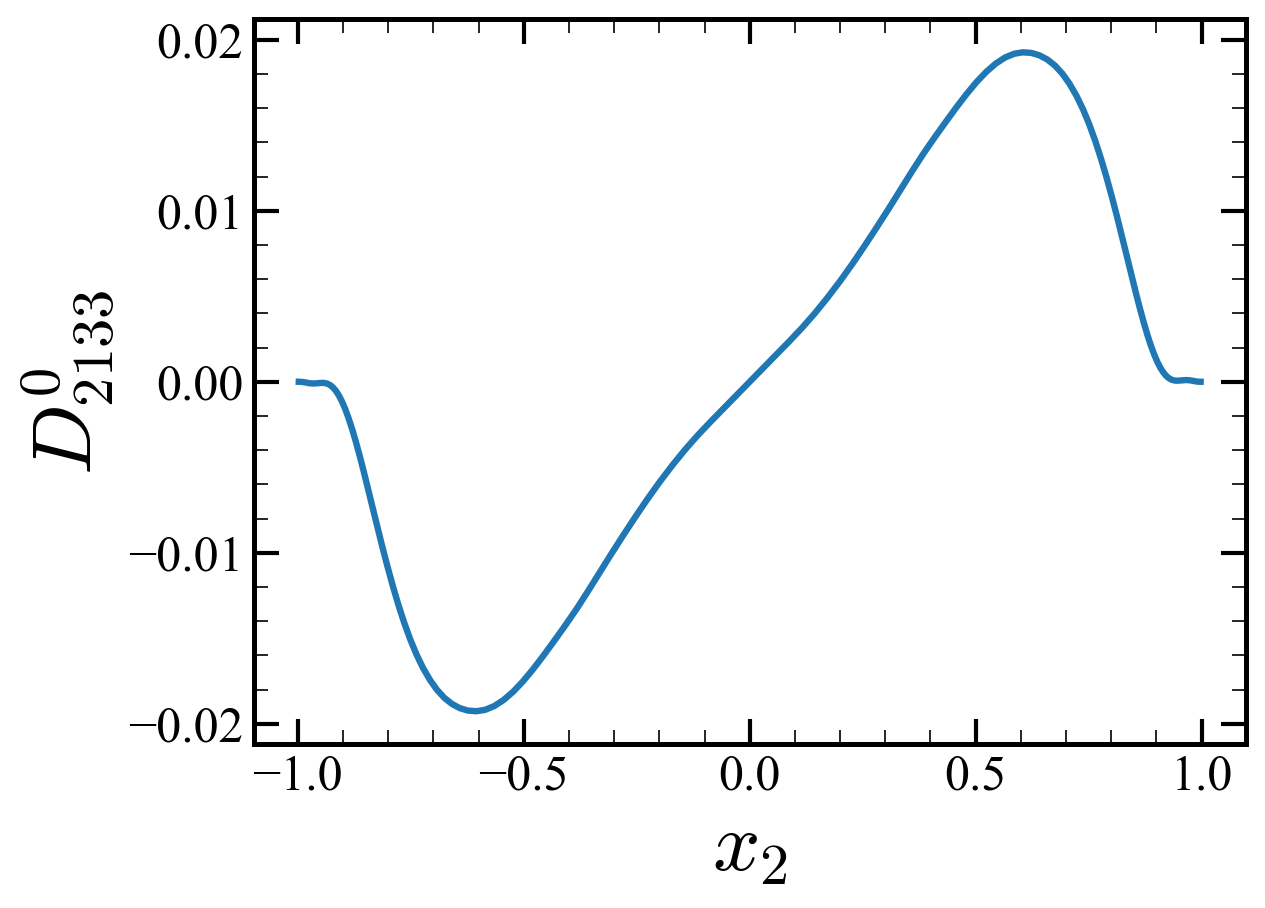}} \\
  \subfigure[$D_{1133}$]{\includegraphics[width=0.22\linewidth]{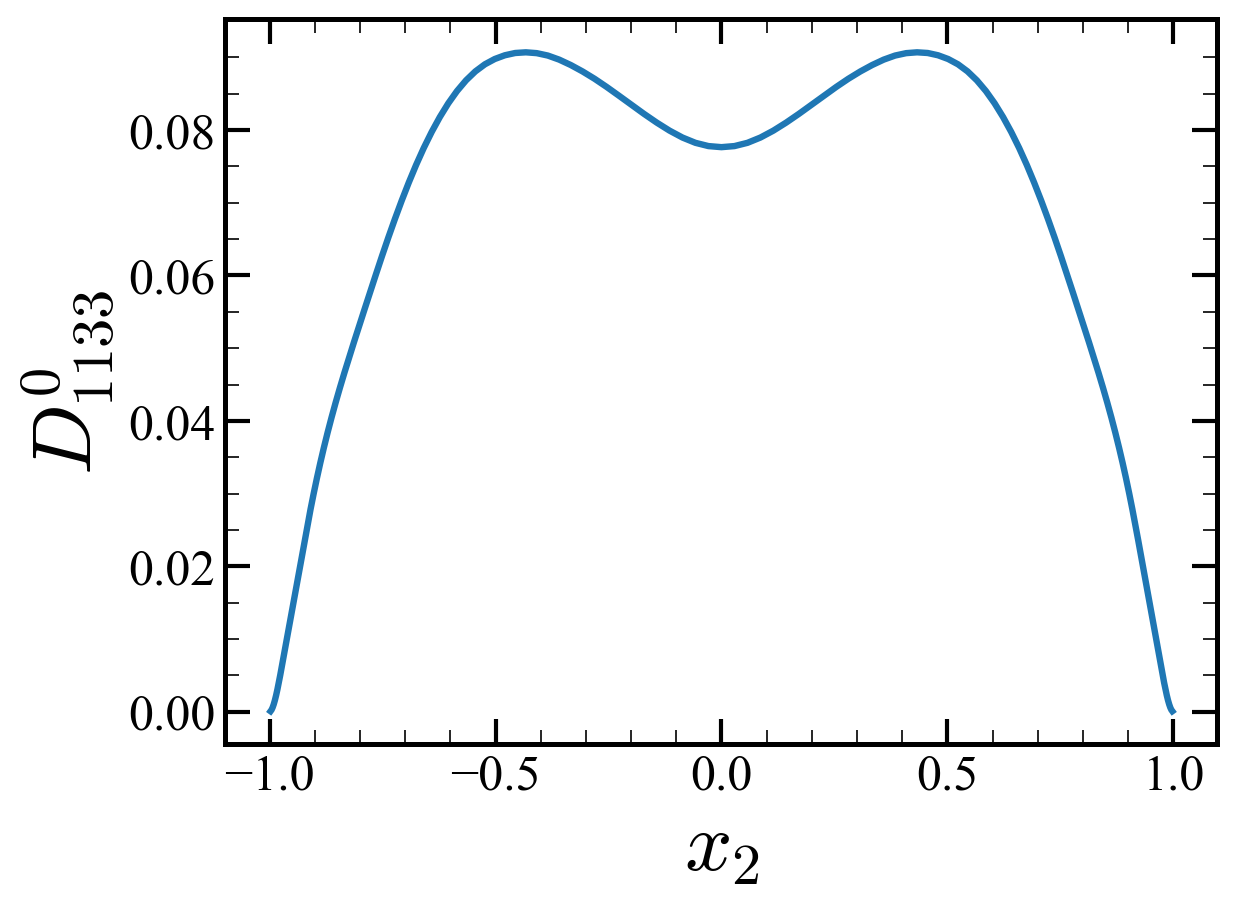}} 
  \subfigure[$D_{2233}$]{\includegraphics[width=0.22\linewidth]{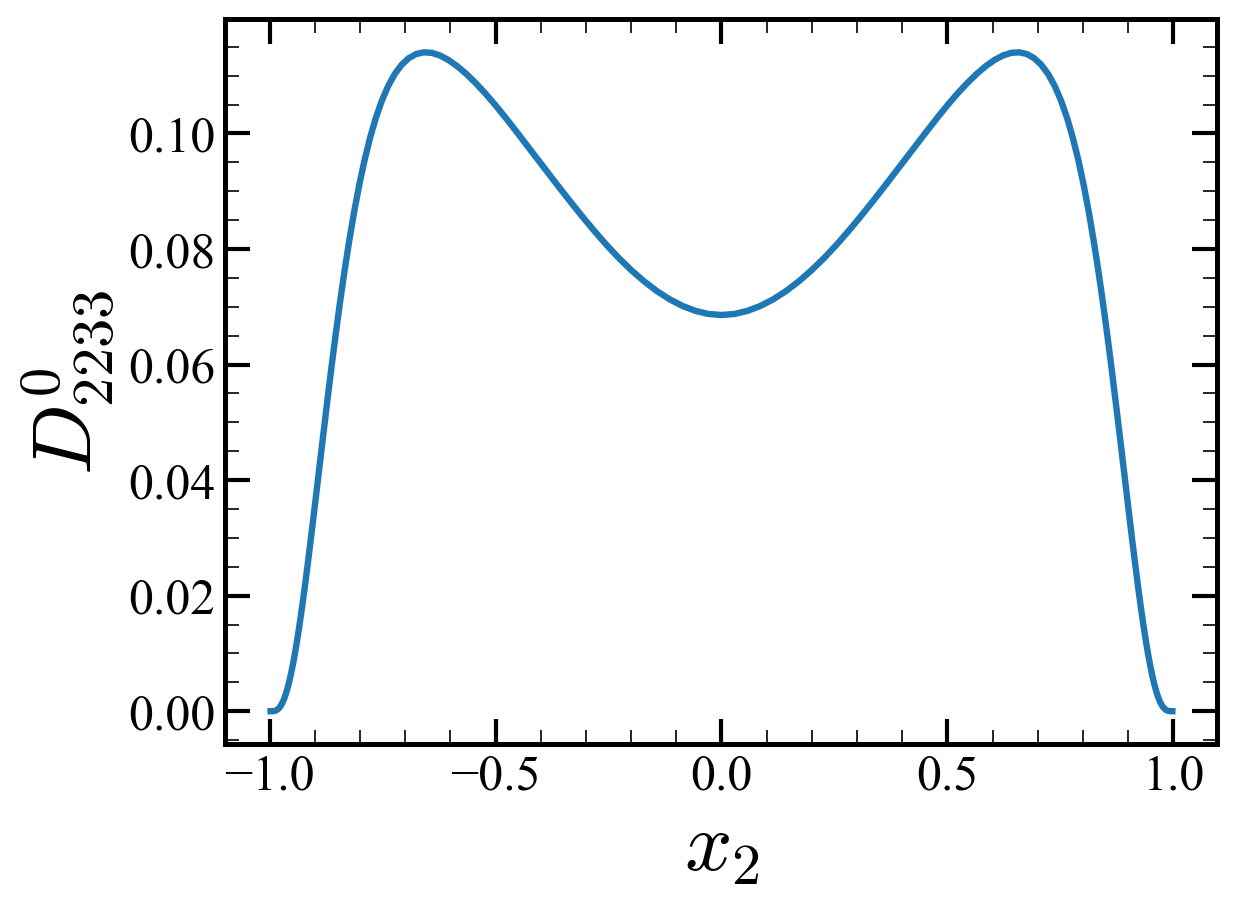}} 
  \subfigure[$D_{3333}$]{\includegraphics[width=0.22\linewidth]{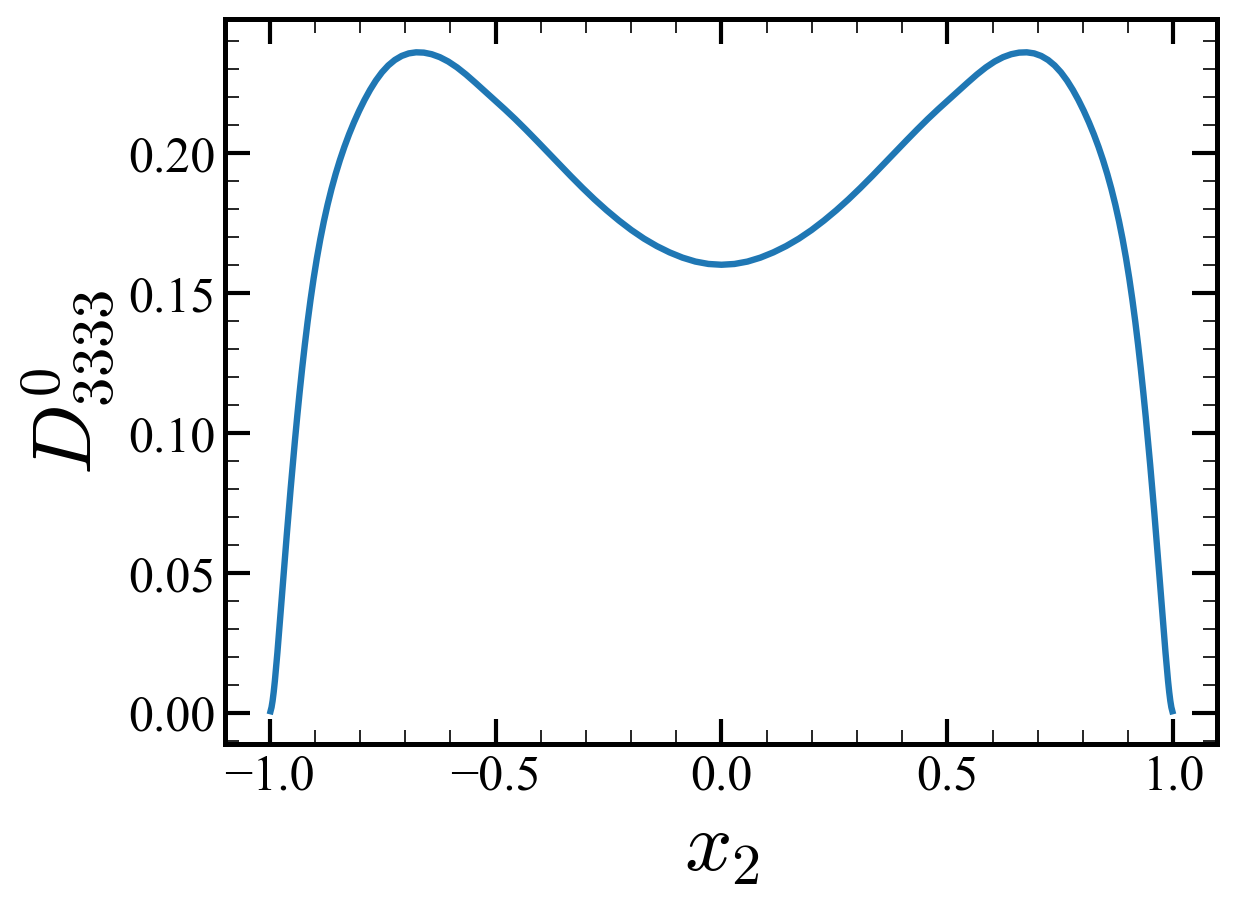}} 
  \caption{Distribution of nonzero $D^0_{ij33}$.}
\label{fig:D0ij33}
\end{figure}

This appendix provides the entire nonzero values of the leading-order eddy viscosity tensor $D^0_{ijkl}$ (Figure~\ref{fig:D0ij11}-\ref{fig:D0ij33}). This corresponds to the eddy viscosity tensor in Equation \ref{eq:leadingApproxGeneral}. The values not shown converge to zero.

\section{Scaling analysis for 2D spatially developing boundary layer}\label{appD}

To determine which components of the eddy viscosity tensor are critical to the RANS of the two-dimensional spatially developing boundary layer, we conduct the following scaling analysis. The flow system with the streamwise length scale $l$ and the wall-normal length scale $d$ is considered where $l\gg d$. Based on the correlation by \citet{White2006}, a typical ratio for $\mathrm{Re}_x \sim O(10^6)$ is $d/l\sim0.02$. Using the leading-order eddy viscosity tensor model, Equation \ref{eq:leadingApproxGeneral}, the Reynolds stress term includes a summation of the eddy viscosity tensor terms. For instance, $\overline{u_2'u_1'}$ is represented as follows:
\begin{eqnarray}
    -\overline{u_2'u_1'}=D_{2111}\frac{\partial U_1}{\partial x_1}+D_{2112}\frac{\partial U_2}{\partial x_1}+D_{2121}\frac{\partial U_1}{\partial x_2}+D_{2122}\frac{\partial U_2}{\partial x_2}
    \label{eq:asymptotics}
\end{eqnarray} 

One needs to consider not only the magnitude of the eddy viscosity tensor element but also the estimated scales of each term in this equation. To evaluate the length scales for the velocity, we set $U_1\sim1$, and the continuity enforces $U_2\sim d/l$. Ignoring $D_{21kl}$ coefficients, the length scales of the four terms on the right-hand side are $1/l$, $d/{l^2}$, $1/d$, and $1/l$. Since $l\gg d$, $D_{2121}$ plays the major role for this Reynolds stress. The next two are the terms that multiply $D_{2111}$ and $D_{2122}$. However, MFM reveals that $D_{2111}$ is one order of magnitude larger than $D_{2122}$. Hence, $D_{2111}$ is the next important eddy viscosity tensor for this Reynolds stress. Likewise, we conducted scaling analysis for all other Reynolds stresses. The analysis informs that $D_{1111}$, $D_{1121}$, $D_{2121}$, and $D_{2221}$ are among the most significant eddy viscosity tensor elements for the case of slowly developing semi-parallel wall-bounded flows.
 
\section{MFM for $D_{ij21}$ measurement}\label{appE}

To compute the eddy viscosity kernel $D_{ij21}\left(x_2,y_2\right)$, we use brute force MFM method using delta function forcing of the velocity gradient at each location. We start from the full kernel eddy viscosity representation in Equation \ref{eq:generalformchannel}. We macroscopically force the mean velocity gradient by ${\partial V_l}/{\partial x_k}=\delta\left(y_2-y_2^*\right)\delta_{k2}\delta_{l1}$, where $\delta(x)$ represents Dirac delta function, $\delta_{ij}$ represents Kronecker delta in index notation, and $y_2^*$ is the probing location of the eddy viscosity. With such forcing, Equation \ref{eq:generalformchannel} becomes the following:
\begin{eqnarray*}
    -\overline{u_i'v_j'}(x_2)&&=\int D_{ijkl}\left({x_2}, {y_2}\right)\left.\frac{\partial V_l}{\partial x_k} \right\vert_{{y_2}}\mathrm{ d}y_2 \\
    &&=\int D_{ijkl}\left({x_2}, {y_2}\right)\delta\left(y_2-y_2^*\right)\delta_{k2}\delta_{l1}\mathrm{ d}y_2 \\
    &&=D_{ij21}\left(x_2,y_2=y_2^*\right)
    \label{eq:bruteforcemethod}
\end{eqnarray*}
Forcing that would maintain the mean streamwise velocity as a Dirac delta function at $y_2=y_2^*$ reveals the eddy viscosity kernel $D_{ij21}\left(x_2,y_2=y_2^*\right)$. By setting $y_2^*$ for all possible locations, we can obtain the eddy viscosity kernel $D_{ij21}$. In numerical implementation of this strategy, instead of dealing with Dirac delta functions, we selected $V_1$ to be a step function with respect to the $y_2$ direction. In discrete space, the MFM is conducted at each discrete point of $y_2^*$ in wall-normal direction with a corresponding heaviside function where the discontinuous point lies at that point. In order to compute the entire kernel $D_{ij21}$, one need to conduct many MFM simulations. More specifically, the number of simulations has to be the number of degree of freedom of the Reynolds-averaged space, i.e. the number of mesh points in the wall-normal direction. For instance, since our RANS space has 144 cell centers, we need 146 MFM simulations, including two for the boundary values.

\bibliographystyle{jfm}
\bibliography{papers}

\end{document}